\documentclass[onecolumn,prd,nofootinbib]{revtex4}
\usepackage{epsf}
\usepackage{dcolumn}
\usepackage{bm}
\usepackage{graphicx}
\def\aj{AJ}%
\def\araa{ARA\&A}%
\def\apj{ApJ}%
\def\apjl{ApJ}%
\def\apjs{ApJS}%
\def\aap{A\&A}%
\def\aaps{A\&AS}%
\def\mnras{MNRAS}%
\def\pasp{PASP}%
\def\nat{Nature}%
\newcommand{\be}{\begin{equation}}
\newcommand{\ee}{\end{equation}}

\def\4he{$^4$He}
\def\3he{$^3$He}
\def\7li{$^7$Li}

\begin{document}

\title{Star formation histories of dwarf galaxies from the Colour-Magnitude
diagrams of their resolved stellar populations}

\author{Michele Cignoni$^{1,2}$ and Monica Tosi$^{2}$\\} \affiliation{
  $^1$Astronomy Department, Bologna University, Via Ranzani 1, I-40127
  Bologna, Italy} \affiliation{$^2$INAF-Osservatorio Astronomico di
  Bologna, Via Ranzani 1, I-40127 Bologna, Italy}

\date{{\today}\\}

\begin{abstract}
In this tutorial paper we summarize how the star formation (SF)
history of a galactic region can be derived from the colour-magnitude
diagram (CMD) of its resolved stars. The procedures to build synthetic
CMDs and to exploit them to derive the SF histories (SFHs) are
described, as well as the corresponding uncertainties. The SFHs of
resolved dwarf galaxies of all morphological types, obtained from the
application of the synthetic CMD method, are reviewed and
discussed. In short: 1) Only early-type galaxies show evidence of long
interruptions in the SF activity; late-type dwarfs present rather
continuous, or {\it gasping}, SF regimes; 2) A few early-type dwarfs
have experienced only one episode of SF activity concentrated at the
earliest epochs, whilst many others show extended or recurrent SF
activity; 3) No galaxy experiencing now its first SF episode has been
found yet; 4) No frequent evidence of strong SF bursts is found; 5)
There is no significant difference in the SFH of dwarf irregulars and
blue compact dwarfs, except for the current SF rates. Implications of
these results on the galaxy formation scenarios are briefly discussed.
\end{abstract}



\maketitle

\section{~Introduction}
\label{intro}
Dwarf galaxies are the most diffused type of galaxies in the Universe
and were probably even more numerous in the past, when they might have
contributed to the population of blue systems overabundant in deep
galaxy counts \citep[e.g.][]{Lilly95,Babul96} and more likely to the
assembling of larger baryonic systems. In spite of having received
less attention than spiral and elliptical galaxies, dwarf galaxies
have probably more cosmological relevance.  For instance, late-type
dwarfs are the preferred targets for cosmologists interested in Big
Bang Nucleosynthesis, because their low metal and helium contents
allow the derivation of the primordial helium abundance from HII
regions spectra with minimum extrapolation
\citep[e.g.][]{Peimbert74,Olive97,Izotov98}. Moreover, their low metallicity and
high gas content make them apparently less evolved than spirals and ellipticals,
thus more similar to what primeval galaxies may have been.

One of the main cosmological interests is related to the possibility that 
today's dwarfs are the survivors of the building blocks of massive
galaxies.  Cold Dark Matter (CDM) cosmology predicts that dwarf 
systems are the first ones to form after the Big Bang, since only dark
matter halos of mass smaller than 10$^8 M_{\odot}$ are able to condense 
from primordial density perturbations. In this
framework, more massive systems are assembled by subsequent merging of these 
protogalactic fragments 
\citep[the hierarchical formation scenario; e.g.][]{White78,Frenk88}, and 
dwarfs have a pivotal role in 
the evolution of massive galaxies. 

Observations do show that galaxies merge in the local Universe and
that big galaxies accrete their satellites. We know the cases of the
Magellanic Stream and of other streams connected to the Sagittarius
dwarf spheroidal (dSph) and other satellites falling on the Milky Way
\citep[e.g.][]{Bellazzini03,Belokurov07}. Andromeda is quite similar
in this respect \citep[e.g.][]{Ferguson02,Ibata04,Ferguson05}, with
streams and clumps just as, or even more than, our own Galaxy.  The
question is whether big galaxies form only by successive mergers of
smaller building blocks, as proposed by the hierarchical formation
scenario, or satellite accretion is a frequent but not necessary and
dominant event, consistent with a downsizing formation scenario.
Downsizing \citep{Cowie96} in principle does not concern the hierarchy
or the epoch of galaxy formation, it simply reflects the observational
evidence that the bulk of stars in more massive galaxies formed
earlier and at a higher rate than those in less massive systems. If
mechanisms are found allowing for these star formation properties in
the bottom-up scenario \citep[e.g.][]{Mouri06,Neistein06,Cattaneo08},
then downsizing is not incompatible either with CDM or with the
hierarchical scenario.  However, downsizing is often seen as the
alternative to hierarchical formation, replacing in this role the
monolithical scenario, where each galaxy forms from the collapse
(dissipative or dissipationless) of its protogalactic gas cloud. In
the monolithical scenario more massive galaxies form much earlier than
less massive ones for simple gravitational arguments \citep{Eggen62},
with timescales for the collapse of the protogalactic cloud originally
suggested to be of the order of 100 Myr, and now more often considered
of the order of 1 Gyr.

One of the effective ways to check whether or not big galaxies are
made {\it only} by successive accretions of satellites like the
current ones is to observe the resolved stellar populations of massive
and dwarf systems and compare their properties with each other.  If
chemical abundances, kinematics and star formation histories of the
resolved stars of massive galaxies are all consistent with those of
dwarf galaxies, then the former can be the result of successive
merging of the latter; otherwise, either satellite accretion is not
the only means to build up spirals and ellipticals or the actual
building blocks are not alike today's dwarfs.

An updated review of the chemical, kinematical and star formation
properties of nearby dwarfs can be found in \cite{Tolstoy09}.  In this
tutorial paper we describe how the star formation history of a
galactic region can be derived from the colour-magnitude diagram of
its resolved stars, and we summarize what people have learnt on the
SFHs of dwarf galaxies from the application of the most popular
approach based on the CMD. In Section II we introduce the method; in
Section III we describe in detail procedures and uncertainties; and in
Section IV we report on the results of its application on the SFH of
dwarf galaxies.  A discussion on how these results may affect our
understanding of galaxy evolution is presented in Section V.

\section{Star formation histories from Colour-Magnitude Diagrams: the method}

The need for understanding the star formation histories of dwarf
galaxies was recognized long ago and over the years many approaches
have been followed to infer the SFH of different galaxies. Among the
many studies devoted to the field, we recall the seminal paper by
\cite{Searle73} and the extensive and detailed analyses performed by
Gallagher, Hunter and collaborators \citep[][and references
  therein]{Gallagher84a,Gallagher84b}, who used various indicators to
estimate the star formation rates at different epochs of large
samples of dwarfs.  The quantum leap in the field occurred two decades
ago, when the power and resolution of new generation telescopes and
detectors allowed people to resolve and measure individual stars even
in the crowded fields of external galaxies and to draw their CMDs.
The CMD of a stellar system is in fact the best information desk on
the system evolution, because it preserves the imprinting of all the
relevant evolution parameters (age, mass, chemical composition,
initial mass function).

Twenty years ago stellar age dating was done with isochrone fitting,
a handy tool for simple stellar populations, such as those of stellar
clusters, but inadequate to interpret the composite populations of
galaxies, where many subsequent generations of stars, with possibly
different initial mass function, metallicity, reddening and distance,
contribute to the morphology of the observational CMD.  With CCD
detectors and new reduction packages for PSF fitting photometry
allowing for the first time to measure accurately individual stars in
Local Group (LG) galaxies, the time had come to develop a reliable tool to
quantitatively derive their SFHs. The best tool is based on CMDs, and
is the extrapolation of the standard isochrone fitting method to the
complicated CMDs of composite stellar populations: the synthetic CMD
method.

\subsection{Building a synthetic population}

The synthetic CMD method allows us to derive all the SFH parameters
within the lookback time reached by the available photometry. To do
this, a synthesizer works with classical ingredients:

\begin{itemize}

\item Star formation law and rate, SFR(t), which regulate the astrated mass at each time t;

\item Initial mass function (IMF), which gives the number N of stars in
  each generation per unit stellar mass interval.  A useful form
  is a power-law
\begin{eqnarray}  
&&dN\propto M^{-\alpha}dM.\label{imf}
\end{eqnarray}
The IMF is usually assumed to be independent of time;

\item Chemical enrichment: due to the galaxy chemical evolution, the
  metallicity of the gas from which stars form changes with time.
  This is described by an age-metalicity relation (AMR) Z(t);

\item Stellar evolution tracks, giving the temperature and
  luminosity of stars of given mass and metallicity at any 
  time after their birth;

\item Stellar atmosphere models, to transform the bolometric magnitudes and temperatures into the observational plane;

\item Binary fraction and mass ratio.

\end{itemize}
\begin{figure}
\centering
\includegraphics[width=17cm]{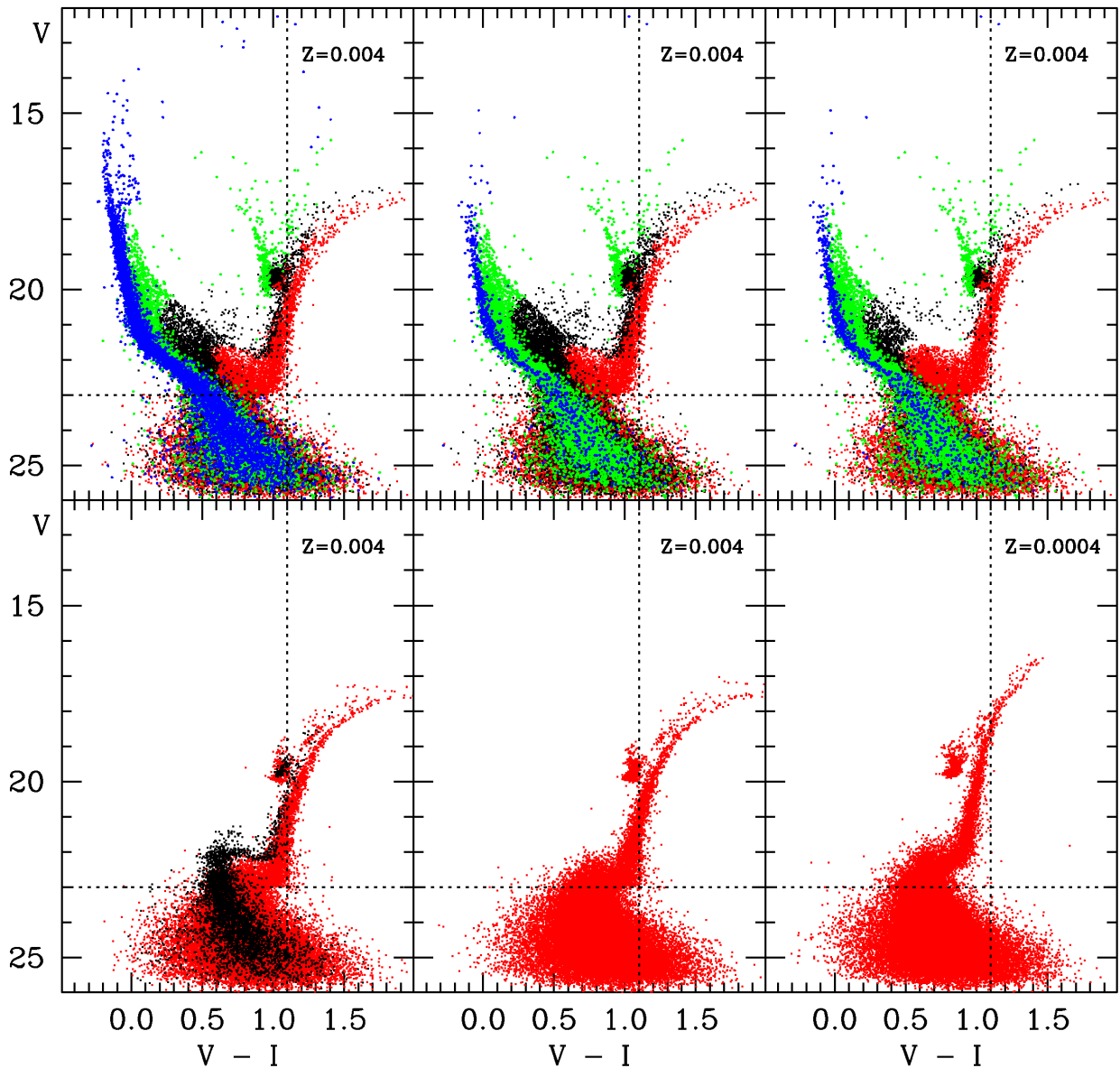}
\caption{The effect of the SFH on the theoretical CMD of a
  hypothetical galactic region with (m-M)$_0$=19, E(B-V)=0.08, and
  with the photometric errors and incompleteness typical of HST/WFPC2
  photometry. All the shown synthetic CMDs contain 50000 stars and are
  based on the Padova models \citep{Fagotto94a,Fagotto94b} with the
  labelled metallicities.  Top-central panel: the case of a SFR
  constant from 13 Gyr ago to the present epoch.  Top-left panel: the
  effect of adding a burst 10 times stronger in the last 20 Myr to the
  constant SFR. The CMD has a much brighter and thicker blue plume.
  Top-right panel: same constant SFR as in the first case, but with a
  quiescence interval between 3 and 2 Gyrs ago; a gap appears in the
  CMD region corresponding to stars 2-3 Gyr old, which are completely
  missing.  Bottom-central panel: SF activity only between 13 and 10
  Gyr ago with Z=0.004.  Bottom-right panel: SF activity only between
  13 and 10 Gyr ago with Z=0.0004: notice how colour and luminosity of
  turnoff, subgiant and red giant branches differ from the previous
  case.  Bottom-left panel: SF activity between 13 and 11 Gyr ago,
  followed by a second episode of activity between 5 and 4 Gyr ago: a
  gap separates the two populations in the CMD, but less evident than
  in the top-right panel case, when the quiescent interval was more
  recent.  }
\label{syn1}       
\end{figure}

The standard procedure is the following. Using a Monte Carlo
algorithm, masses and ages are extracted according to the IMF and the
SF law (e.g. constant or exponentially decreasing with time,
proportional to some power of the gas density, etc.). The metallicity
follows suitable AMRs. The extracted synthetic stars are placed in the
CMD by interpolation among the adopted stellar evolution tracks of the
assumed metallicity.  In order to take into account the presence of
unresolved binary stars, a chosen fraction of stars are assumed to be
binaries and coupled with a companion star. The fake population is put
at the distance of the galaxy we want to analyze, simultaneously
correcting for reddening and extinction. Finally, photometric errors,
incompleteness and blending factors, as accurately estimated from
artificial star tests on the actual photometric images of the examined
field, are applied to the synthetic CMD.

  Different combinations of the parameter choices provide the whole
  variety of CMDs observed in real galaxies.  As an example,
  Fig.\,\ref{syn1} shows the CMDs resulting from 6 representative,
  although simplistic, cases.

The six panels of  Fig.\,\ref{syn1} illustrate the
effect of different SFHs on the synthetic CMD of a hypothetical galactic region with 
number of resolved individual stars, photometric errors, blending 
and incompleteness factors typical of a region in the SMC 
imaged with HST/WFPC2. The top three
panels show examples of CMDs typical of late-type galaxies, with ongoing or
recent star formation activity. 
If the SFR has been constant for all the galaxy
lifetime, the CMD of the region is expected to have the morphology of the
top-central panel, with a prominent blue plume mostly populated by main-sequence
(MS) stars and an equally prominent red plume resulting from the overposition
of increasingly bright and massive stars in the red giant branch (RGB), 
asymptotic giant branch (AGB) and red supergiant phases. 
At intermediate colours, for decreasing brightness, stars in the 
blue loops, red clump and subgiant phases are visible, as well stars at
the oldest MS turnoff (MSTO) and on the faint MS of low mass stars. Stars of all ages
are present, from those as old as the Hubble time to the brightest ones a few
tens Myr old.

If we leave the SFH unchanged except for the addition of a burst ten
times stronger concentrated in the last 20 Myr, the CMD (top-left
panel) has a much brighter and more populated blue plume, now
containing many stars a few Myr old.  In the top-right panel the same
constant SFR as in the first case is assumed, but with a quiescent
interval between 3 and 2 Gyrs ago: a gap is clearly visible in the CMD
region corresponding to isochrones with the age of the missing stars.

The three bottom panels of Fig.\,\ref{syn1} show CMDs typical of
early-type galaxies, whose SF activity is concentrated at earlier
epochs. If only one SF episode has occurred from 13 to 10 Gyr ago,
with a constant metallicity Z=0.004 as in the top-panel cases, the
resulting CMD is shown in the bottom-central panel. If the SF has
occurred at the same epoch, but with a metallicity ten times lower,
the evolutionary phases in the resulting CMD (bottom-right panel) have
colours and luminosities quite different from the previous
case. Finally, the bottom-left panel shows the case of two bursts, the
first from 13 to 11 Gyr ago and the second from 5 to 4 Gyr ago. The
gap corresponding to the quiescent interval is visible in the CMD,
although not as much as the more recent gap of the top-right panel.

Once a synthetic CMD is built, the challenging part of the method is
the comparison with the observational CMD. The best values of the
parameters (IMF, AMR, SFR, binary fraction, reddening and distance
modulus) are found by selecting the cases providing synthetic CMDs
with morphology, colours, number of stars in the various evolutionary
phases and luminosity functions consistent with the observed
ones. Independently of the specific method, any approach is
unavoidably statistical and does not provide a unique solution for the
SFH of the examined region.  Nevertheless, the synthetic CMD method is
quite powerful, thanks to the wealth of independent constraints
available in a good CMD, and it strongly reduces the range of possible
scenarios.

In the following, we will describe what are the major uncertainties in the
method. Before diving into details, however, it is crucial to understand which 
parts of a CMD are the most reliable clocks.

\subsection{Stellar ages from a CMD}

What can be learnt from a CMD? All the evolutionary sequences are
witness of the same SFH, but some sequences are specifically sensitive to
age more than to any other ingredient (e.g. metallicity,
convection, etc..). In order to track the history of a galaxy, it is
necessary to select the safest age indicators. Because
different evolutionary phases populate different CMD regions, one must then
know which parts of the CMD are more informative.

The best indicators share a useful feature: the age is related to the
luminosity, which depends on the burning rate and on the available
fuel. The MS is the archetype of this class of phases, since in this
stage the stars obey a mass-luminosity relation $L\propto M^{n}$
(where n varies from 3 to 4). This relationship has strong
implications. Adopting a mean value $n=3.5$ and considering that the
available fuel is proportional to the stellar mass M, the time spent
in MS is proportional to $M^{-2.5}$: massive stars live for short
times (50 Myr for a $8\,M_{\odot}$), mapping \emph{only} the recent
SFH, while objects smaller than $1.5\,M_{\odot}$ can survive for many
Gyr, mapping the recent as well as the ancient star formation
history. From the point of view of the CMD, this one-to-one
correspondence of luminosity and mass/age allows, for a given
metallicity, a direct conversion of the MS information into the SFH.

Beyond the MS a mass-luminosity relation does not hold anymore, and
the luminosity is rather sensitive to the core mass growth. The phases
between the MS and the red giant are so fast (thermodynamical evolution)
that the probability of observing their stars is low (compared to that for 
nuclear phases). This causes the so-called Hertzsprung gap, i.e. the observed
lack (or paucity) of stars in the evolutionary phase right after the
MS. However, for stars smaller than about $2-2.5\, M_{\odot}$, the evolutionary
times are long enough (because the degeneracy pressure prevents a
rapid core contraction) to define another useful age indicator: the
\emph{sub-giant branch} (SGB). Like the MS turn-off, the SGB fades as 
the the age increases.

Later evolutionary phases, namely the RGB, the horizontal branch (HB),
the red clump (RC) and the asymptotic giant branch, with the exception
of the blue loop (BL) stage, are questionable age indicators. In fact,
the CMD position of such objects mostly reveals the age through the
color, which can be influenced by several factors. As an example,
aging the stars makes the RGB redder, but theoretical uncertainties
like the color transformations and the super-adiabatic convection can
lead to higher color shifts. The RGB is undoubtedly more sensitive to
the metallicity. On the other hand, all these phases are indisputable
signature of stars older than a limiting age and extremely useful in
the age dating of galaxies too distant to have the MSTO reachable by
any photometry: HB stars are always older than 10 Gyr, RGBs at least
1-2 Gyr old, and AGBs older than 100 Myr. Moreover, as thoroughly
discussed by Greggio \citep{Greggio02}, the relations existing between
the number counts of post-MS stars and their mass helps in
constraining the SFH.

Among core helium burning stars, the HB and the RC phases are composed
by stars of initial mass smaller than about 2 solar masses, whose luminosity 
depends on the helium core mass and is quite independent of the stellar
mass. In particular, the HB color frequently shows a correlation with
metallicity (the ``first parameter''), while age is only one
of the possible secondary parameters. Quite different is the behavior
of intermediate mass stars (over $2\, M_{\odot}$): during the core
helium burning these objects describe a large loop in color (the so-called
blue loops) and their luminosities are critically sensitive to the
mass; this is because the core mass is connected to the extension of
the convective cores in the previous MS structures.  Thanks to this
feature, for ages of 100-500 Myr, the luminosity of the loops fades
with age and the BL is an excellent age indicator.

After the core helium burning phase, low and intermediate mass stars
experience the AGB phase. As for the RGB, a relation between the
luminosity and the core mass holds. In addition, many phenomena occur
(e.g. mixing and extra mixing of convective layers, thermal pulses,
etc.) which are not yet understood in detail, and this leaves
considerable uncertainty.

\begin{figure}[]
\includegraphics[width=8.cm,height=8cm]{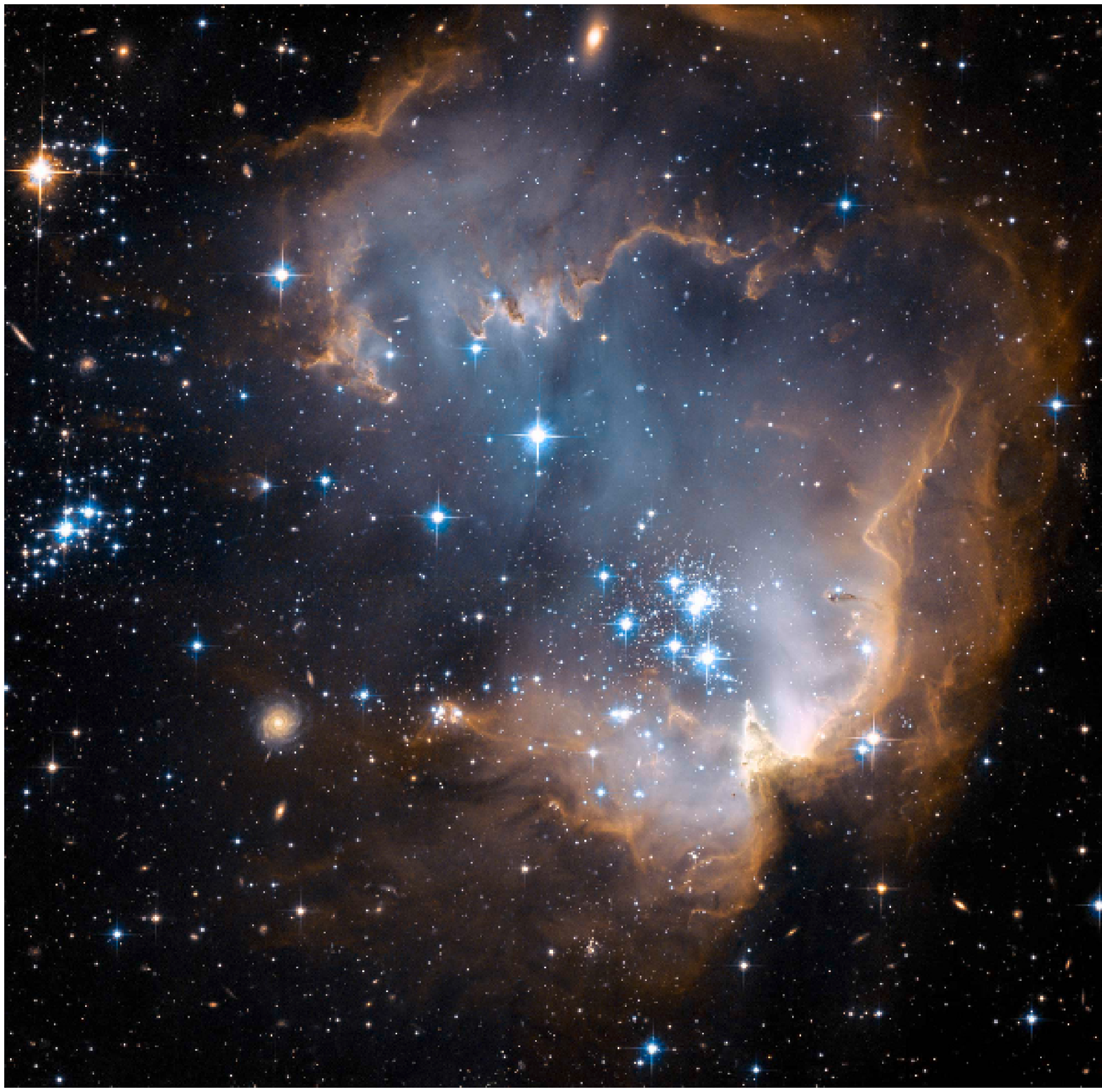}
\includegraphics[width=8.cm,height=8cm]{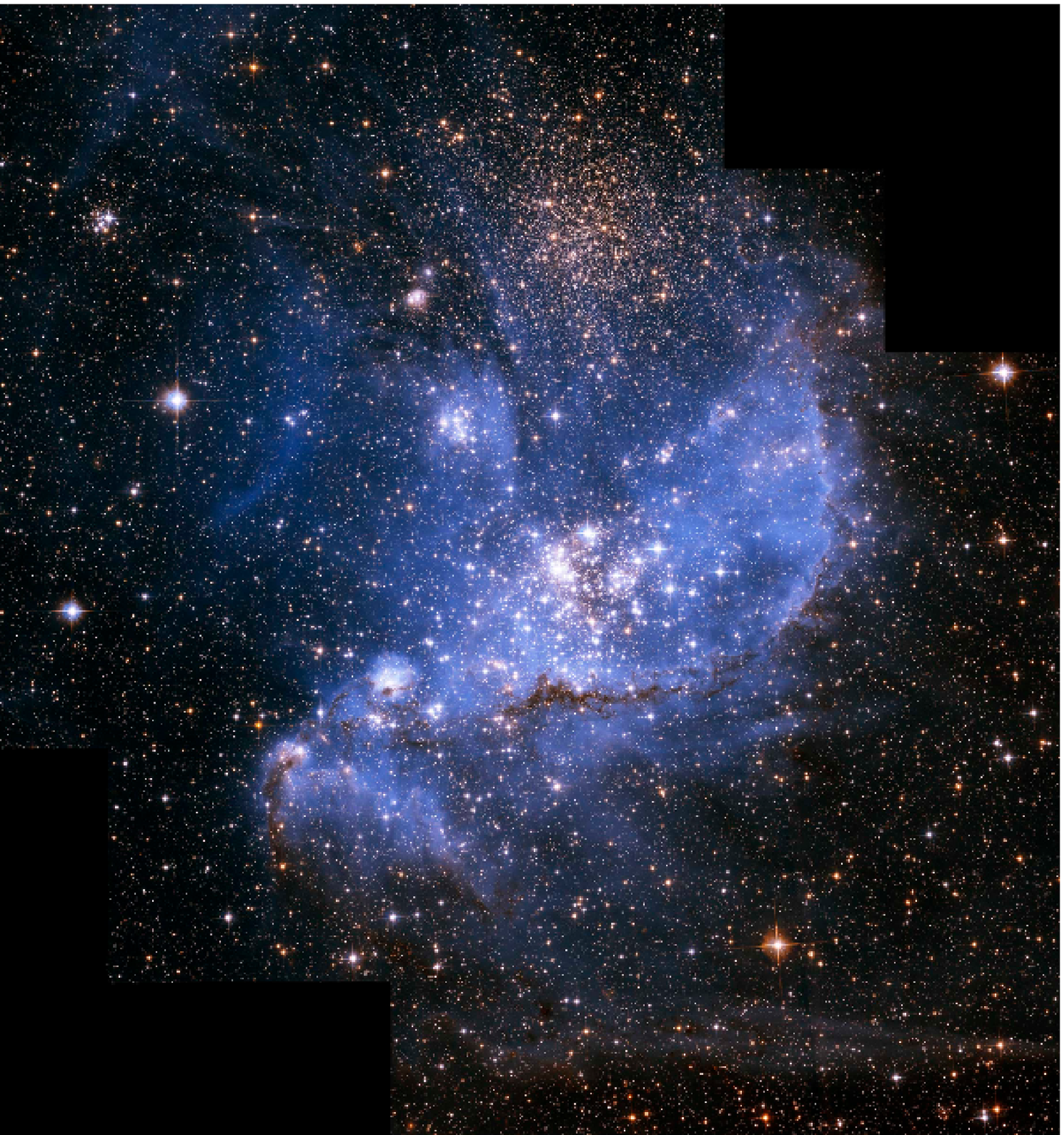}\\
\vspace{1cm}
\includegraphics[width=8.cm,height=11cm]{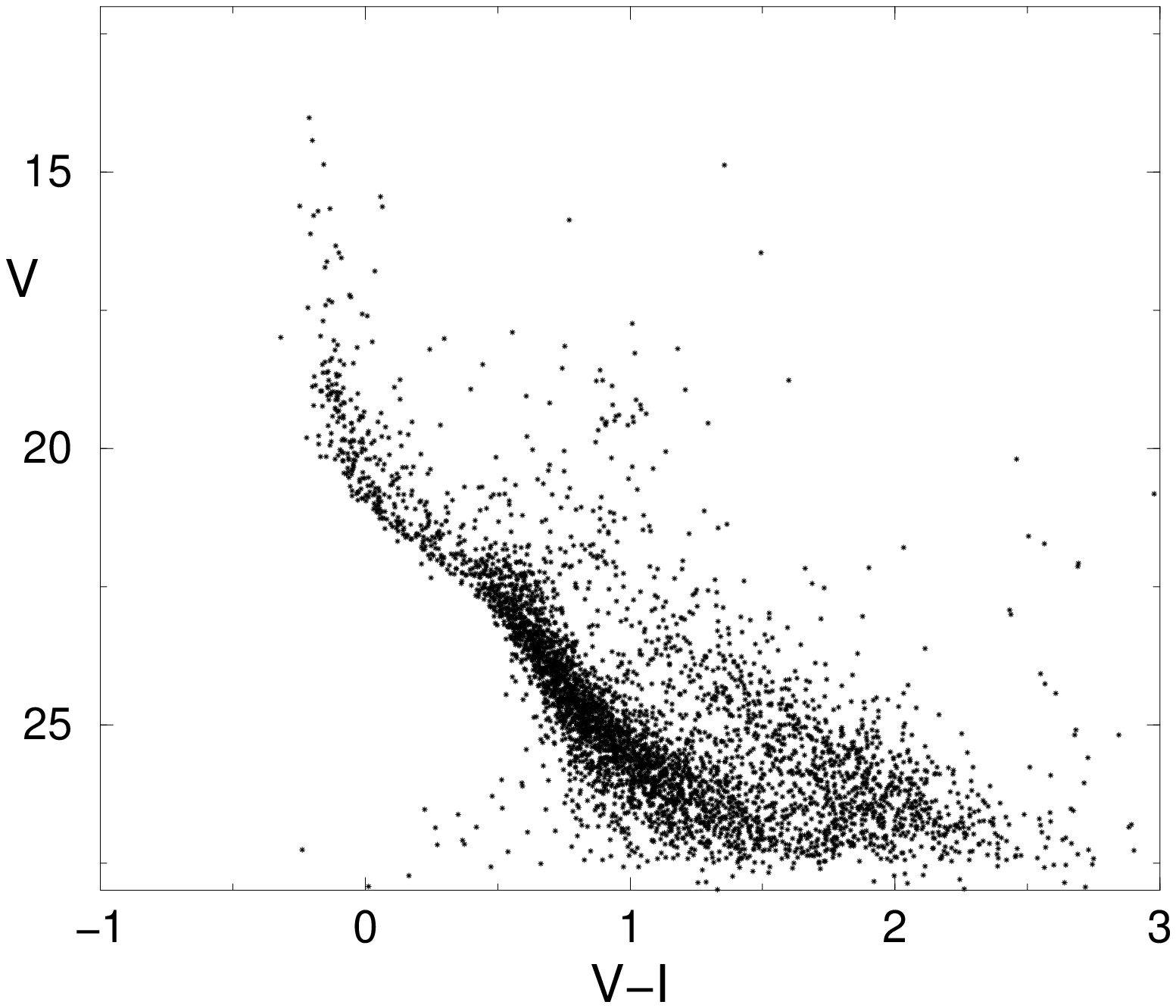}
\includegraphics[width=8.cm,height=11cm]{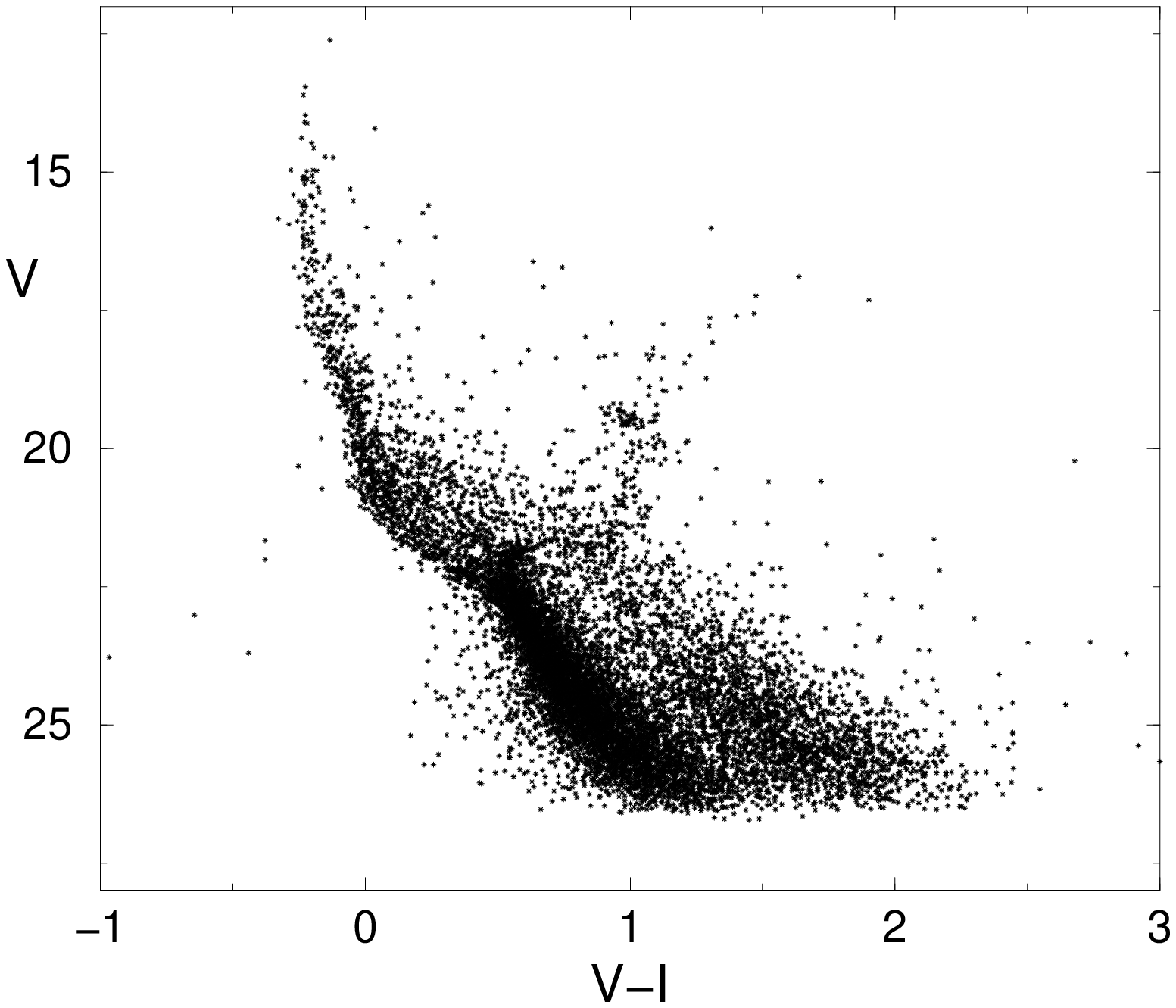}
\caption{Top panels: HST/ACS true color images of NGC602 (left), a
  very young cluster in the wing of the SMC and NGC346 (right), a
  populous young cluster in the main body of the SMC. Images credit
  NASA, ESA and A. Nota (STScI, ESA). Bottom panels: CMD of NGC602
  (left panel) and of NGC346 (right panel). PMS stars are clearly
  visible at the right of the Main Sequence.}
\label{pms}
\end{figure}

The very first and the final stages of stellar evolution, namely the
pre-main sequence (PMS) and the white dwarf (WD) regimes also deserve some
comments. As for the former, prior to reaching the MS, the star's
energy source is a contraction on thermodynamical timescales (tenths of
Myr). While aging, pre-main sequence stars fade and become
hotter. Figure \ref{pms} shows the HST/ACS images and the corresponding CMDs 
of two star forming regions in the Small Magellanic Cloud, NGC602 (left-hand 
panels, \cite{Carlson07}) and
NGC346 (right-hand panels, \cite{Sabbi07}). The well-defined sequence
well separated from the canonical main sequence, which appears on the
right-hand side of these CMDs, is composed by PMS stars.
Were not this phase affected by several theoretical and
observational uncertainties, the PMS would be a powerful clock for the 
most recent Myrs \citep[see e.g.][and references therein]{Cignoni09}. 

On the other hand, WDs represent the final fate of all stars with masses
below $8\,M_{\odot}$. These stars share a useful feature: the peak of
the WD luminosity function fades with age. Unfortunately, the presence
of theoretical uncertainties (e.g. crystallization processes, nuclear reaction
rates, convection, mass loss, and initial chemical composition)
together with an intrinsic low luminosity ($10<M_{V}<18$) tend to
invalidate its reliability.

In conclusion, MS and SGB stars are certainly the most reliable age
indicators. If we add that in these phases faint objects live
longer, a deeper CMD gives a better chance to robustly trace the past
star formation history.
It is important to underline that quantitative and
qualitative indicators can be combined and typically complement each
other. In other words, when a deep CMD is not available, evolved stars
can be used very profitably to recover the SFH, although with higher uncertainty
and within shorter lookback times.

\subsection{Deep is better}
\label{deep}

\begin{figure}[]
\includegraphics[width=8.5cm,height=11cm]{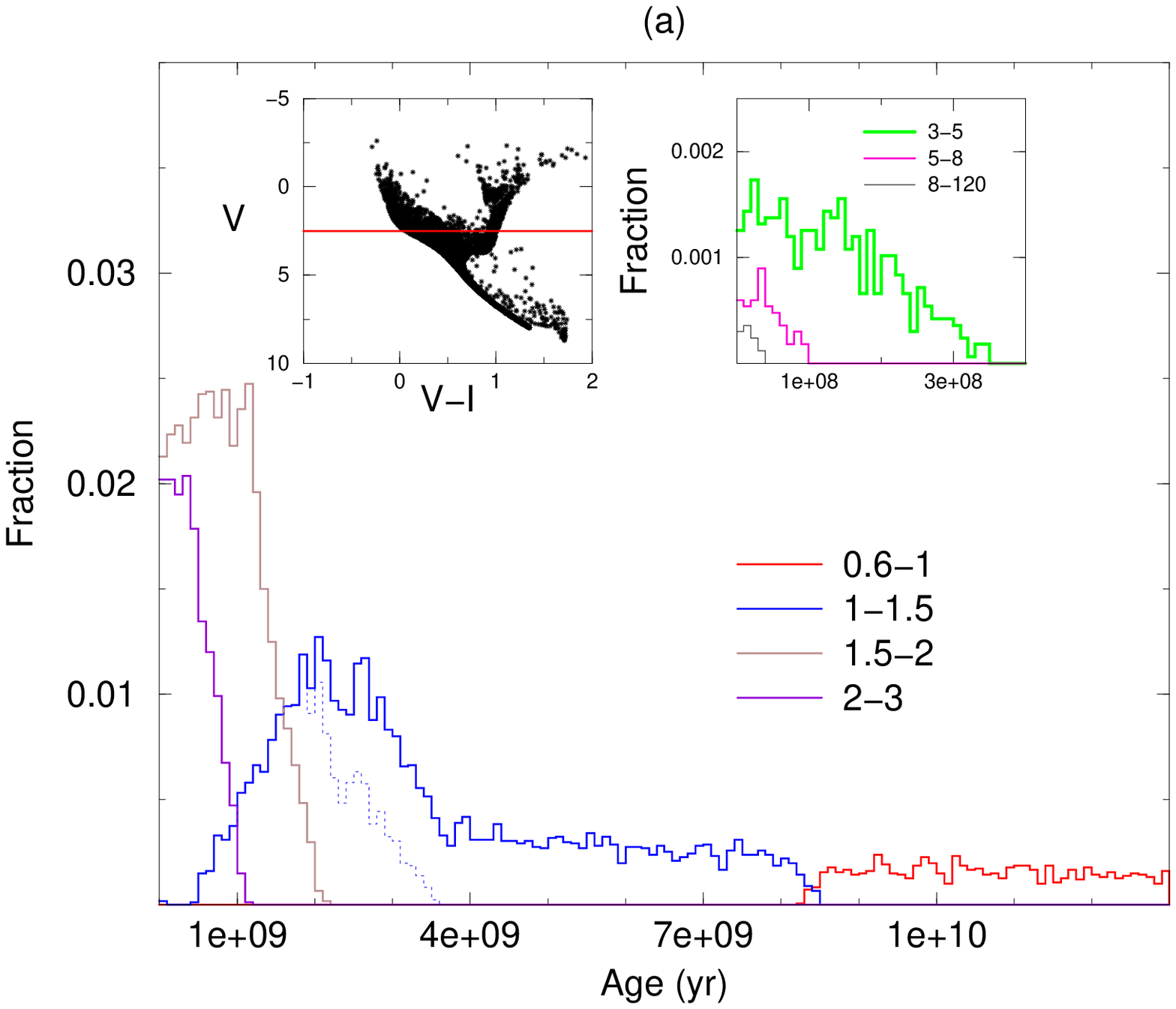}
\includegraphics[width=8.5cm,height=11cm]{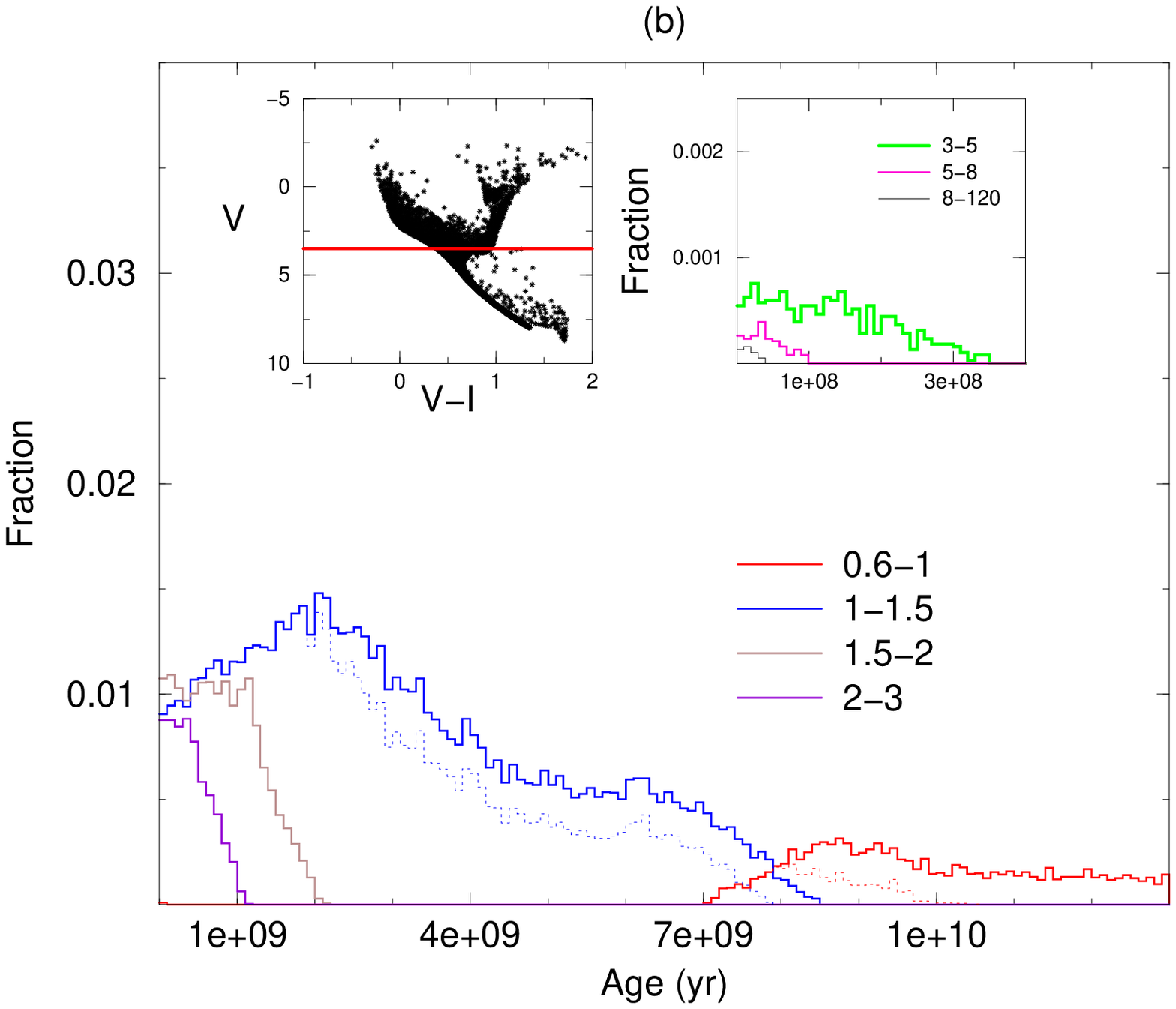}\\
\includegraphics[width=8.5cm,height=11cm]{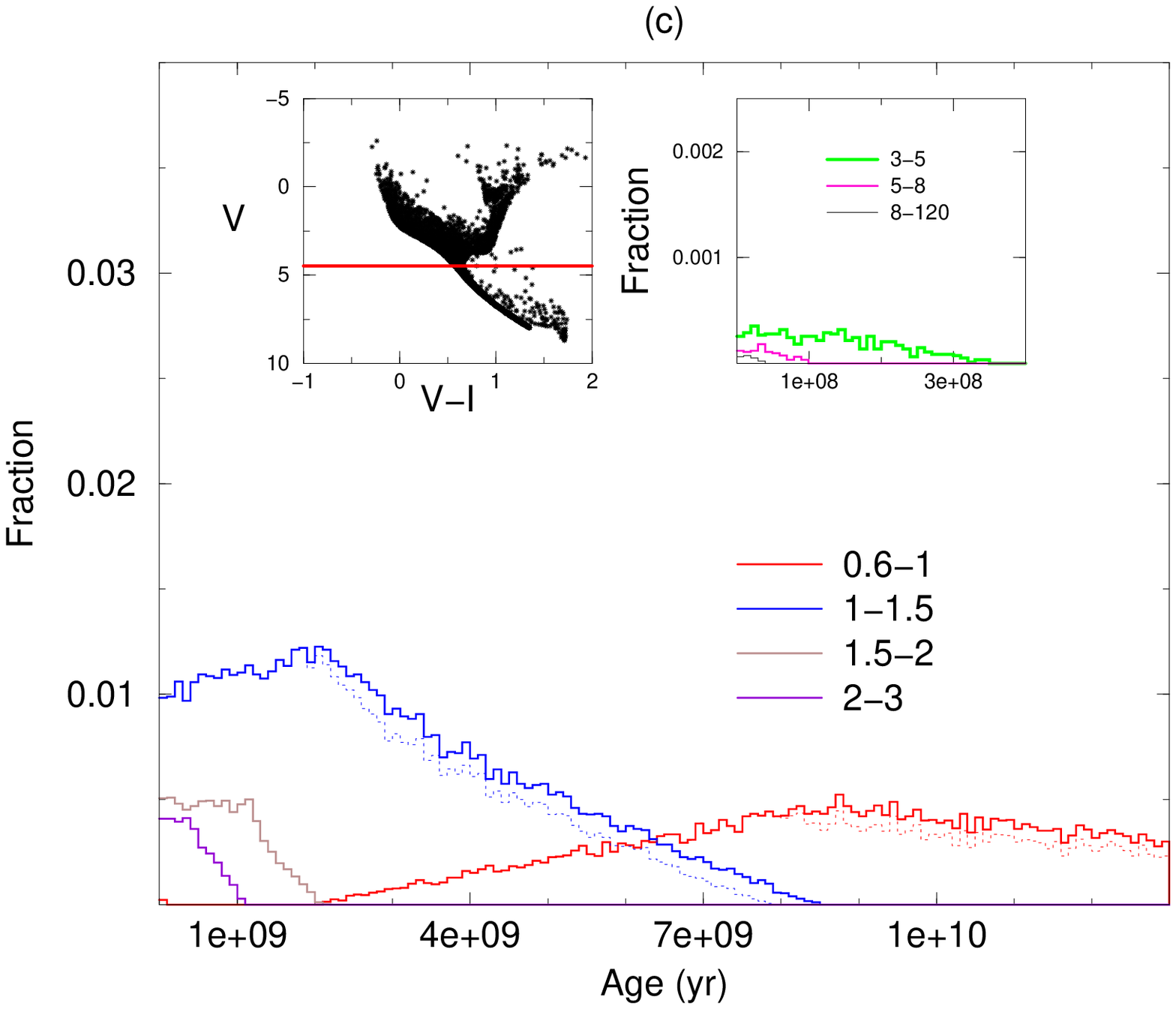}
\includegraphics[width=8.5cm,height=11cm]{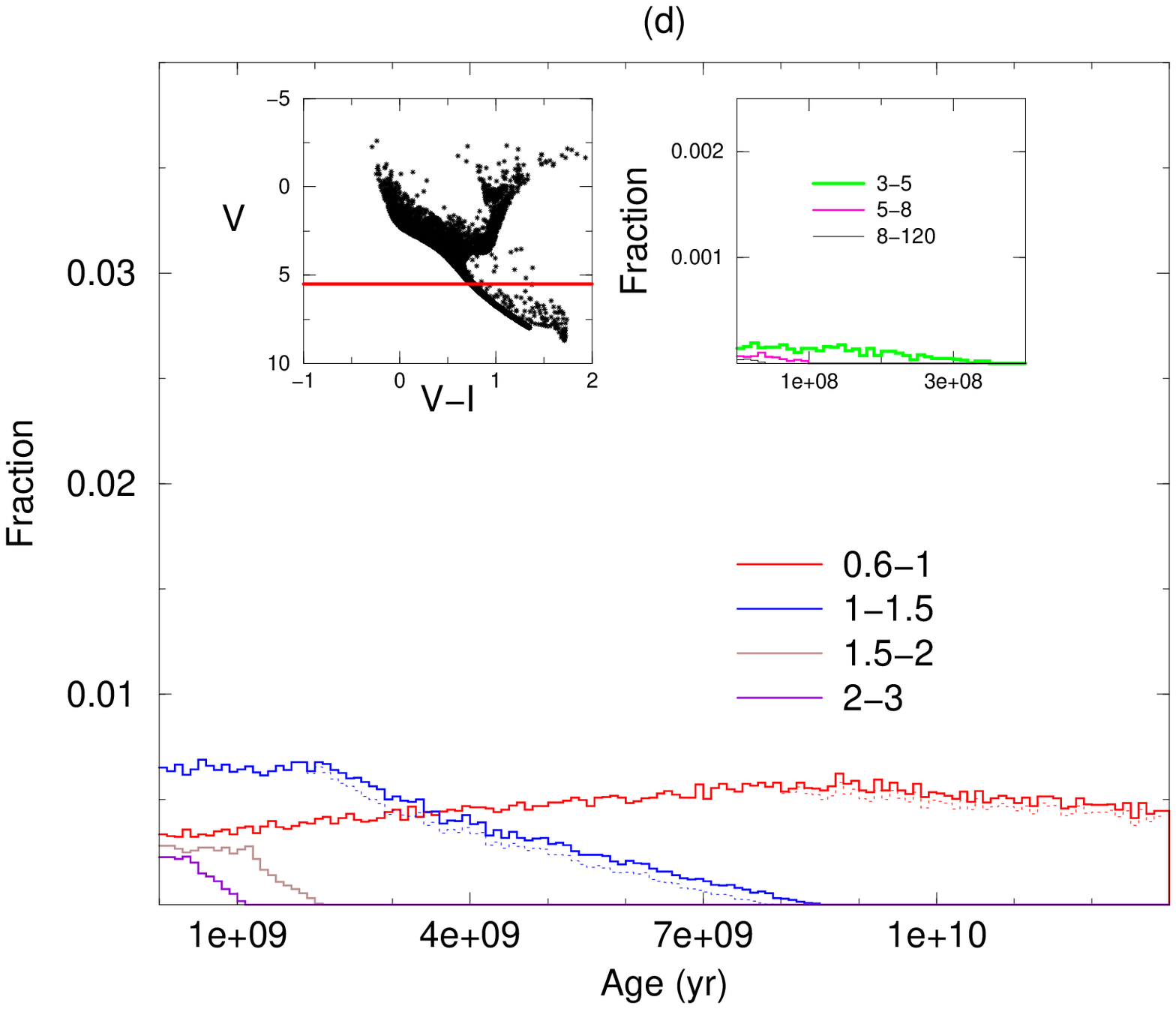}
\caption{Fractional age distributions of stars generated with a
  constant SFR (0-13 Gyr) for four different completeness limits. In
  different colors (solid lines) are plotted the contributions from
  different mass ranges. In panel (a) only stars with absolute
  magnitude brighter than 2.5 are plotted, while in panels (b), (c)
  and (d) this limit is respectively $M_{V}=3.5$, $M_{V}=4.5$ and
  $M_{V}=5.5$. The dotted lines represent the contribution of PMS, MS,
  SGB. In each panel, the left and right sub-figures show respectively
  the adopted simulated population (with the limiting magnitude marked
  by a red line) and a blow-up with the contribution of intermediate
  and massive stars in the last 300 Myr. }
\label{grotte} 
\end{figure}

The presence of a completeness limit (due to both the intrinsic
crowding and distance of the examined galaxy and to the instrumental
capabilities) hinders the possibility to exploit all the information
contained in a CMD. To visualize this effect, we built age-frequency
plots for various stellar mass ranges, assuming different completeness
limits (see Figures \ref{grotte}) and using an artificial population
generated from the Padova stellar models \citep{Fagotto94a} with
$Z=0.004$, no binaries, and constant SFR. In order to be as general as
possible, all the results are shown using absolute magnitudes. In all
panels of Fig. \ref{grotte} we plot with different colors the fraction
of stars of various mass ranges visible in the CMD above the assumed
completeness limit as a function of their age.
 
 In Figure \ref{grotte}-a the completeness limit is set to
 $M_{V}=2.5$: we see that on the MS only stellar masses higher than
 $1.5\,M_{\odot}$ are brighter than this limit and usable witnesses of
 the last $2$ Gyr.  On the other hand, the long-lived nature of lower
 masses guarantees to trace long periods of star formation, but not
 the recent SFR. The explanation comes from the limit $M_{V}=2.5$
 itself: it cuts off the MS, so for stars below $1.5\,M_{\odot}$ we
 see only later (i.e. brighter) evolutionary phases, represented by
 the RGB and the central helium burning (the dotted lines for the mass
 ranges $0.6-1.0 \,M_{\odot}$ and $1.0-1.5 \,M_{\odot}$ represent the
 contribution of PMS, MS and SGB stars).

Moving the limit to $V=3.5$ (Figure \ref{grotte}-b), the mass interval
$1-1.5\,M_{\odot}$ is visible in the CMD and informative of the star
formation history between now and $\sim 8$ Gyr ago. However, only
masses over $1.2-1.3\,M_{\odot}$ are on the MS. Lower masses or,
equivalently, older epochs, must refer to RGB and He burning
phases. The situation improves when the completeness limit is set at
$M_{V}=4.5$ (\ref{grotte}-c), and the MS phase is visible for all
stars down to $\sim 0.8 \,M_{\odot}$. This limit represents a very
good level for studying the history of a resolved galaxy, since it
guarantees age sensitive tracers (MS and SGB stars) covering the
entire Hubble time (13 Gyr).

Finally, Figure \ref{grotte}-d shows the age plots for a completeness
limit $M_{V}=5.5$: at this luminosity, the zero age main sequence is
reached for sub-solar masses, whose lifetimes are longer than the age
of the Universe. With respect to the $M_{V}=4.5$ case, the advantages
here are: 1) a much more reliable photometry of the turn-off and SGB
stars; 2) further information on the IMF, thanks to a better coverage
of the lowest/faintest mass intervals, where the IMF slope may
significantly change \citep[e.g.][and references
  therein]{Kroupa01,Chabrier03}; 3) a better constraint on Z(t), given
the mild influence of the SF law on the CMD position of low mass
stars.

\begin{figure}[]
\includegraphics[width=8.5cm,height=11cm]{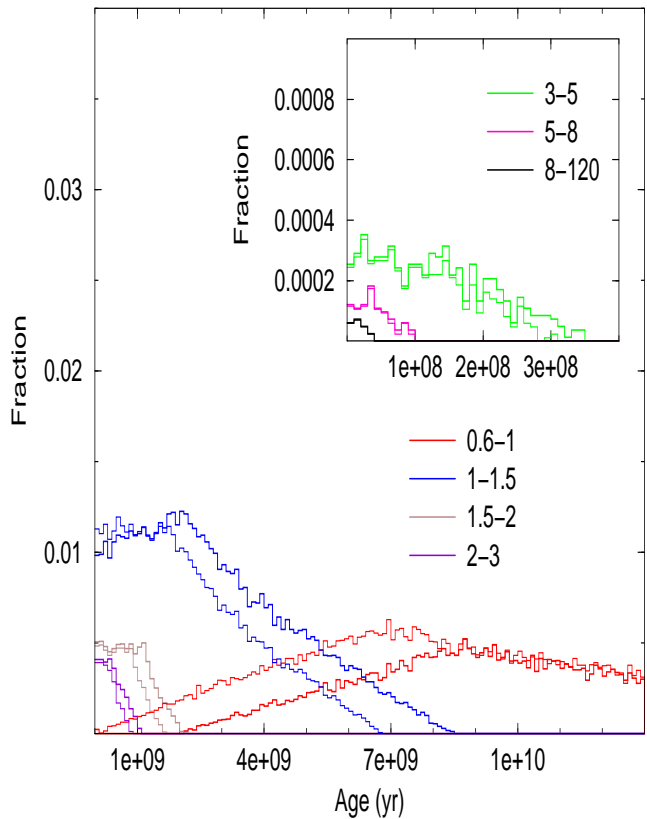}
\caption{Fractional age distributions for stars brighter than $M_{V}=4.5$ at two
  different metallicities. The thick line represents $Z=0.004$, while
  the thin line represents $Z=0.0004$.}
\label{grotte2} 
\end{figure}

These results depend on the assumed chemical composition. This is
important when one considers that in galaxies some chemical enrichment
must always be at work. Following this paradigm, the oldest stars in a
galaxy are expected to be metal poor. Changing the metallicity has two
main effects on the model; namely, changes in the evolutionary
lifetimes and changes in the stellar luminosity, which in turn can
sensibly modify the relation between CMD and SFH. To investigate this
phenomenon, in Figure \ref{grotte2} the frequency-age plot for the
completeness limit $M_{V}=4.5$ is shown for two different
metallicities, Z=0.004 (thick lines in the figure) and Z=0.0004 (thin
lines). Lowering the metallicity accelerates the evolution, and the
age distribution for all the mass intervals (except
$0.6-1\,M_{\odot}$) is shifted by, at least, 1 Gyr with respect to the
Z=0.004 cases. The age plot for masses $0.6-1\,M_{\odot}$ has a
different genesis: part of these stars live more than 13 Gyr, so the
evolutionary effect is not visible. In contrast, the mass range
$0.6-1\,M_{\odot}$ emphasizes the luminosity effect: a lower
metallicity pushes MS stars over the completeness limit $M_{V}=4.5$,
injecting younger stars in the age distribution, that now involves
ages between 0 and 13 Gyr. In practice, a lower metallicity mimics
what happens with a more favorable completeness limit.

This implies the following rule: in order to safely use the CMD
for an estimate of the oldest star formation history we need to
resolve all the stars down to $M_{V}=4.5$. Since this magnitude can be reached 
only in the closest galaxies, this implies that in most cases the information 
on the earliest 
SF activity  is either completely lacking or very uncertain.

\section{Deducing the SFH: guidelines}

\subsection{A changing landscape}

The first procedures to derive the SFH of nearby galaxies from
synthetic CMDs were developed by the Bologna and the Padova groups
about 20 years ago
\citep[][]{Ferraro89,Tosi91,Bertelli92,Greggio93,Marconi95}, with the
latter then combining with the Canary group
\citep[][]{Gallart96a,Gallart96b,Aparicio97a}. These works used
luminosity functions, color distributions and the general CMD
morphology to constrain the underlying SFH. In particular, the ratio
of star counts in several regions of the CMD was used to determine
both the SFR and the IMF \citep[][]{Bertelli92}.  The drawback of
these procedures is the lack of a robust statistical criterion to
evaluate the best solution and the corresponding uncertainties. On the
other hand, these authors made an optimal use of all the CMD phases
and took into careful account all the properties and uncertainties of
stellar evolution models, thus avoiding blind statistical approaches,
which can lead to misleading results.

Later on, several methods have been proposed to statistically compare
simulated and observed CMDs. In this framework, some groups have
derived the SFHs of galaxies in the LG, \citep[e.g.][]{Tolstoy96b,
  Aparicio97a,Dolphin97,Gallart99, Cole99, Hernandez99,
  Holtzman99,Harris01}.  Others, see e.g. \citep[][]{Vergely02,
  Cignoni06,Schroder03,Hernandez00b}, have tackled the question of the
SFH in the solar neighborhood. To the same class of investigators we
can assign also the study by \cite[][]{Naylor06} focused on star
clusters.  In all these works, the emphasis is transferred from the
stellar evolution properties to the problem of selecting the most
appropriate model through decision making criteria: here, the
likelihood between observed and model CMD is evaluated on statistical
bases. There are subtle differences among different groups, reflecting
how these authors define the likelihood and how they solve it. The
advantages are mainly three: the possibility to exploit each star of
the CMD, and not only few strategic ratios; the evaluation of the
uncertainty on the retrieved SFH, which is robust; the explorability
of a wide parameter space. However, a blind statistical approach is
not risk free. Although significant advances have been made in stellar
evolution and atmosphere theories, several processes (only to cite the
most infamous ones, the HB morphology, RGB and AGB features,
convection in general) remain poorly understood and affect the
statistical tests. If some parts of the CMDs have a low reliability
and others are statistically weak, but very informative (like the
helium burning loops), any blind algorithm may miss something
crucial. In this case, a careful inspection of the CMD morphology, in
particular the ratios of stellar number counts in different
evolutionary phases is the irrenounceable and necessary complement to
the statistical approach. Finally, whatever the adopted procedure, the
absolute rate of star formation must be obtained normalizing the best
model to the observed number of stars.

 The major differences among
the various procedures concern the approach to select
the best solutions and the treatment of metallicity variations. In
2001 the predictions of the synthetic CMD method from about ten
different groups were compared with each other, showing that, within
the uncertainties, most procedures provided consistent results
\citep[the {\it Coimbra Experiment}: see][and references
  therein]{Skillman02}.

In the following section we describe the main steps to rank the
likelihood among CMDs.


\subsection{To grid or not to grid}

 In order to decide if a synthetic CMD is a good representation of the
 data, the observed and the model star-counts can be compared in a
 number of CMD regions. In \citep[][]{Gallart99} these regions are
 large and strategically chosen to sample stars of different ages or
 specific stellar evolutionary phases and to take into account
 uncertainties in the stellar models: this solution guarantees an
 optimal statistics, but has the drawback of underexploiting the fine
 structure of the CMD. Another possibility is to choose a fine grid of
 regions (see e.g. \citep[][]{cigno06,Dolphin02}), counting how many
 predicted and observed stars fall in each region: the temporal
 resolution is higher, but the Poisson noise is the new drawback. An
 intermediate solution is to build a variable grid, coarser where the
 density of stars is lower and finer when the density is higher
 \citep[see e.g.][]{Vergely02, Aparicio09}. At this level, ad-hoc
 weighting of some regions can be introduced both to emphasize CMD
 regions of particular significance for the determination of age, and
 to mask those regions where stellar evolutionary theory is not
 robust.

 Other authors avoid to grid the CMD: for instance, in
 \citep[][]{Tolstoy96b} each model point (apparent magnitude and
 color) is replaced with a box with a Gaussian distributed probability
 density (the photometric error). The total likelihood of a model is
 the product of the probabilities of observing the data in each
 box. The idea of this method is equivalent to use blurred isochrones
 (each point is weighted by the Gaussian spread), so that the
 photometric uncertainty is embodied in the theoretical model.

\subsection{Maximum Likelihood}

The next step is to choose a criterion for the comparison between
synthetic and observational CMDs. For any grid scheme, once binned,
data and synthetic CMDs are converted in color-magnitude
\emph{histograms}. So, the new problem is to quantify the similarity
among 2D-histograms. One possibility is to minimize a $\chi^2$
likelihood function: when the residuals (differences among theoretical
and observed star-counts in the CMD regions) are normally distributed,
all models that have a $\chi^2$ greater than the best fit plus one are
rejected. However, when the distribution is not normal, a $\chi^2$
minimization leads to a wrong solution. This motivates the use of the
Poisson likelihood function instead of the least-squares fit-to-data
function.  In order to determine the uncertainties around the best
model, a valid alternative is to use a bootstrap test: the original
data are randomly re-sampled with replacements to produce
pseudo-replicated data sets. This mimics the observational process: if
the observational data are representative of the underlying
distribution, the data produced with replacements are copies of the
original one with local crowding or sparseness. The star formation
recovery algorithm is performed on each of these replicated data
sets. The result will be a set of ``best'' parameters.  The confidence
interval is then the interval that contains a defined percentage of
this parameter distribution.

One aspect deserves closer inspection: the minimization of a merit
function of residuals ($\chi^2$ or Poisson likelihood) is a global
measure of the fit quality. The first side effect is purely
statistical: low density regions of the CMD may be ignored in the
extremization process, whereas well populated phases (as the MS) are
usually well reproduced. This is a problem of contrast: low density
regions are Poisson dominated, thus, they are much easier to match
with respect to the well populated regions.

 A Monte Carlo method can be used with great success to evaluate this
 bias: building synthetic CMDs from the best set of parameters and
 re-recovering the SFH can allow to remark any statistical
 discrepancy. Then, a straightforward solution is to enhance the
 significance of the discrepant regions of the CMD (with appropriate
 weights).

  Another problem is connected with the theoretical ingredients we
  have used in the models: first of all, stellar evolution models are
  not perfect and computations by different groups show systematic
  differences \citep[see e.g.][for a review]{Gallart05}. Model
  atmospheres are often unreliable for cool and metal rich
  stars. Moreover, our models are only covering a part of the possible
  parameter space and some degree of freedom (additional
  metallicities, mass loss, overshooting, etc..)  may have been
  neglected. In this case, our best model is only the best (in a
  relative sense) of the explored parameter space, not necessarily a
  good one.


 In order to deal with these undesirable effects, the residuals can be
 placed in the CMD, identifying all the regions where the discrepancy
 between observed and predicted star-counts is larger. If the
 residuals are larger and concentrated in some part of the CMD, we may
 understand what is the reason for the discrepancy and take it into
 account. For instance, a poor fit in the red giant branch, less
 populated than the main sequence but morphologically well defined in
 color (and for this reason often neglected in $\chi^2$ minimization),
 may suggest a wrong metallicity, a different mixing length parameter
 or wrong color transformations.

\subsection{Wondering in the parameter space}
\label{wonder}
 The main drawback of Maximum Likelihood (ML) approaches is the
 computational burden. Algorithms that find the ML score must search
 through a multidimensional space of parameters, using for instance
 derivative methods, like Powell's routine, or non-derivative ones,
 like the downhill simplex routine, or genetic approaches \citep[see
   e.g.][]{Aparicio09}.

These techniques are not guaranteed to find the peak, but work
relatively well for a limited number of parameters. Traditionally,
this question has been tackled by constructing synthetic CMDs from a
SFH built as a series of contiguous bursts and finding the amplitudes
of each burst that give the maximum probability to have produced the
data: the synthetic CMD is now a linear sum of the partial CMDs
produced from a single realization for each burst. In this way a huge
parameter space can be explored: rather than calculating a complete
CMD for each SFR(t), the partial CMDs can be linearly combined to
build a CMD for any SFR(t). In order to reduce the Poisson noise, the
partial CMDs are simulated with many more stars than observed
(typically, 100 times more).

  The computer time spent for building the final
 CMD is only that needed to go through a finite number of models,
 simply equal to the number of combinations of Z(t) relations, IMF
 slopes, reddenings, and distance moduli \emph{times} the number of
 age bins in the solution.

\begin{figure}[]
\centerline{ \epsfxsize= 14 cm \epsfbox{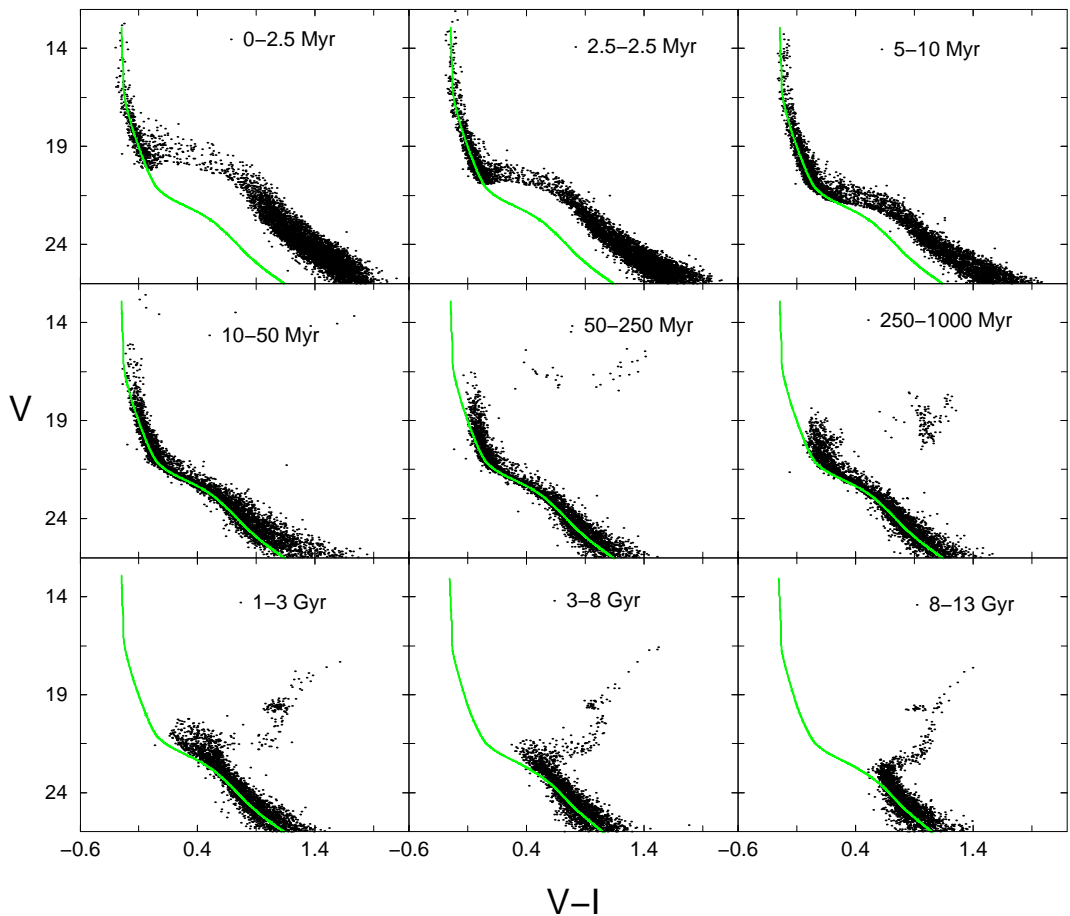}}
\caption{Partial theoretical CMDs. Each one is generated with a
  constant star formation rate, for the age interval pointed out in
  the label. The bright part of the main sequence is dominated by high
  mass stars so the CMD time step has to be shorter.}
\label{partials}
\end{figure}

Age Bins: Disentangling a stellar population showing both very recent
(Myr) and very old (Gyr) episodes of star formation is not
straightforward. Only low mass stars survive from ancient
episodes because their evolutionary timescales are very long: small
CMD displacements, for example due to photometric errors, can bias
their age estimates up to some Gyr. In this case, increasing the time
resolution, besides being time-consuming, may produce unrealistic star
formation rates due to misinterpretations. Hence, the choice of
temporal resolution must follow both the time scale of the underlying
stellar populations and the data scatter (photometric errors,
incompleteness, etc.). A practical way out is to use a coarser
temporal resolution for the older epochs, which: 1) allows us to avoid
SFH artifacts at early epochs; 2) reduces the Poisson noise; 3)
reduces the parameter space.  Figure \ref{partials} shows a possible
time stepping \citep[see][]{Cignoni09}. Finally, it's worth noting
that the choice of each set of age bins will prevent to identify any SF
episode shorter than the bin duration: for instance, the 1 Gyr lull
(between 2 and 3 Gyr ago) in the star formation history, as simulated
in the top-right panel of Figure \ref{syn1}, will result in a lower
(half) activity in the 7th age bin (1-3 Gyr).

In the next sections we describe some numerical experiments
illustrating the reliability of a typical ML algorithm. In particular,
the sensitivity of such algorithm to several physical uncertainties is
outlined. These examples expand the discussions and results by
\cite{Cignoni06}. The experiments described here are based on
different set of tracks, mass spectrum and photometric
errors/completeness, but the results are the same as in
\cite{Cignoni06}, thus showing that they are independent of these
assumptions.  Other instructive examples can be found in
\cite{Gallart99}, \cite{Gallart05} and \cite{Kerber09}.

\subsection{A practical example}
\label{example}
To describe how a ML procedure works, let's build a fake galaxy assuming for 
sake of simplicity a constant star formation rate between now
and 13 Gyr ago and a metallicity fixed at $Z=0.004$. We put it at the
distance  of the closest dwarf irregular galaxy, the Small
Magellanic Cloud (SMC), $(m-M)_{0}=18.9$, and adopt the SMC mean foreground 
reddening $E(B-V)=0.08$. To minimize statistical fluctuations the Monte
Carlo extractions are iterated until we have 30000 stars brighter
than $V=23$, which roughly corresponds to about 100000 stars in the
entire CMD. Photometric errors and incompleteness as obtained
in actual HST/ACS SMC campaigns \citep{Cignoni09} are
convolved with the synthetic data, producing a realistic artificial
population. This fake galaxy will be used as reference set in all the following
exercises.

To recover its SFH, we have gridded the CMD in small bins of color and
magnitude (0.1 mags large) and we have minimized a Poisson
likelihood. The time stepping for the partial CMDs is the following
(going backward in time from the present epoch to 13 Gyr ago): 100
Myr, 400 Myr, 500 Myr, 1 Gyr, 2 Gyr, 3 Gyr, 3 Gyr, 3 Gyr. A bootstrap
technique is implemented to determine the final uncertainties.

\begin{figure}[]
\epsfxsize= 8 cm \epsfbox{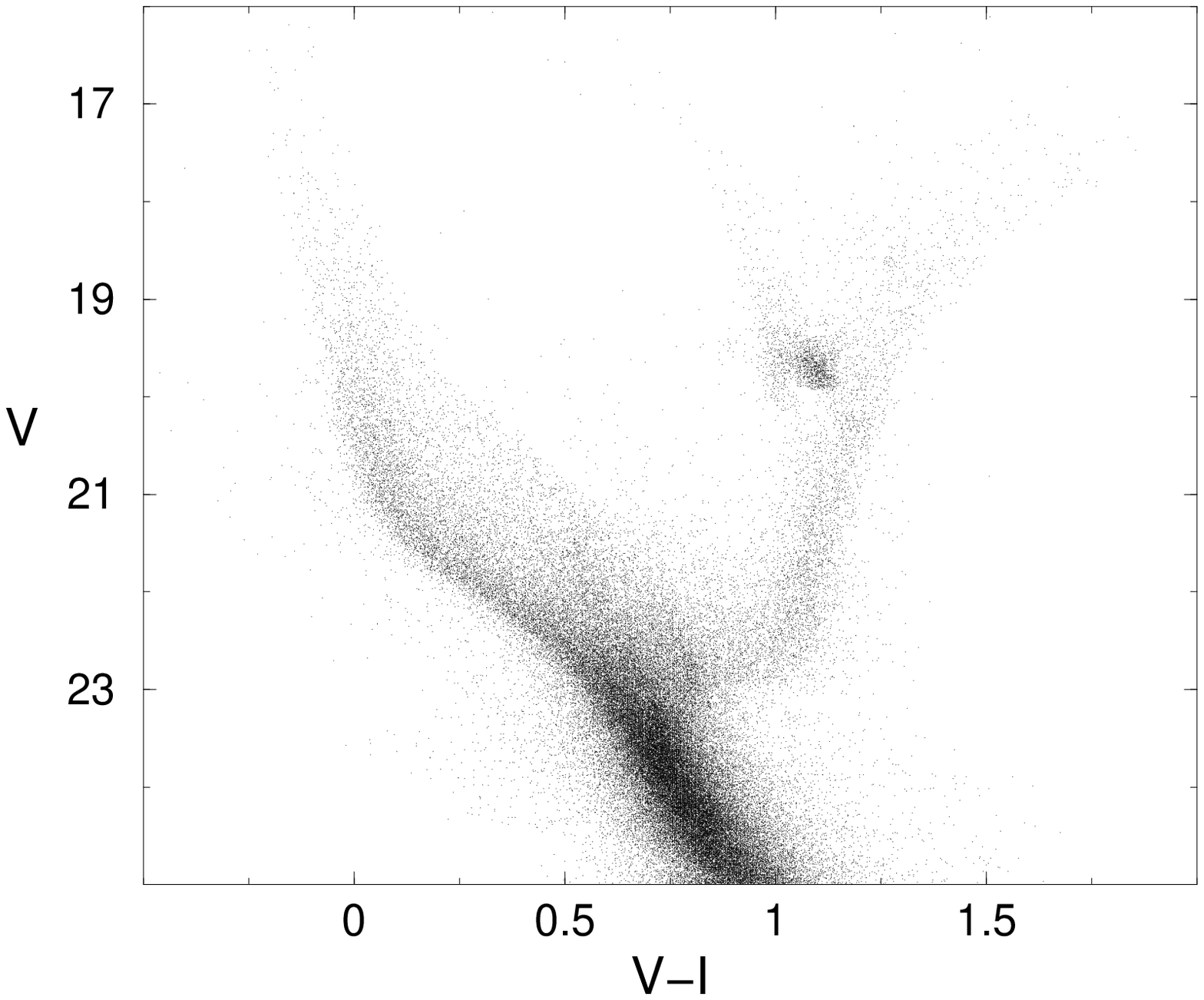}
\epsfxsize= 9 cm \epsfbox{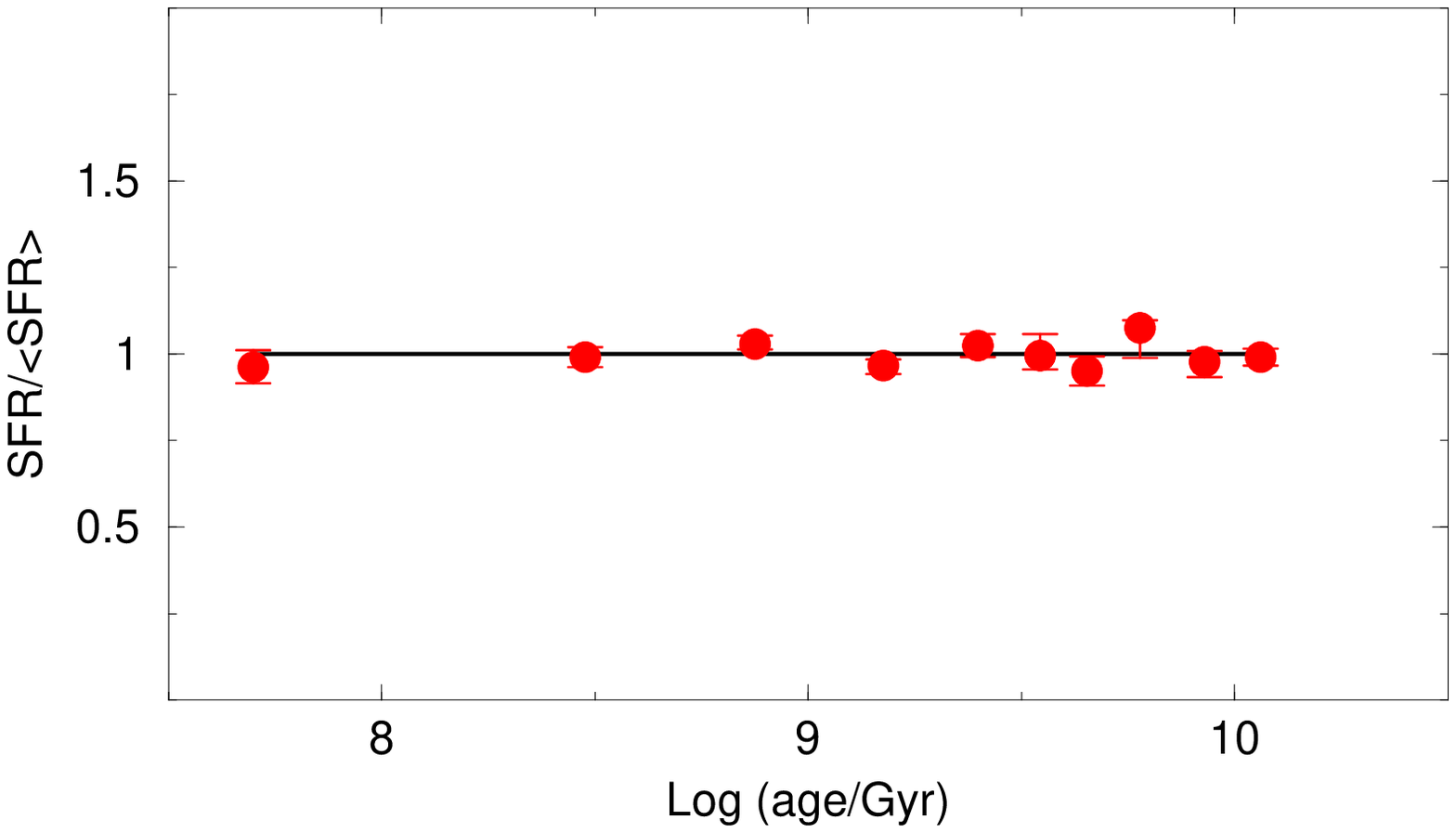}
\caption{Starting case for the experiments on the uncertainties on the
  retrieved SFH due to different factors. In the left-hand panel we show
  an artificial population of stars generated with constant SFR,
  Salpeter IMF, constant metallicity Z=0.004 (\citep[][]{Fagotto94b}),
  $(m-M)_{0}=18.9$, $E(B-V)=0.08$ and the HST/ACS errors and completeness of
  the photometry in the SMC field NGC602 (\citep[][]{Cignoni09}). In
  the right-hand panel the input (solid line) and the recovered (dotted)
  SFH are compared. }
\label{best}
\end{figure}

Figure \ref{best} shows on the left-hand panel the CMD of our
reference fake galaxy, and on the right-hand panel the SFH recovered
from it, using only stars brighter than V=23 (i.e., M$_V$=3.85), for a
self-consistency check. As expected the retrieved SFR is fitted by a
constant value. We will use this basic experiment as starting point
for a series of exercises aimed at describing the major uncertainties
affecting the synthetic CMD method.

\subsection{Uncertainties affecting the synthetic CMD procedures}

Contrary to real cases, in the reference case of Figure \ref{best} we
have all the information: all parameters are known and the data are
complete down to 23 (${M_{V}=3.85}$). Real galaxies are far
from this ideal condition. Inadequate information or uncertainty about
the assumed parameters can influence the identification of the best
SFH. The major sources of uncertainty are primarily the IMF, the
binary fraction and the chemical composition. From the observational
point of view, the completeness level is another important
factor. Moreover, population synthesis methods make a number of
simplifications to reduce significantly the computational load;
e.g. reddening constant across the data, same distance for all stars,
linear age-metallicity relation, etc. Dropping these simplifying
assumptions considerably complicates all analyses of the CMD
properties.

There are two main strategies to face the complexity of the
problem. One is to increase the number of free parameters in the
model. For example, \citep[][]{Aparicio97b} recover simultaneously
distance, enrichment history and SFR of the local dwarf LGS 3. The
other is to reduce the data complexity by means of additional
information; for instance, the metallicity may be estimated from appropriate
spectroscopy and multi-band observations may help to disentangle the
reddening.

In the following sub-sections we test the reliability of the star
formation recovery when the uncertainties related to each parameter are taken 
into account individually. A word of caution is necessary for the interpretation
of these exercises: in
each case we show how the recovery of the SFH of the reference fake galaxy is
affected by forcing the procedure to adopt a specific (and in most cases wrong) 
value for the tested parameter. This is aimed at emphasizing the effect of that
parameter. In the derivations of the SFH of real galaxies, 
the parameter values are all unknown (which complicates the derivation), but the
selection procedure is allowed to cover all the meaningful ranges of
values and can therefore distinguish which combinations allow to maximize the
agreement with the data.

\subsection{Completeness}

The numerical experiment seen in section \ref{example} represents an
ideal situation: 1) the SMC is one of the closest galaxies, 2) HST/ACS
currently provides the top level photometry in terms of spatial
resolution and depth. At the distance of the SMC (60 kpc), ACS
photometry can be 100\% complete to $V\approx 24-25$. Farther galaxies
and/or images from ground based telescopes have larger photometric
errors and more severe incompleteness. For comparison, the optical
survey conducted at the Las Campanas Observatory 1 mt Swope Telescope,
to trace the bright stellar population in the Magellanic Clouds, is
50\% complete to $V\sim 21-22$ (see \citep[][]{Harris04}).

\begin{figure*}[]
\epsfxsize= 8.5 cm \epsfbox{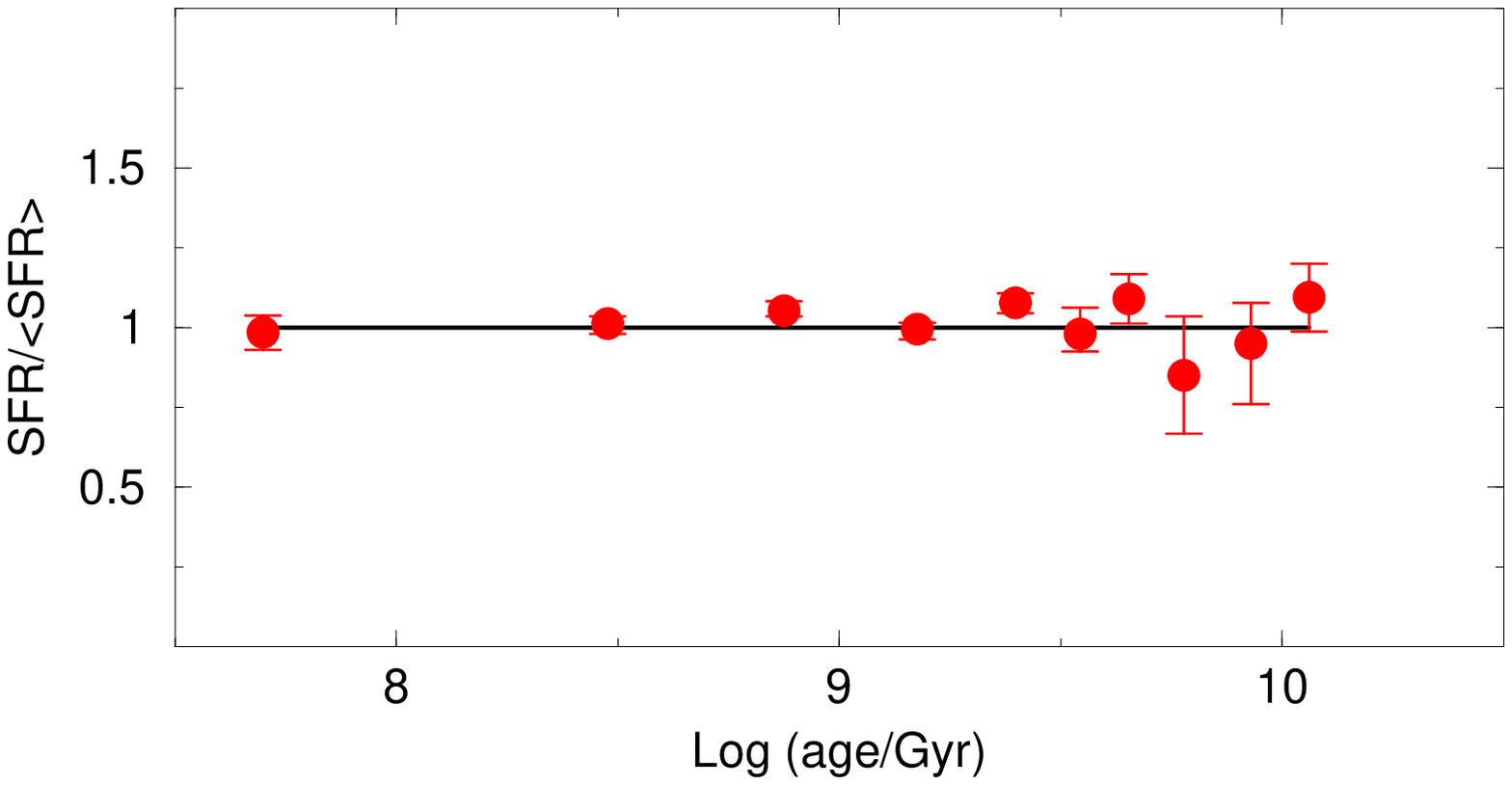}
\epsfxsize= 8.5 cm \epsfbox{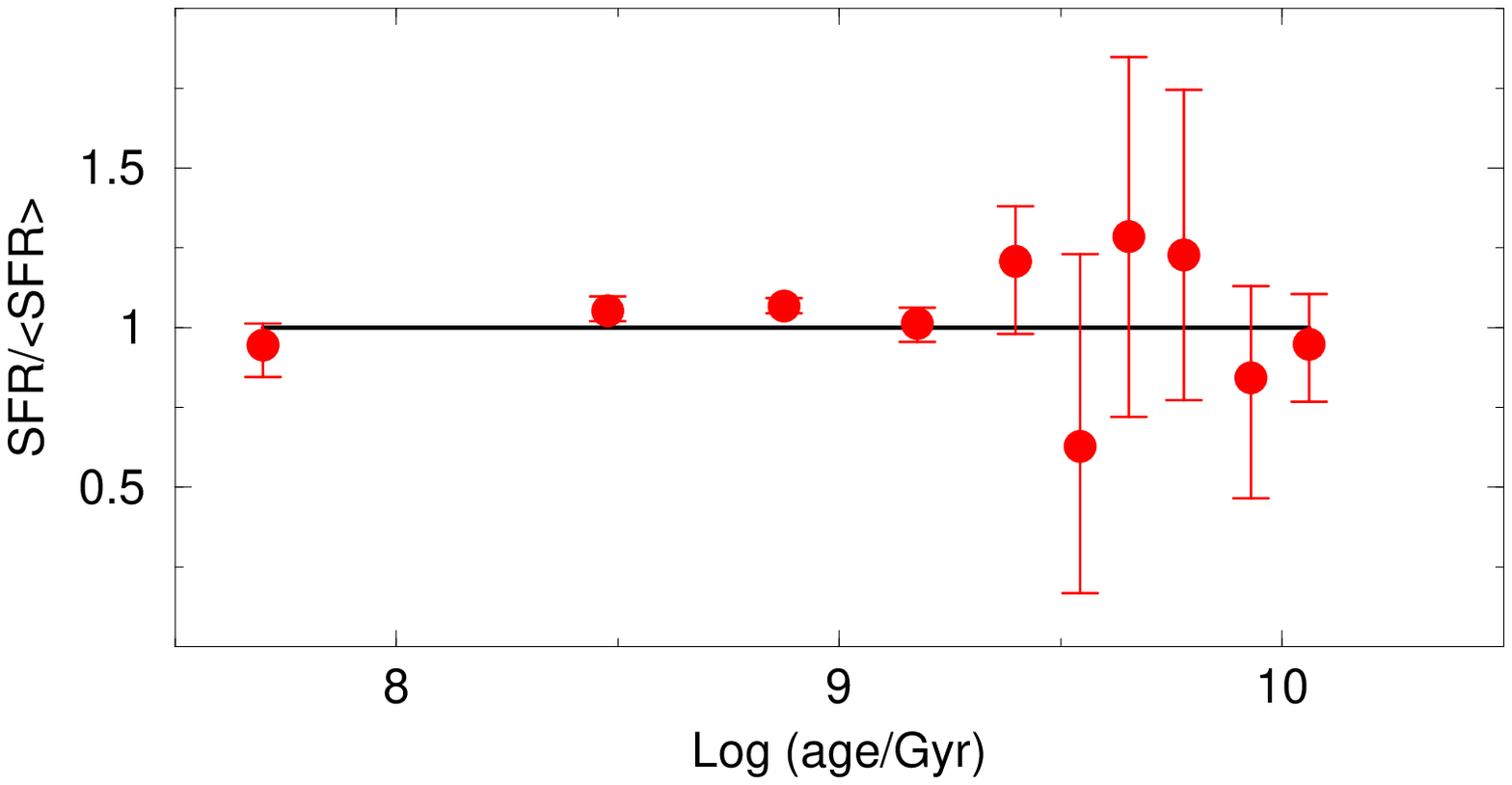}
\caption{Effect of different completeness limits: in the left panel
  the SFH is recovered using only stars brighter than V=22. In this
  case, most of the information is still retrieved. In the right
  panel, this limit is lowered at V=21: as expected, most of the old
  star formation (older than 1 Gyr) is much more uncertain.}
\label{compl}
\end{figure*}

To demonstrate the importance of the completeness limit, we perform the
star formation recovery using only stars brighter than $V = 21$ and
$V=22$.  The results are displayed in Figure \ref{compl}.
It is evident that the deeper the CMD, the higher the chance to
properly derive the old star formation activity. Compared to the
$V=23$ case, where the SFR recovery is accurate and precise at any
age, the quality already drops when the completeness limit is at
$V=22$: the larger error bars at old epochs reflect the fact that the
only signature of the oldest activity comes from evolved stars, less
frequent and much more packed in the CMD than the corresponding MS
stars. Rising the limiting magnitude at $V=21$ further worsens the
result, and the recovered SFH is a factor of 2 uncertain for ages
older than 1-2 Gyr.

 These results are actually optimistic: we have analyzed different
 levels of completeness with the same photometric errors (HST/ACS),
 but this is an utopic situation. More distant galaxies have a
 less favorable completeness limit, and also the photometric error
 is larger. In these cases, an additional blurring occurs.

\subsection{IMF}

A large body of evidence seems to indicate that: 1) the stellar IMF
has a rather universal slope, 2) above $1\,M_{\odot}$ the IMF is well
approximated by a power law with Salpeter-like exponent \citep{Salpeter55}, 
3) below $1\,M_{\odot}$ the IMF flattens.

According to \cite{Kroupa01}, the average IMF (as derived from
local Milky Way star-counts and OB associations) is a three-part power
law, with exponent $\alpha=2.7\pm0.7$ for $m>1\,M_{\odot}$,
$\alpha=2.3\pm0.3$ for $0.5\,M_{\odot}<m<1\,M_{\odot}$,
$\alpha=1.3\pm0.5$ for $0.08\,M_{\odot}<m<0.5\,M_{\odot}$. Other
authors, in the past, have proposed different (although somewhat
similar) slopes for the IMF of various stellar mass ranges
\citep[e.g.][]{Tinsley80,Scalo98,Chabrier03}. Given these
uncertainties, it is necessary to evaluate how this impacts the
possibility to infer the SFH. In fact, we have the degeneracy condition
that false combinations of IMF and SFH can match as well the
present day mass function (the current distribution of stellar masses)
of MS stars. To quantify it, three fake populations were
generated with different IMF exponents ($\alpha$ = 2, 2.35 and 2.7), but
the SFH searched using always 2.35. The results
are shown in Figure \ref{imf}.

\begin{figure*}[]
\epsfxsize= 8.5 cm \epsfbox{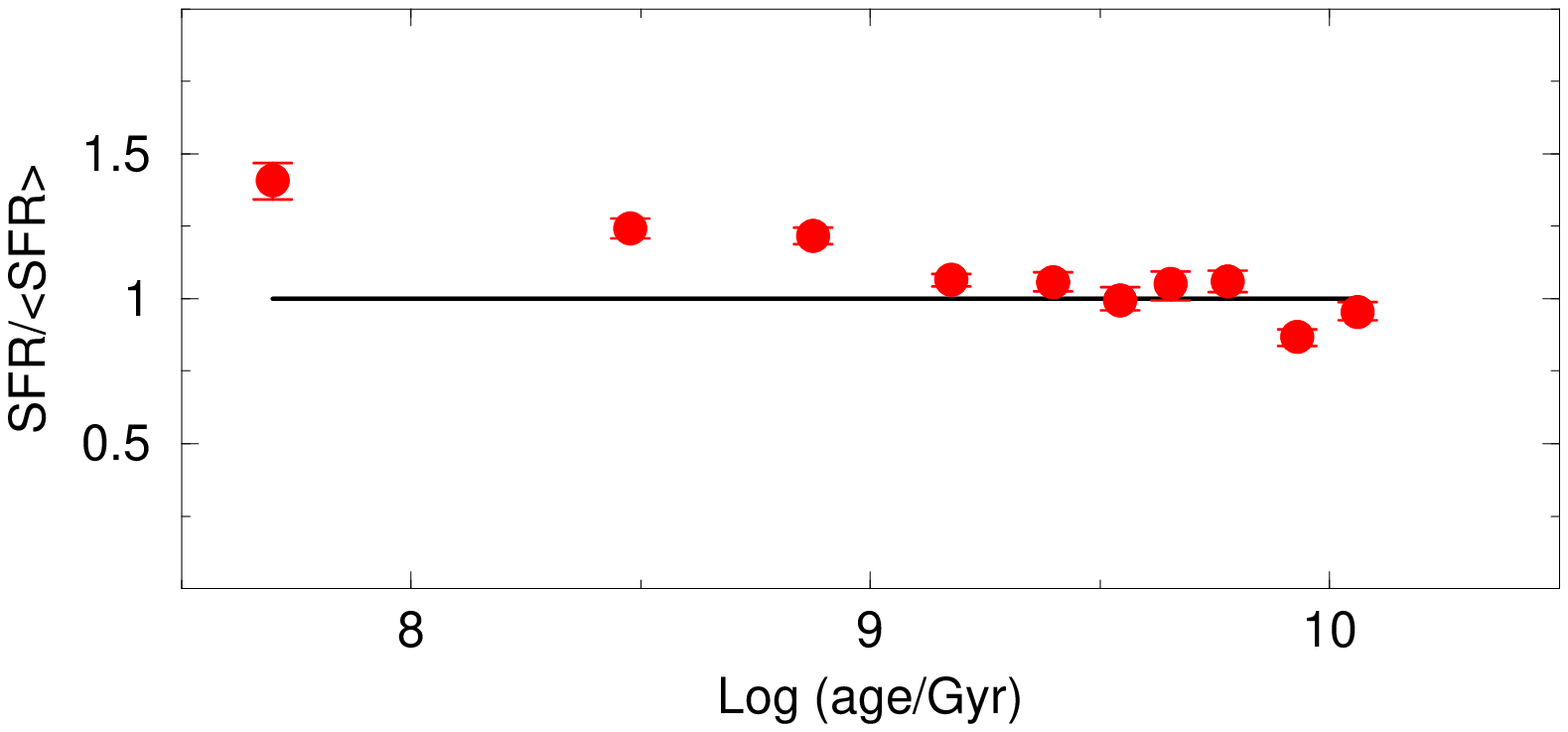}
\epsfxsize= 8.5 cm \epsfbox{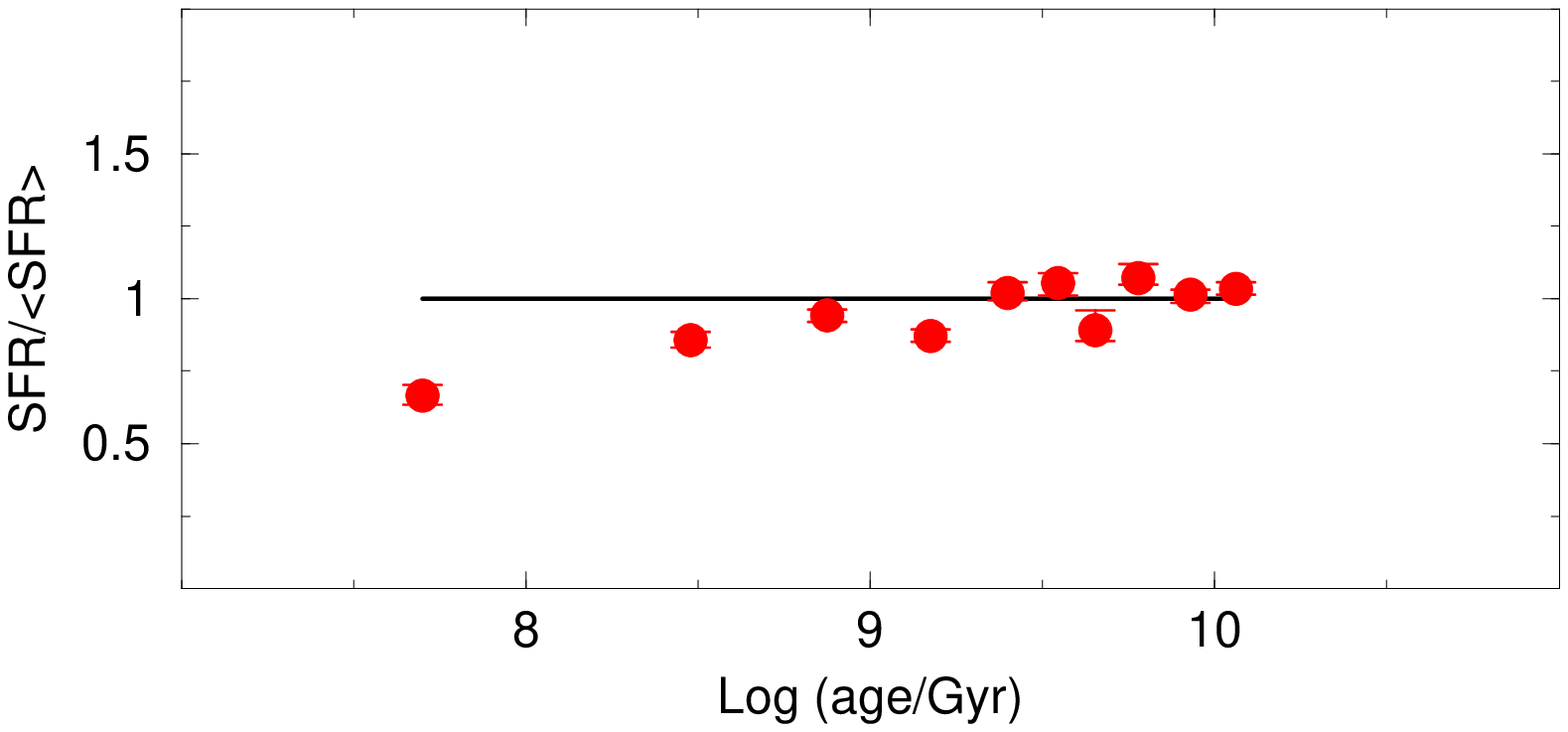}
\caption{Effect of different IMFs. SFH recovered assuming Salpeter's
  IMF exponent (2.35) for synthetic stellar populations actually
  generated adopting different IMF (labeled in each panel) and a
  constant SFR.}
\label{imf}
\end{figure*}

To interpret these results, one must recall that recent steps of star
formation are still populated by the entire mass spectrum, while old
steps see only low mass stars because the more massive stars born at
those epochs have already died. For old stars a steeper IMF is almost
indiscernible from a more intense star formation. In fact, even if
evolved stars are included in the SFH derivation, the mass difference
between a star at the RGB tip and those at the MS turn-off is only few
hundredths of solar masses: too small for allowing the identification
of any IMF effect.

 For young stars the situation in different. Any attempt to reproduce
 with an IMF steeper than that of the reference population the number
 of stars on the lower MS requires a stronger SF activity, but this
 (wrong) solution leads to overestimating the number of massive
 stars. Hence, for young stars, IMF and SFH are not
 degenerate. However, an automatic optimization algorithm, if not
 allowed to search better solutions including also the IMF among the
 free parameters, inevitably faces the impossibility to accommodate the
 number of both massive and low mass stars, by choosing a compromising
 recent SFH giving higher weight to the more populated (although less
 reliable) CMD region.

As shown in Fig.\ref{imf} the automatic fit tends to overestimate the
age of any population whose IMF is actually steeper than the adopted
one. And vice versa, for a flatter IMF.  Again, we remark that the
automatic solution is only the best solution in a parameter space
where the IMF is fixed, not necessarily a good one: if the CMD of the
recovered SFH is compared with the reference CMD, we immediately
recognize that the ratio between low and massive stars is wrong. In
other words, to figure out whether our "best" solution is actually
acceptable, it is always crucial to compare all its CMD results with
the observed one.

\subsection{Binaries}
Another source of uncertainty is the percentage of stars in unresolved
binary systems and the relative mass ratio. The presence of a given
percentage of not resolved binary systems affects the CMD
morphology. The aim, here, is to see if these effects can destroy or
alter the recovered information on the SFH. In order to perform this
analysis, we build fake populations using different prescriptions for
the binary population (10\%, 20\% and 30\% of binaries with random
mass ratio), but the SFH is searched ignoring any binary population
(i.e. assuming only single stars). Our models do not include binary
evolution with mass exchange, thus we assume that each star in a
double system evolves as a single star.

\begin{figure*}[]
\epsfxsize= 8.5 cm \epsfbox{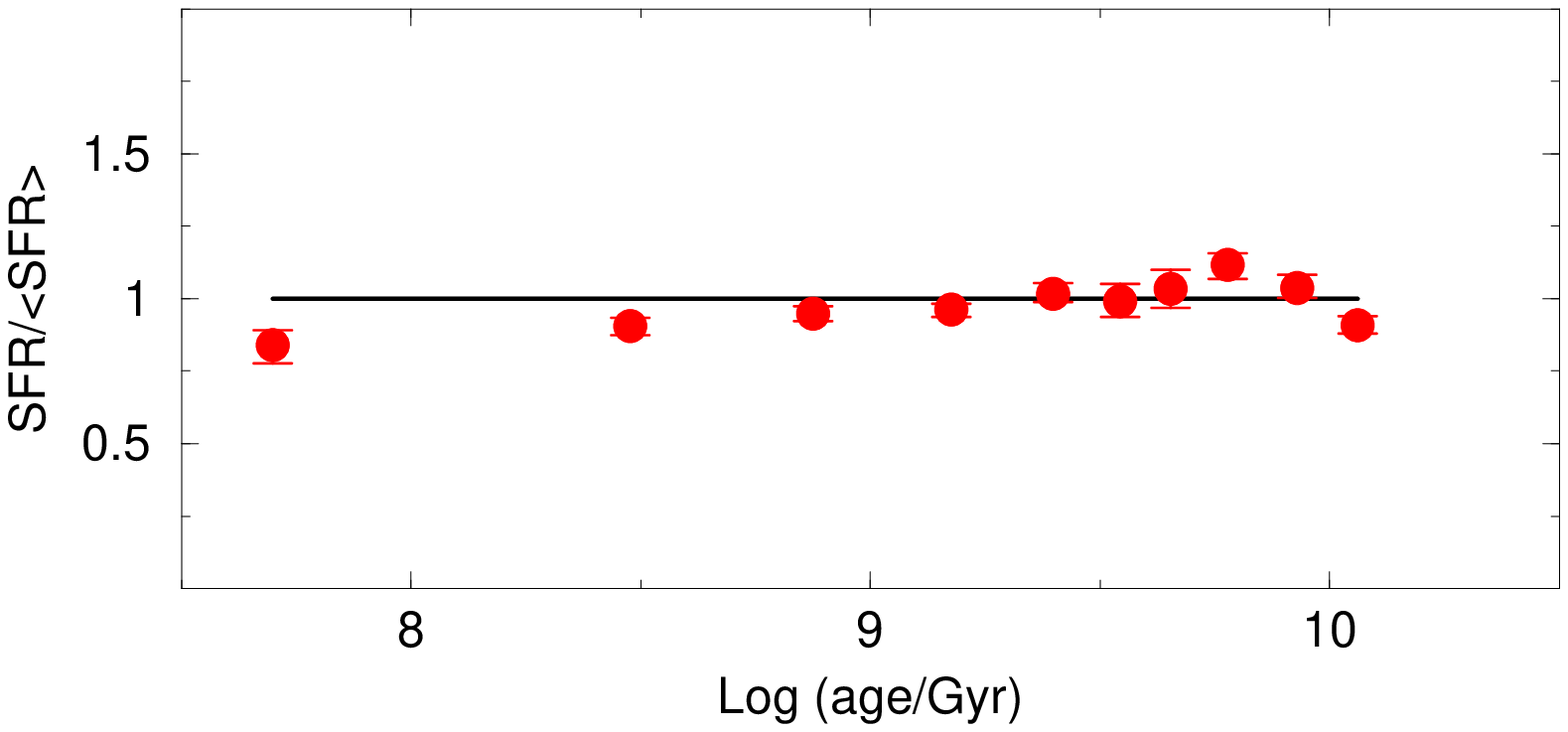}
\epsfxsize= 8.5 cm \epsfbox{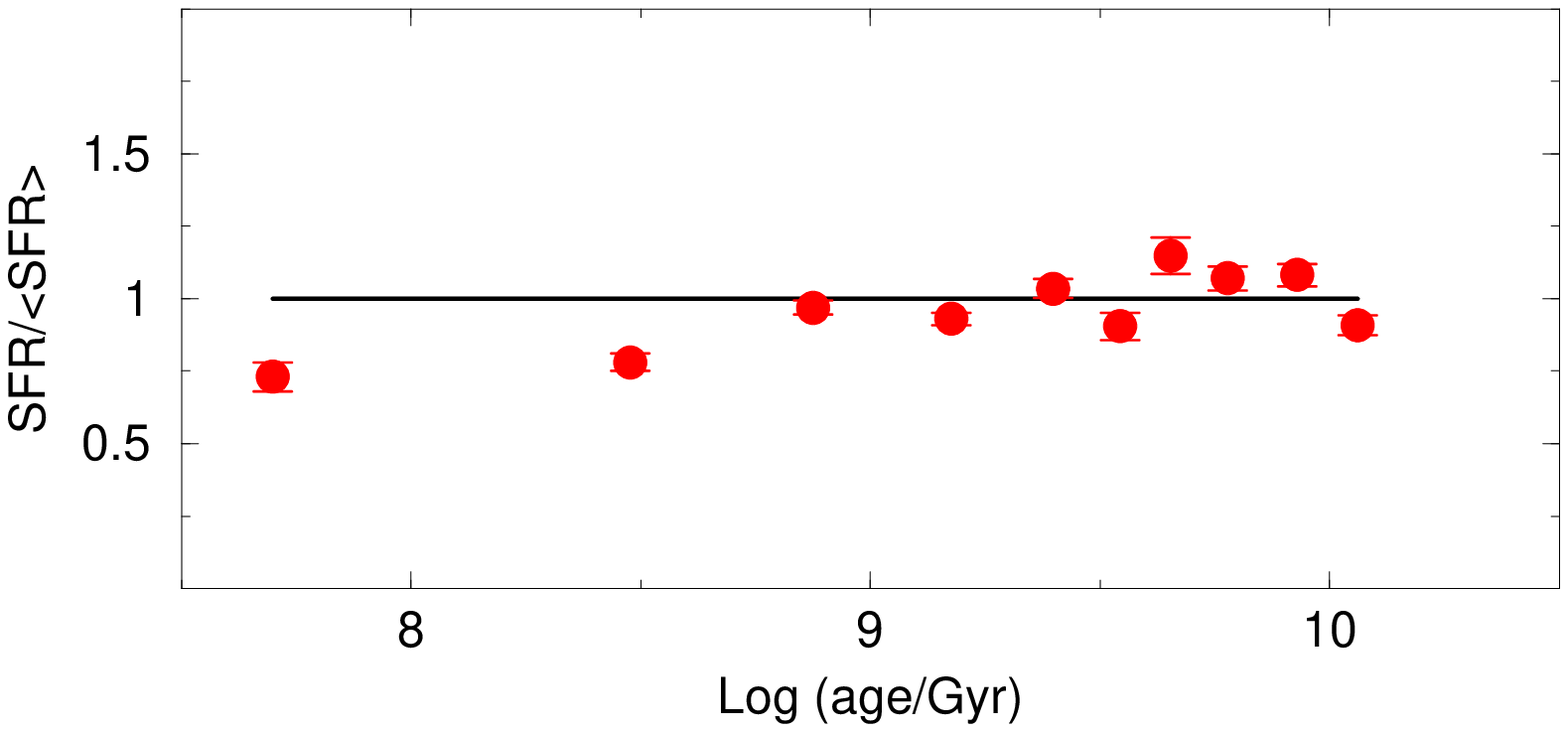}
\epsfxsize= 8.5 cm \epsfbox{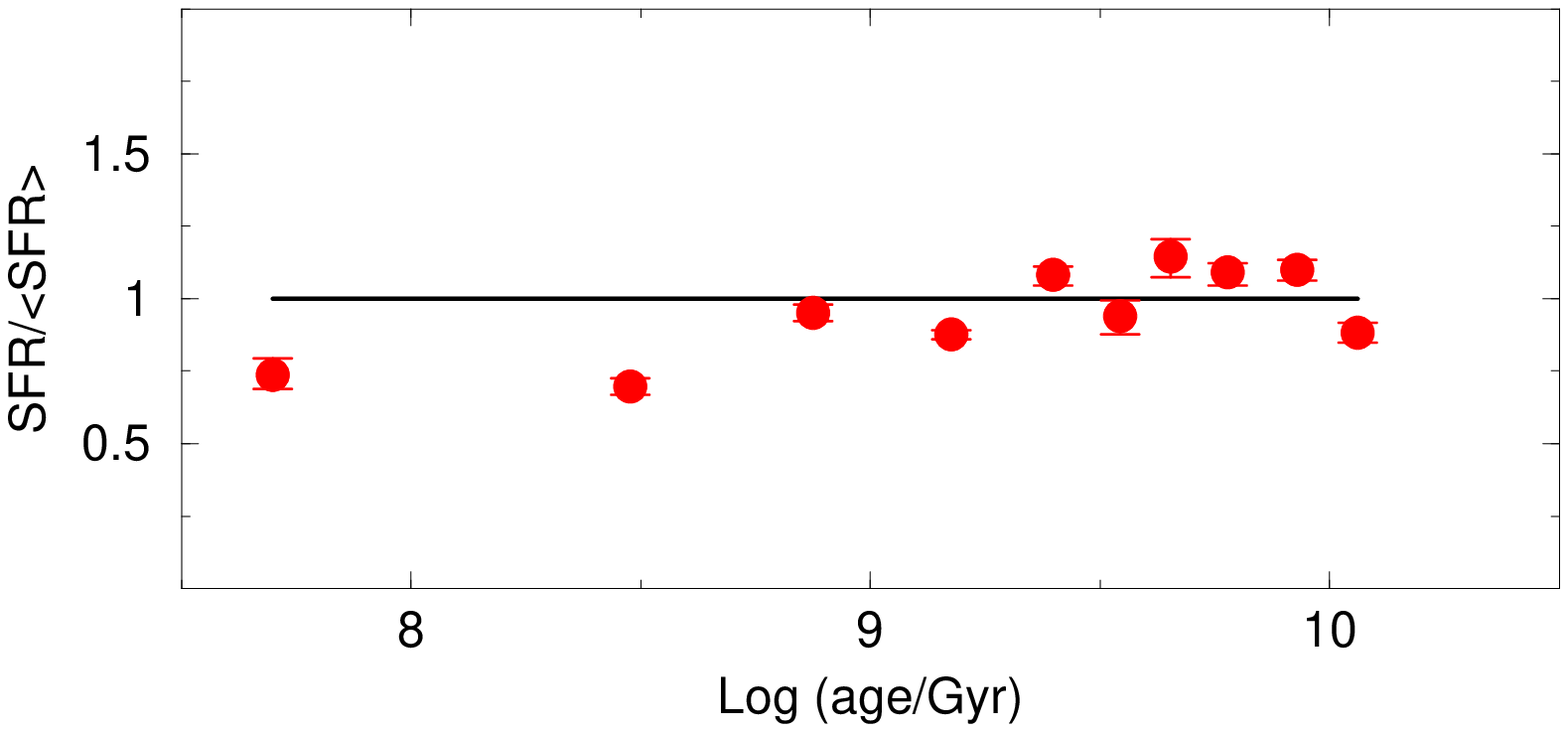}
\caption{Effect of unresolved binaries. Three fake populations are
  built with different percentage of binary stars (10\%, 20\% and
  30\%). The SFR is recovered using only single stars.}
\label{bina}
\end{figure*}

Figure \ref{bina} shows the results: as for the IMF, also in this case
a modest systematic effect is visible. This is because the stars which
are in binary systems are brighter and redder than the average single
star population (for an in depth analysis see
e.g. \citep[][]{Hurley98b}). For the recent SFH, this corresponds to
moving lower MS stars from a star formation step to the contiguous
\emph{older} step: in this way, the most recent star formation step is
emptied of stars, mimicking a lower activity. Intermediate SF
epochs are progressively less affected, because some stars get in and
some stars get out of the step bin. For the oldest epochs the
situation is opposite: the binary effect is to move stars towards
younger bins. Here the SGB, that is the main signature of any old
population, is brighter because of the binaries, mimicking a younger
system.


\subsection{Metallicity and metallicity spread}
 
The precise position of a star on the CMD depends on the chemical
composition, namely the mass fraction of hydrogen, helium and metals
(X, Y, Z respectively). The Z content mainly changes the radiative
opacity and the CNO burning efficiency: the result of a decreasing Z
is to increase the surface temperature and the luminosity of the
stars. This has two consequences of relevance for us: a) metal poor
stars have a shorter lifetime compared to the metal-rich ones (because
over-luminous and hotter), b) a metal-poorer stellar population is
bluer, but can be mistaken for a younger but metal-richer population.

\begin{figure*}[]
\epsfxsize= 8.5 cm \epsfbox{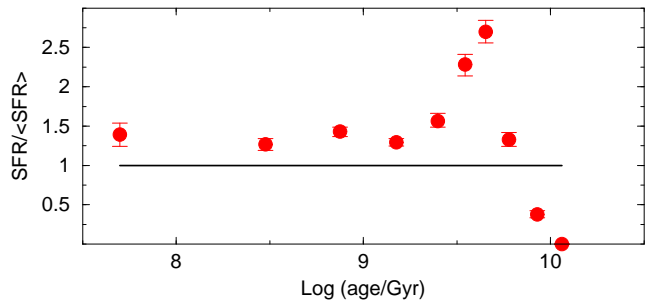 }
\caption{ Sensitivity test to metallicity. The reference fake galaxy
  has a variable composition: $Z=0.004$ for stars younger than 5 Gyr,
  $Z=0.002$ for older stars. The red dots represent the SFR as
  recovered using a single metallicity $Z=0.004$ for all epochs.}
\label{metals}
\end{figure*}

To test these effects, the first stars (ages older than 5 Gyr) in our
reference fake population are built with a slightly different metallicity
($Z=0.002$) than the younger objects, which have the usual $Z=0.004$. Then,
we recover the SFH by adopting a model with $Z=0.004$ independently of
age. The results are shown in Figure \ref{metals}: neglecting that the
oldest population of our galaxy was slightly metal poorer, systematic, non 
negligible discrepancies appear in the recovered SFH. It is the
classical age-metallicity degeneracy: to match the blue-shifted
sequences of old metal poorer stars, our models with wrong metallicity
must be younger. Note that the overall trend of the young SFR is not
significantly biased, while the old SFR is now significantly different.

This result is a strong warning against any blind attempt to match the
CMD with a single (average) metallicity, especially considering that
many galaxies exhibit a pronounced age metallicity-relation,

During and at the end of their life stars pollute the surrounding
medium, so we expect that more recently formed stars have higher
metallicity and helium abundance than those formed at earlier epochs.
The progressive chemical enrichment with time results from the
combined contribution of stellar yields, gas infall and outflows,
mixing among different regions of a galaxy. Observational studies have
shown that several galaxies reveal a metallicity spread at each given
age.

\begin{figure*}[]
\epsfxsize= 6 cm \epsfbox{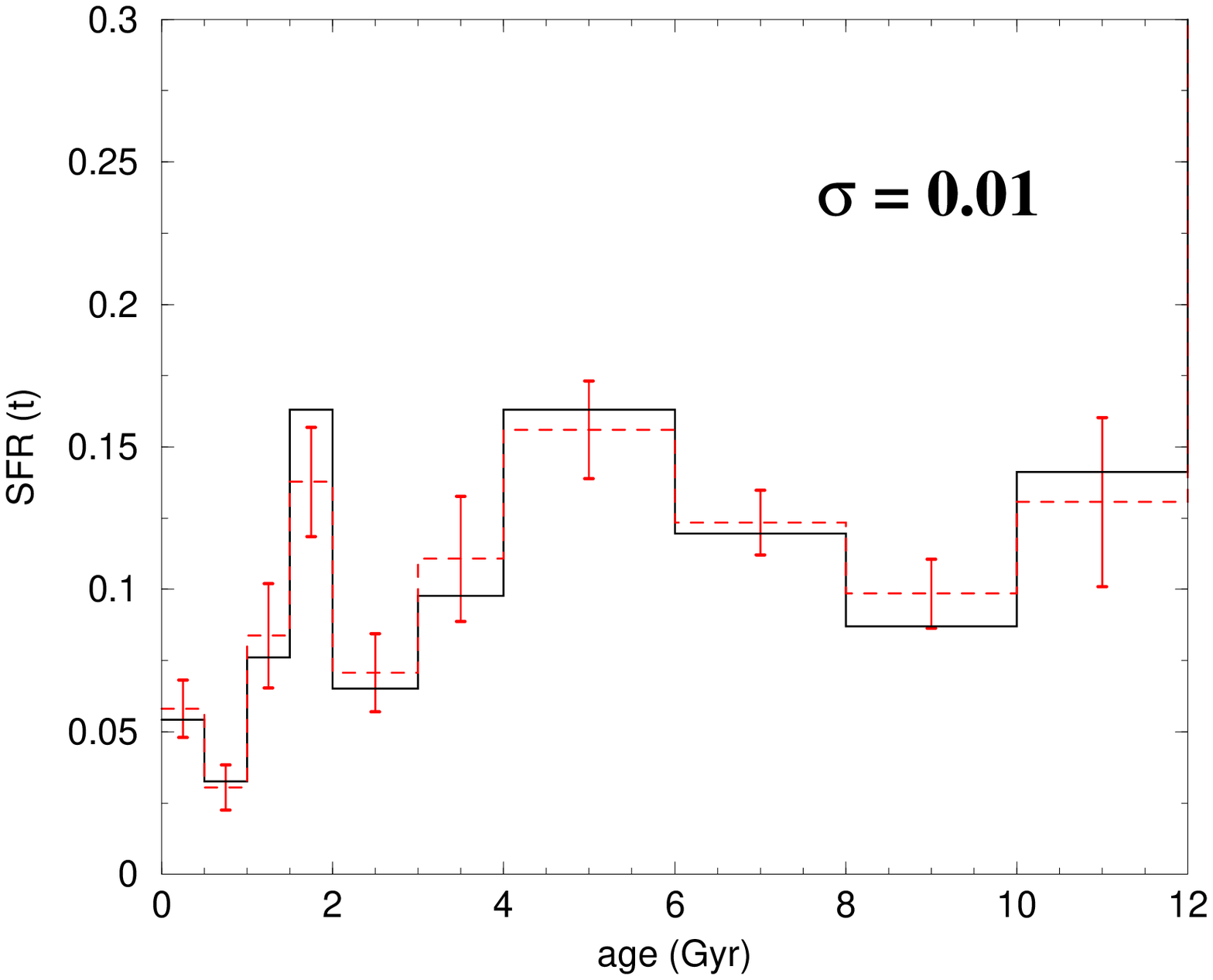}
\epsfxsize= 6 cm \epsfbox{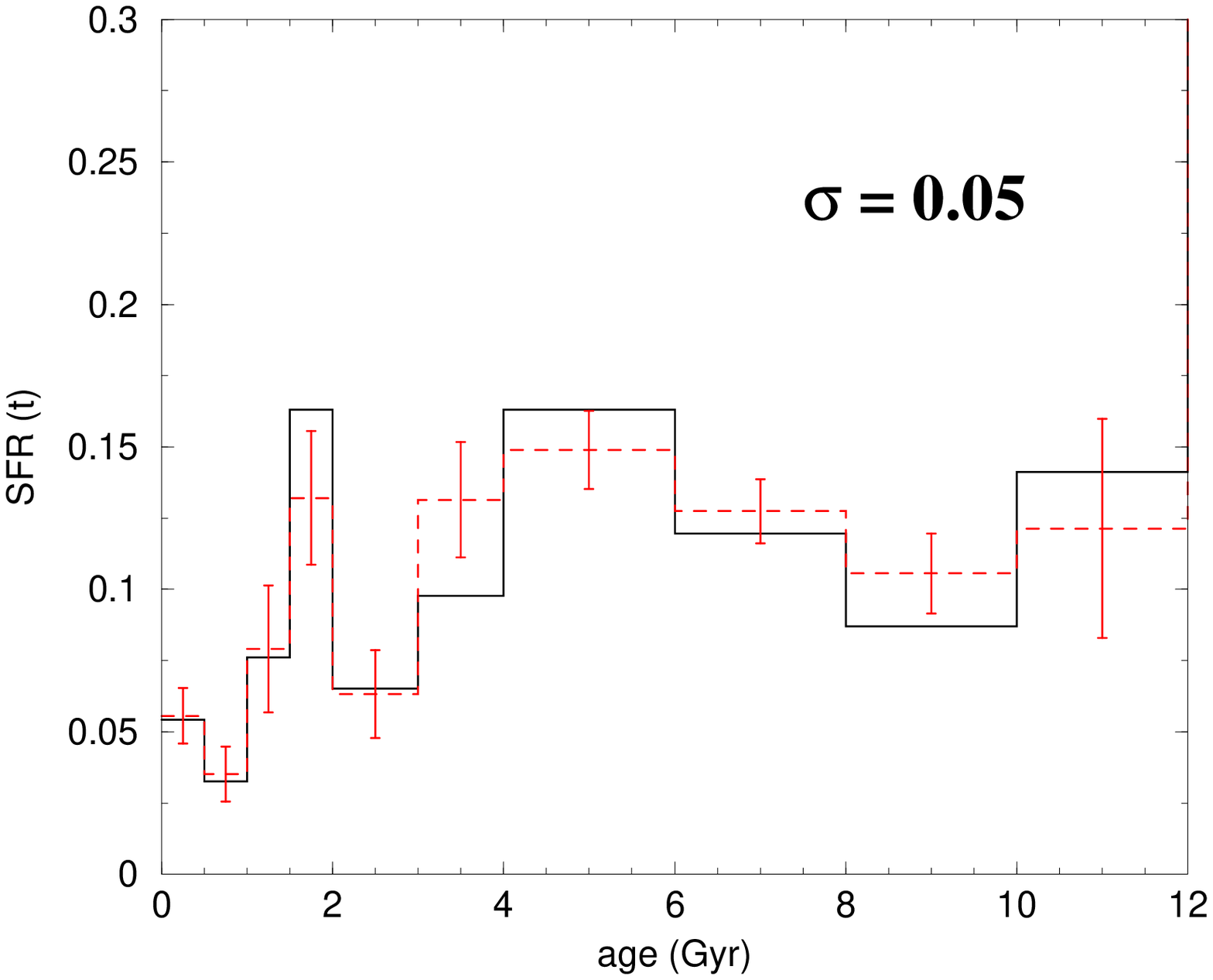}
\epsfxsize= 6 cm \epsfbox{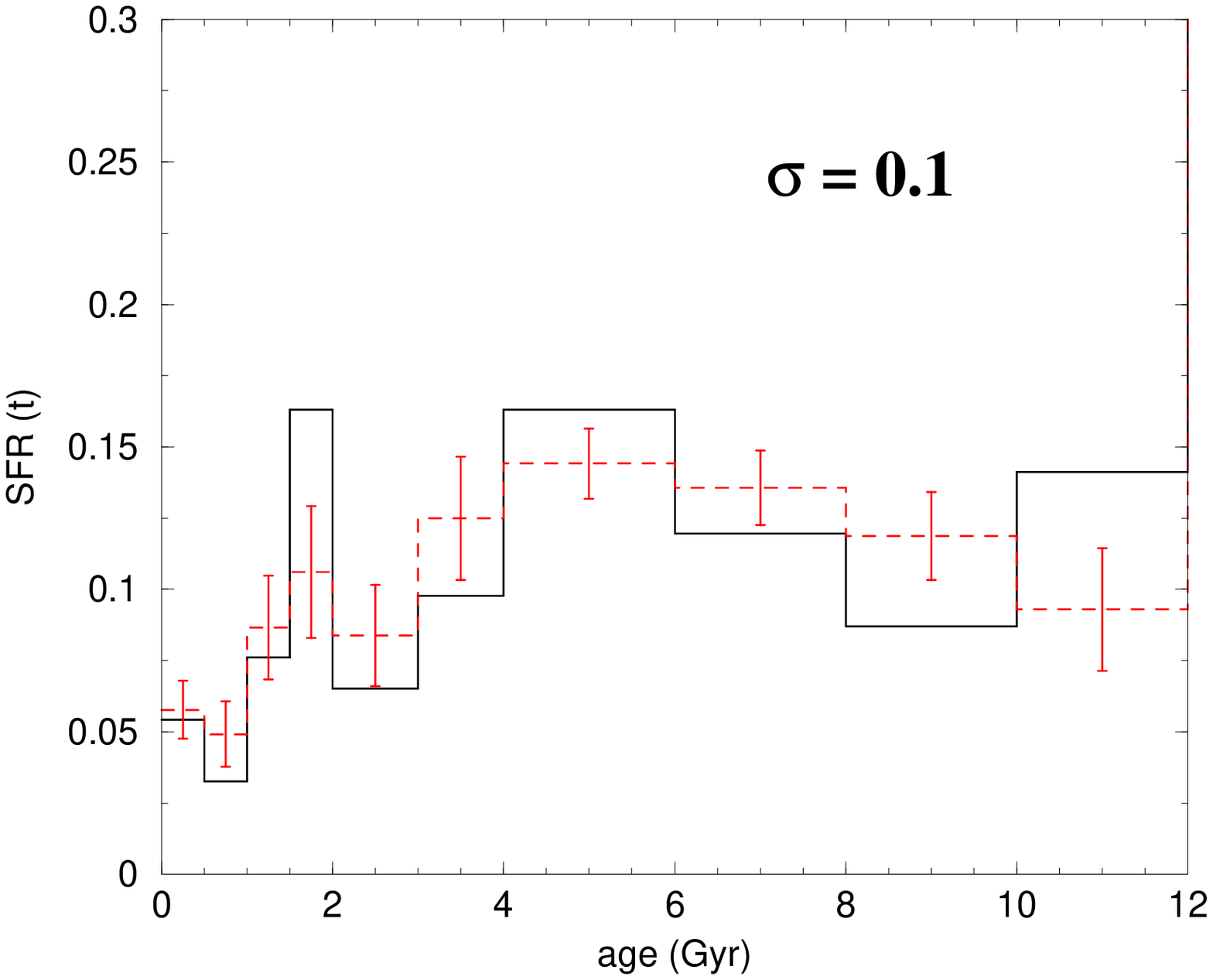}
\epsfxsize= 6 cm \epsfbox{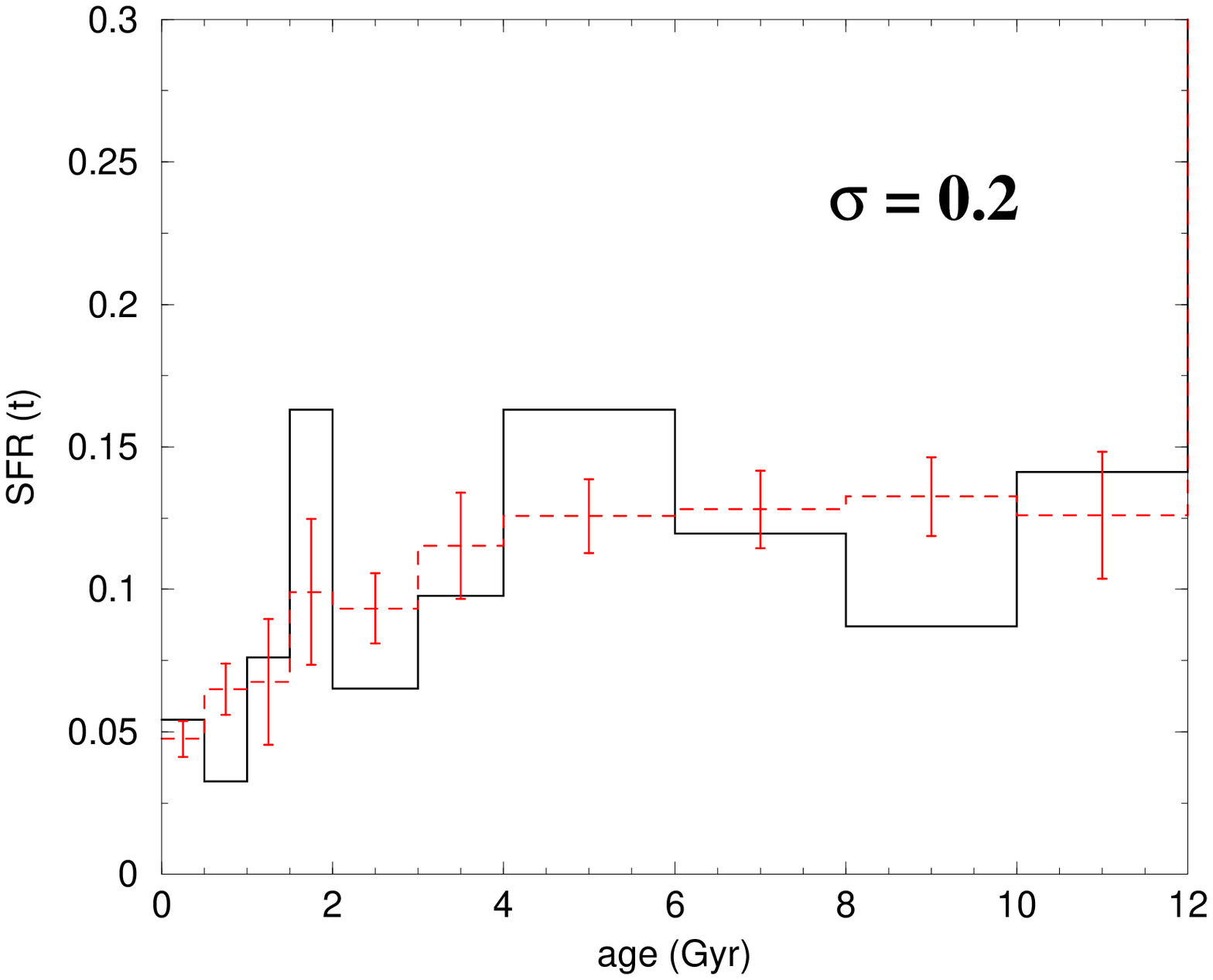}
\caption{Sensitivity test to the metallicity dispersion. Solid line:
  SFH assumed for the fake population. Dashed line: recovered SFH. The
  $\sigma$ value indicates the dispersion in $[Fe/H]$, used for the
  reference artificial data. The model has the same mean metallicity,
  but no dispersion.}
\label{disp}
\end{figure*}

In \citep[][]{Cignoni06} the SFH sensitivity to a metallicity
dispersion is tested: there several fake populations were generated with
a mean metallicity $Z=0.02$ plus a variable dispersion from
$\sigma=0.01$ dex to $\sigma=0.2$ dex in $[Fe/H]$. Then, the SFH was
searched adopting in the model the same mean metallicity of the
artificial data, but \emph{without metallicity spread}. The results
are shown in Fig. \ref{disp}. Above $\sigma=0.1$ dex, the retrieved
SFH differs significantly from the true one: this numerical experiment
points out that the metallicity dispersion can be a non negligible factor.

\subsection{Helium content} 

The SFH retrieval must be also tested against possible variations of
the helium abundance for a given metallicity. The abundance of He and
metallicity influence the stellar structure through the molecular
weight. Increasing Y corresponds to increasing the molecular weight
and affects the hydrostatic equilibrium. The pressure decreases and
the star shrinks (producing heat), reaching a new equilibrium
characterized by a smaller radius and a higher central temperature. As
a consequence, the efficiency of the central burning increases and
this makes the star brighter and hotter.

Helium absorption lines appear only in the spectra of very hot
stars. Hence, the traditional procedure to infer the helium abundance
of a standard stellar population cannot be via direct measures and is
instead based on the correlation (which is assumed linear) between the
helium mass fraction $Y$ and the metal abundance $Z$.

\begin{figure}
\epsfxsize= 8 cm \epsfbox{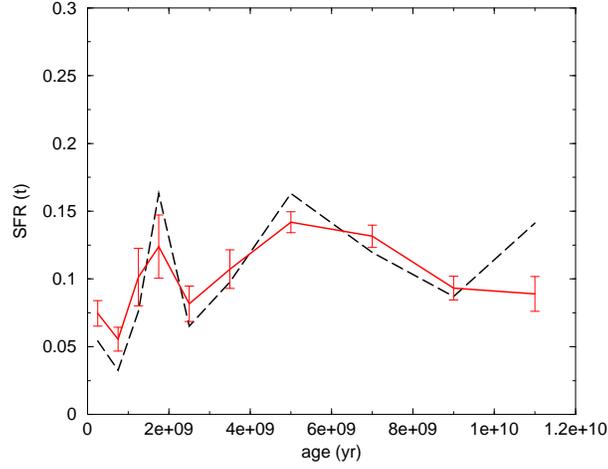}
\caption{Experiment of the helium effect on the SFH: reference fake
  population and model have the same metallicity $Z=0.004$, but
  different helium content ($Y=0.23$ for the model and $Y=0.27$ for
  the artificial data). The black dashed line is the input SFH. The
  red solid line is the recovered SFH.}
\label{hewrong}
\end{figure}

In order to explore the effects of a wrong choice of the helium
content we have built a reference fake population with $Z=0.004$ and
$Y=0.27$ and we have tried to recover its SFH using a $Z=0.004$ model
coupled with $Y=0.23$, which is obviously a very extreme assumption
and therefore provides a stringent upper limit to the possible Y
effects on the SFH. The result is shown in Figure \ref{hewrong}: there
are a number of small variations, but the general morphology is well
reproduced.
The small features of the recovered SFR are blended (the peaks are
attenuated) but the overall trend is still recovered.

\subsection{Reddening and distance}

\begin{figure*}[]
\epsfxsize= 8.5 cm \epsfbox{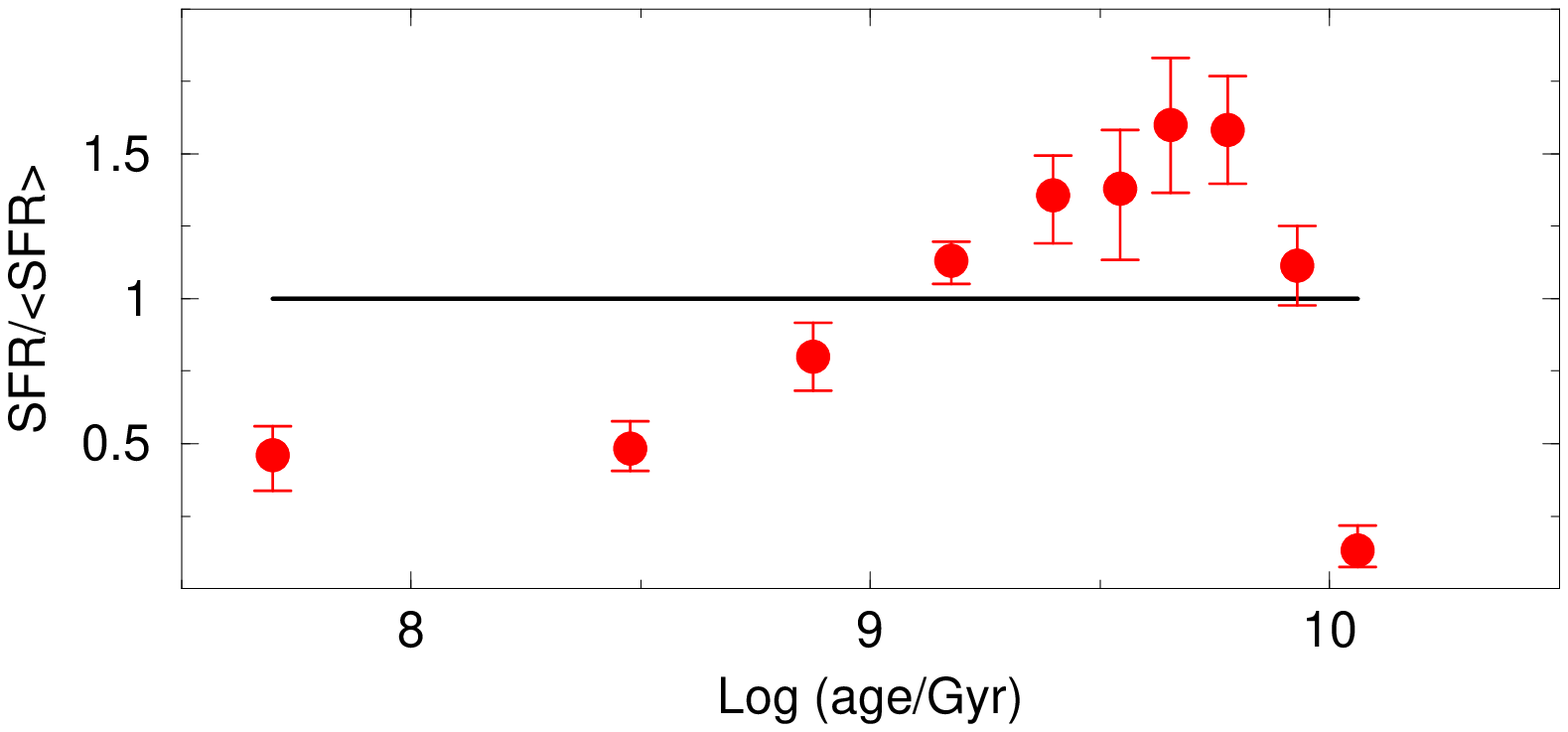}
\epsfxsize= 8.5 cm \epsfbox{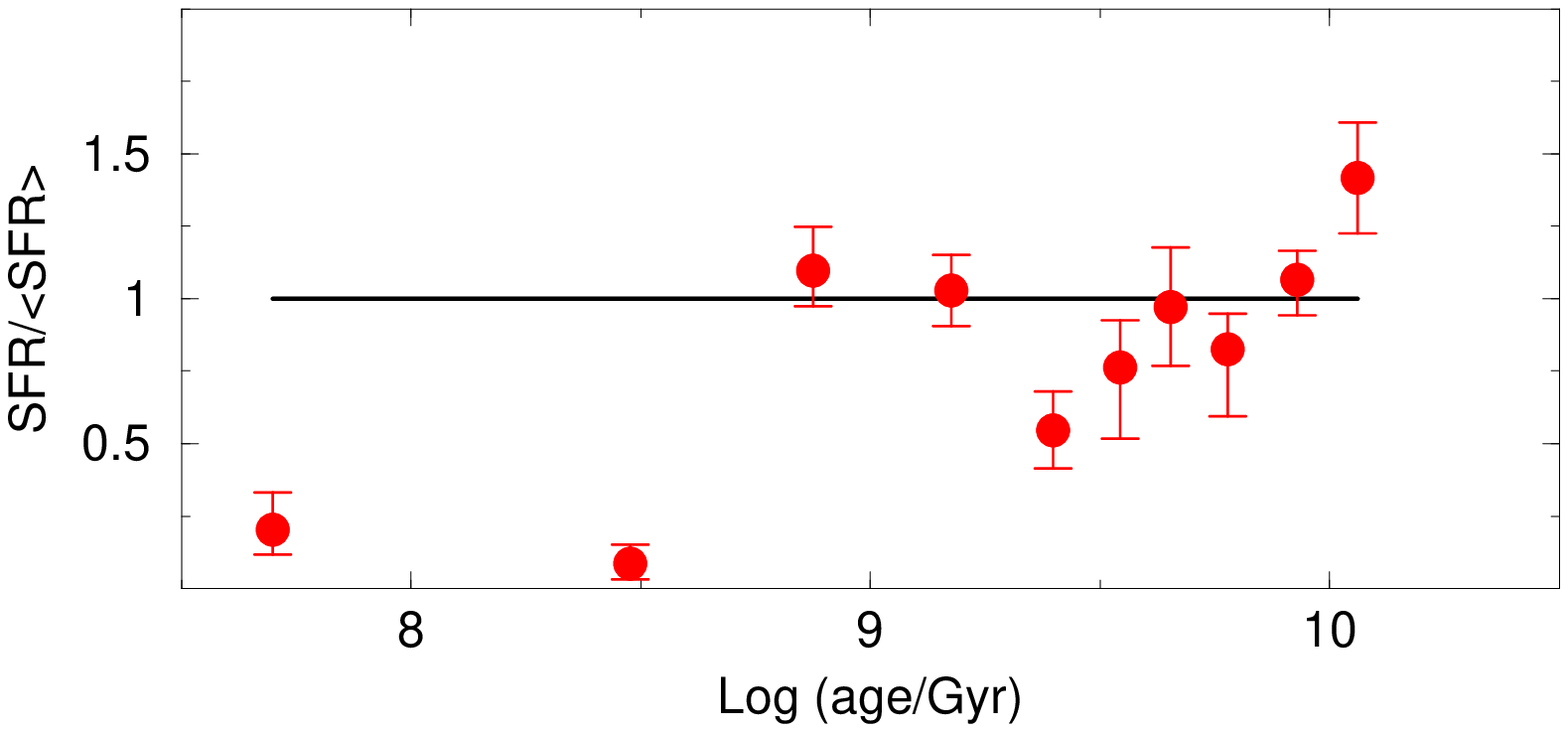}
\caption{Sensitivity test to reddening and distance. Red dots
  represent the recovered SFH. In the left panel the reference fake
  population is built with a distance spread $(m-M)_{0}=18.6$, while
  the model used to retrieve the SFH adopts $(m-M)_{0}=18.9$. In the
  right panel the reference fake population is built with reddening
  $E(B-V)=0.16$, while the model used to retrieve the SFH adopts
  reddening $E(B-V)=0.08$.}
\label{dist_red}
\end{figure*}

In order to test how wrong choices of reddening and distance
invalidate the possibility to recover the right SFH, we consider two
extreme situations: first, the fake galaxy is put at $(m-M)_{0}=18.6$,
with the models still assuming $(m-M)_{0}=18.9$; second, the fake
galaxy is reddened using $E(B-V)=0.16$ mag, while the models adopt
$E(B-V)=0.08$. The results are shown in Figure \ref{dist_red}: it is
evident that the SFH is not recovered, and shows large deviations from
the reference case at any epoch. However, it is worth to remind that
we have derived the best models from a blind $\chi^2$ minimization: it
is clear that at the distance of SMC, with the beautiful HST
photometry available, both a visual inspection of the CMD morphology
(in particular the blue MS envelope) and the evaluation of the
$\chi^2$ probability would be effective to reject all the models with
wrong distance and reddening.  The problem is more challenging for
more distant galaxies, where the only observables are massive stars
and reddening and distance are difficult to constrain.

Other potential problems are related to differential reddening and
line-of-sight depth. There are several indications that some galaxies
or portions of them are affected by differential reddening. For
instance, young stars can be still surrounded by relics of their
birthing cocoon material and suffer an additional amount of
absorption. The first signature of a differential reddening is the red
clump morphology when it appears elongated and/or tilted \citep[see
  e.g.][]{Olsen99}. Another effect is a smeared appearance of color
and magnitude in any stellar evolution sequence of the CMD.

A finite line-of-sight depth is a natural expectation, at least for
nearby galaxies where the physical extension can be a non negligible
fraction of their distance from us.  As an example, according to
\citep[][]{Stanimi04}, the SMC may have a depth of 14-17 kpc,
corresponding to a spread in magnitude of few tenths of
magnitude. This effect may alter the evolutionary information from the
clump and the RGB. A correct description of the stellar populations
thus requires analyses involving the spatial structure. Both
differential reddening and line-of-sight effects can be dealt with as
additional free parameters.

\begin{figure*}[]
\epsfxsize= 8.5 cm \epsfbox{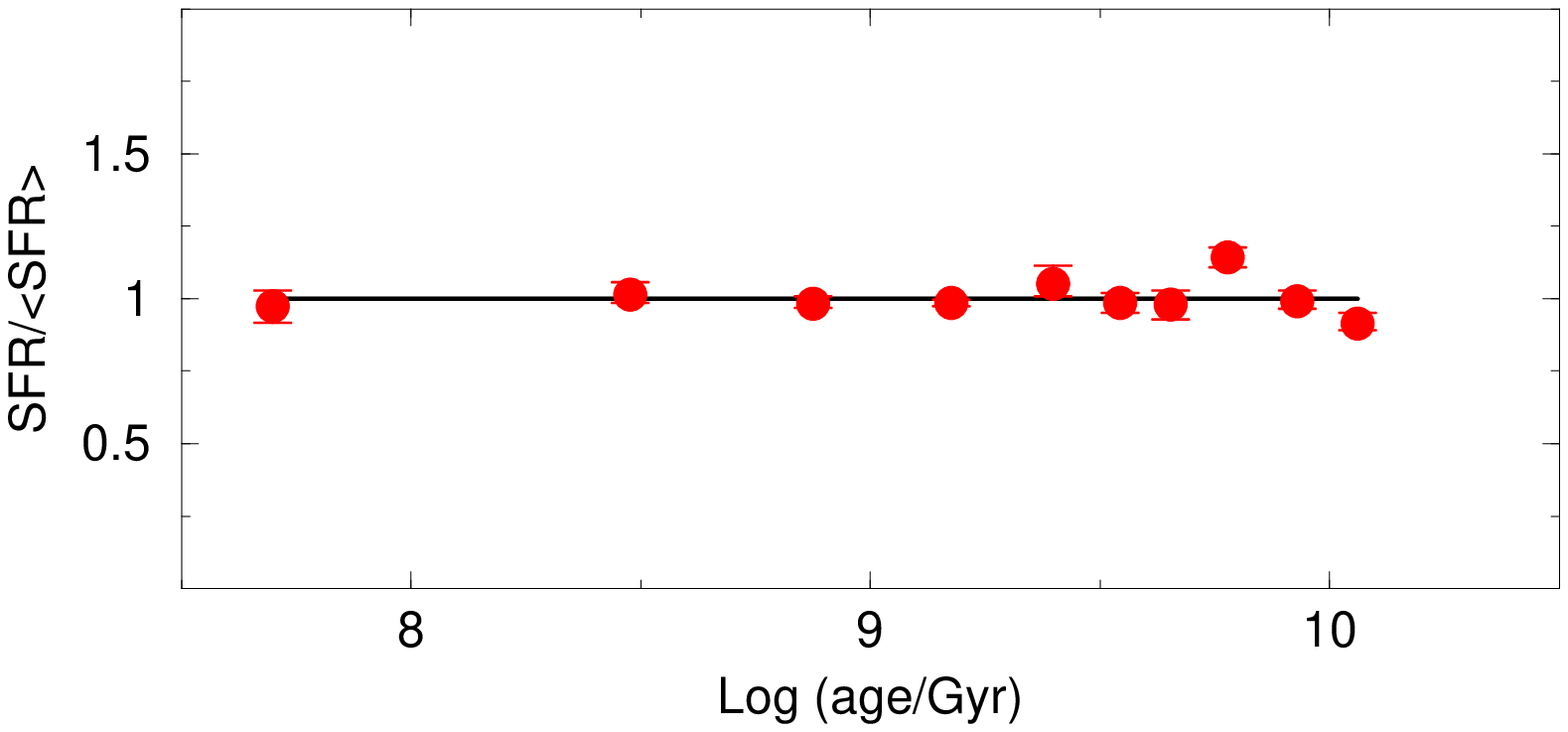}
\epsfxsize= 8.5 cm \epsfbox{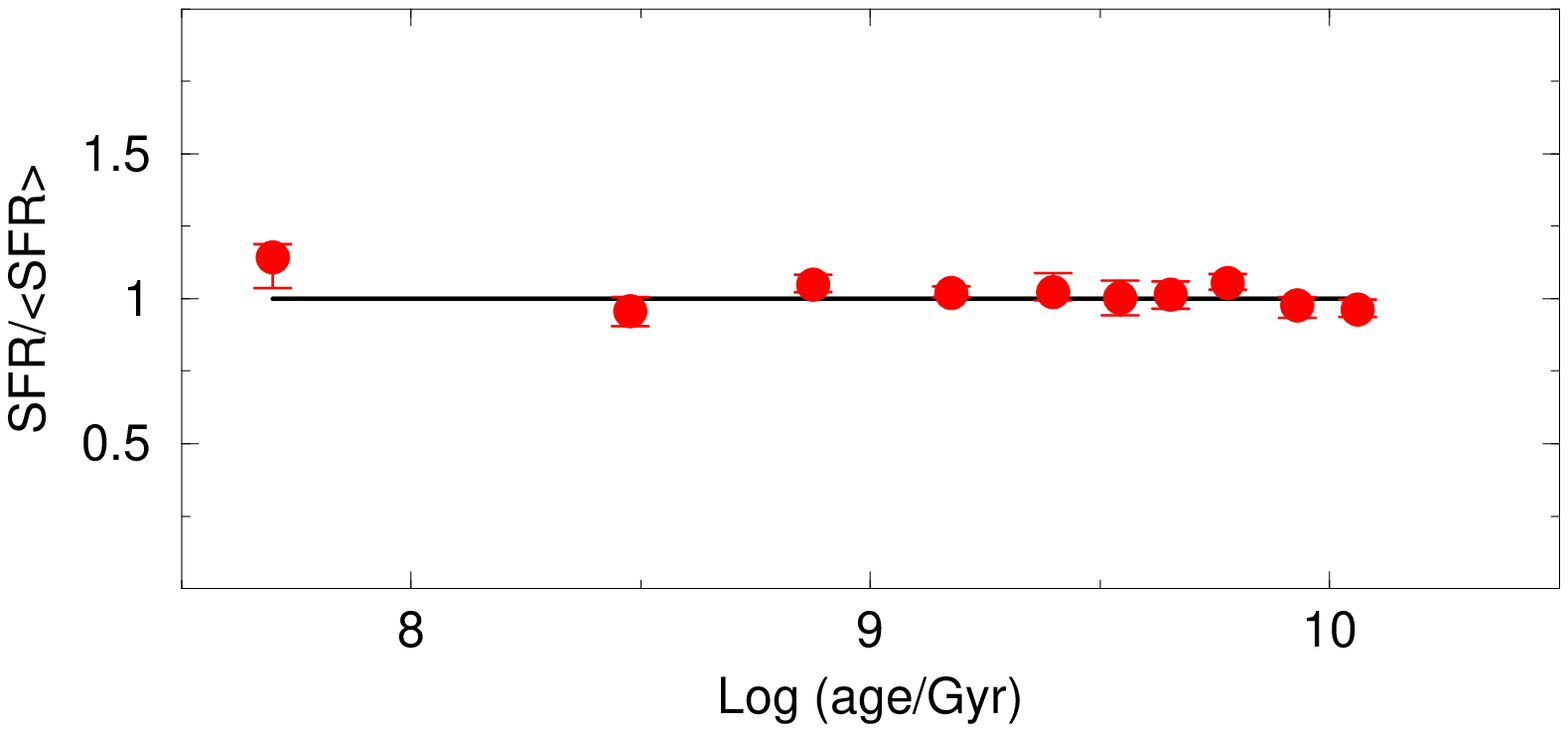}
\caption{Sensitivity test to differential reddening and line-of-sight
  spread. Red dots represent the recovered SFH. In the left panel the
  reference fake population is built with a distance spread
  $(m-M)_{0}=18.9\pm0.2$, while the model used to retrieve the SFH has
  a single distance $(m-M)_{0}=18.9$. In the right panel the reference fake
  population is built with a differential reddening
  $E(B-V)=0.08\pm0.04$, while the model used to retrieve the SFH has a
  single reddening $E(B-V)=0.08$.}
\label{dist_red_spread}
\end{figure*}

In order to check how the recovered SFH is influenced by differential
reddening and line-of-sight spread, we built two fake populations with
the following features: the first has a reddening dispersion
$E(B-V)=0.08\pm0.04$, the second has a distance spread
$(m-M)_{0}=18.9\pm0.2$. Figure \ref{dist_red_spread} shows the result when
the SFH is searched using the canonical combination
$E(B-V)=0.08$ and $(m-M)_{0}=18.9$. The reference SFH is still fully
recovered. The reason is in the random nature of these uncertainties
that produce a blurred CMD, but not a systematic trend.

\subsection{Additional issues}

The previous experiments do not exhaust all the possible sources of
uncertainty. Rather, they represent the best understood and manageable
ones: among others, convection, atmospheres and color transformations,
stellar rotation, mass exchange in binary systems may be relevant as
well. Moreover, we have explored each single bias separately, while in
a real galaxy several uncertainties may be at work simultaneously with
different intensities. In this case, the final effect may not be the
simple summation of the previous results: some of the effects may
compensate each other or conspire to build an uncertainty larger than
the sum of the individual uncertainties.

For example, in the metallicity experiment, when the reference old
population was metal poorer than in the search procedure, the
retrieved SFH resulted younger. However, this is true \emph{only} if
the best SFH is searched with fixed reddening. Letting the reddening
vary could lead to a different SFH (and reddening).

To summarize, the final uncertainty on the recovered SFH strongly depends on 
which parameter space is explored. Generally speaking, when the best photometric
conditions are achieved and all the parameter space is properly covered, within
the reached lookback time the
error on the epochs of the SF activities is around ten percent of their age and
that on the SFR is of the order of a few. With poorer photometry or with coarser
procedures the uncertainties obviously increase, but the qualitative scenario is
usually reliably derived.

\section{Star formation histories of resolved dwarfs from synthetic CMD
analyses: results}

In spite of the uncertainties described above in the identification of
the {\it best solution} for the SFH, the synthetic CMD method is
extremely powerful in reducing the range of acceptable scenarios,
i.e. the range of values of the various parameters. As demonstrated
every time different synthetic CMD procedures have been applied to the
same galaxy region, all the solutions come out consistent with each
other (see e.g. the {\it Coimbra experiment} on the LMC bar
\cite{Skillman02}, and IC1613 \cite{Skillman03}). We can therefore
dare drawing some general conclusions from the results obtained so far
with this method.

\begin{figure}[htbp]
\begin{center}
\epsfxsize=6.5in
\epsfbox{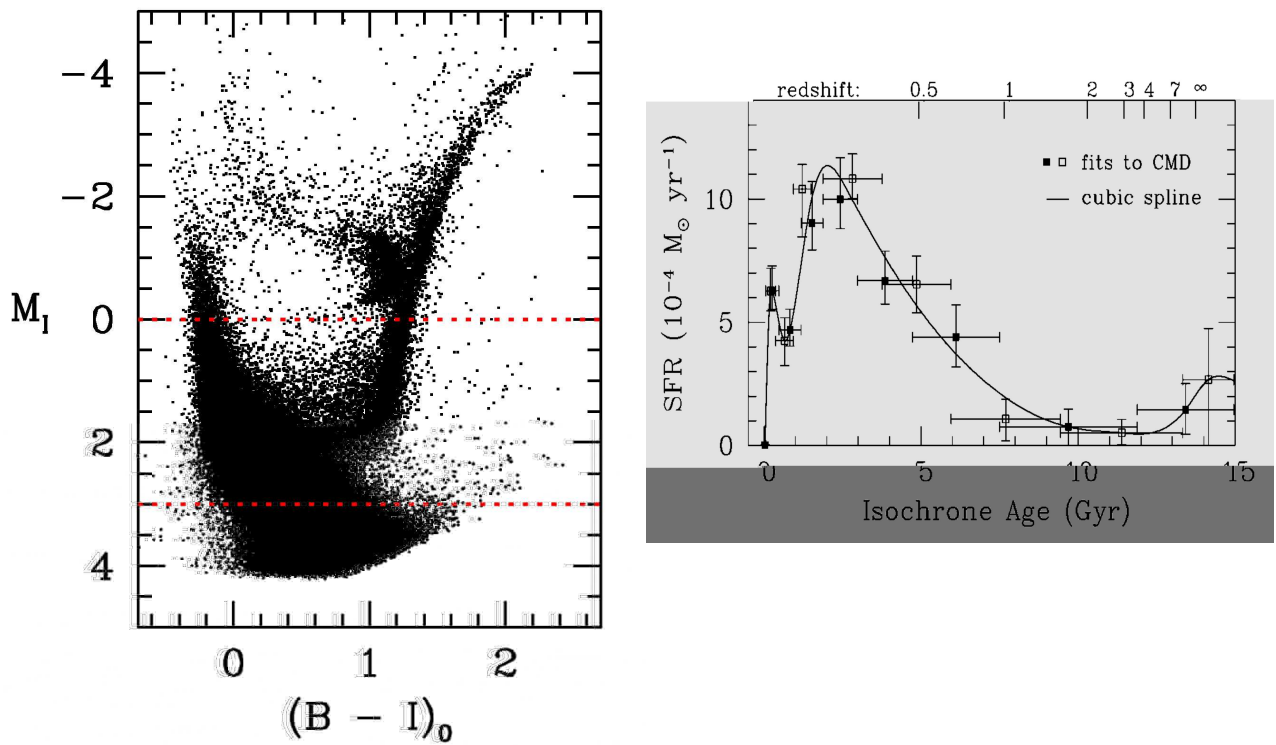}
\end{center}
\caption{CMD and SFH of Leo~A as derived by \cite{Cole07} from HST/ACS data.
Notice the impressive depth and tightness of the CMD, allowing to infer the SFH
even at the earliest epochs. Courtesy A. Cole.}
\label{leoa}
\end{figure}

Since the dawn of its application, the method immediately proved its
power.  First, it was found that the SFH differs significantly from
one galactic region to the other even in tiny systems such as WLM, the
first dwarf irregular (dIrr) in the Local Group to which the method
was applied \citep{Ferraro89}.  As soon as a few other nearby
irregulars were studied, it turned out that, contrary to the common
belief of the time, the SF activity in late-type dwarfs within the
lookback time spanned by the available photometry has occurred in long
episodes of moderate intensity, separated by short quiescent phases,
rather than in short episodes of strong intensity, separated by long
quiescent intervals
\citep{Tosi91,Greggio93,Marconi95,Gallart96a,Tolstoy96a}. In other
words a {\it gasping} \citep{Marconi95} rather than a {\it bursting}
regime.

Nowadays, one can resolve individual faint/old stars in galaxies of
the LG and its immediate vicinities, and infer their SFHs over long
lookback times. In the (still few) cases when the oldest MSTO is
reached, the SFH can be derived over the entire Hubble time, as
already achieved in some regions of the LMC
\citep{Holtzman99,Olsen99,Dolphin00b,Smecker02,Javiel05,Gallart08}, of
the SMC \citep{Dolphin01b,Noel07,Cignoni09} and in Leo~A
\citep{Cole07}.  As an example, Fig.\ref{leoa} shows the CMD obtained
by \cite{Cole07} from ACS imaging of the dIrr Leo A, located at 800
kpc from us \citep{Dolphin02}, and the resulting SFH. In Leo~A the
star formation activity was present, although quite low, at the
earliest epochs, and 90\% of the activity occurred in the last 8 Gyr,
with the main peak around 2 Gyr ago and a secondary peak a few
hundreds Myr ago. This SFH is very similar to that of fields in the
LMC, SMC, IC1613 \citep[][]{Hidalgo09} and other late-type galaxies
and we can consider it typical of dIrrs: a rather continuous star
formation since the earliest epochs, but with significant peaks and
gasps. Notice that the main SFR peak in dIrrs rarely occurs at the
most recent epochs.

The high spatial resolution of HST cameras also allows to spatially resolve the
SF activity, at least within relatively recent epochs. For instance, 
\cite{Dohm98} and \cite{Dohm02} have 
measured the SF activity over the last 0.5 Gyr in 
all the sub-regions 
of the dIrrs Gr8 and Sextans A, close to the borders of the LG.  
The resulting space and time distribution of the SF, with lightening and fading of 
adjacent cells, once again shows a gasping regime, and is
intriguingly reminiscent of the predictions of the stochastic self-propagating
SF theory proposed by \cite{Seiden79} almost 30 years ago.

\begin{figure}[htbp]
\begin{center}
\epsfxsize=3.5in
\epsfbox{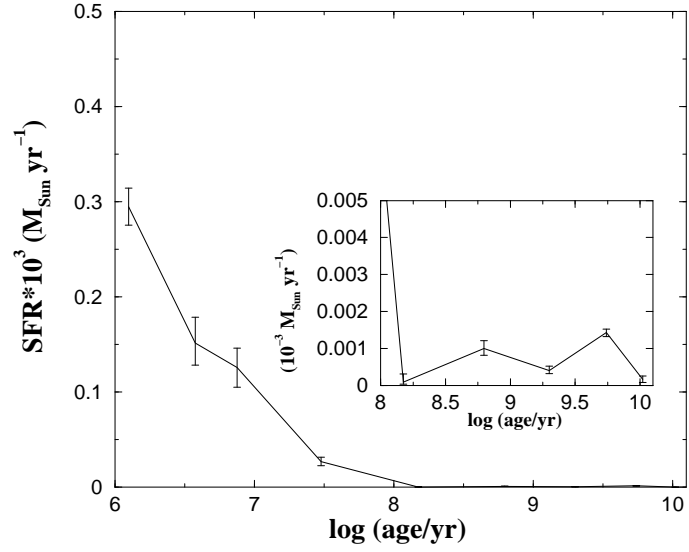}
\caption{ SFH of the ACS field centered on the SMC young cluster NGC602 
as derived by \cite{Cignoni09}. The oldest part of the SFH
  is zoomed-in in the upper right inset. The ACS image and the CMD of the field
  are shown in the left-hand panels of Fig.\ref{pms}.}
\label{sfh_602}
\end{center}
\end{figure}


In the Magellanic Clouds, the conditions are clearly optimal, thanks to their
proximity. 
Fig.~\ref{sfh_602} displays the SFH of the ACS field centered on the very young 
cluster NGC602 in the Wing of the SMC. It shows 
that the cluster has formed most of its
stars around 2.5~Myr ago, while the surrounding field has formed stars
continuously since the earliest epochs.
All the studies on the MC fields have found that the SFHs of their different 
regions
differ from one another in the details (e.g. epoch of activity peaks, enrichment
history, etc.) but are always characterized by a gasping regime. 
In the LMC a clear difference has been found between the SFH of field
stars and of star clusters, the latter showing a long quiescence phase absent in
the field. This difference is not found in the SMC.

\begin{figure}[htbp]
\begin{center}
\epsfxsize=6.5in
\epsfbox{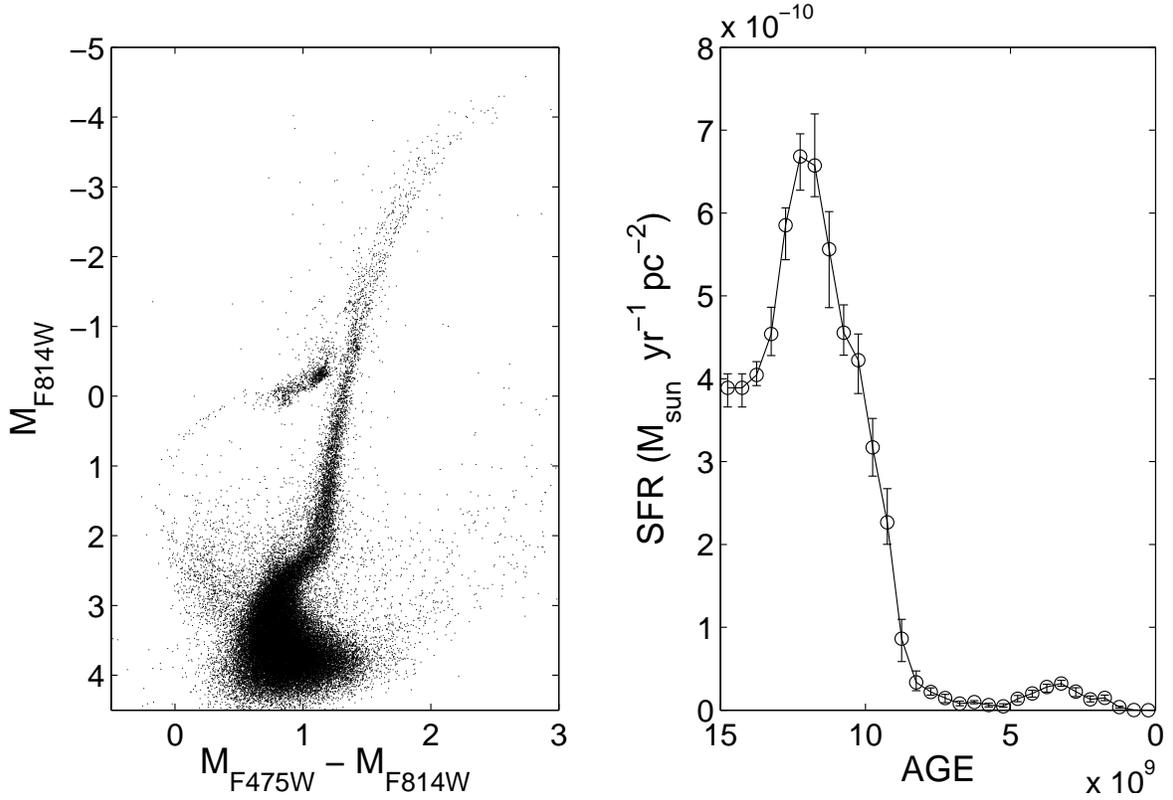}
\end{center}
\caption{CMD and SFH of Cetus as derived by Monelli et al. (in preparation) 
from HST/ACS data in the framework of the L-CID project \citep{Gallart07}.
Notice again the impressive depth and tightness of the CMD, allowing to infer 
the SFH even at the earliest epochs. Courtesy M. Monelli and C. Gallart.}
\label{cetus}
\end{figure}

To find SFHs peaked at earlier epochs one needs to look at early-type
dwarfs: dwarf ellipticals (dEs), dSphs and even transition-type dwarfs
clearly underwent their major activity around or beyond 10 Gyr ago
\citep[][]{Hidalgo09}. The latter also have significat activity at
recent epochs \citep[e.g.][]{Young07}.  The former have few (or no)
episodes of moderate activity in the last several Gyrs
\citep[e.g.][]{Smecker96,Hurley98,Gallart99,Hernandez00,Dolphin02,Dolphin05}. 
A beautiful example of CMDs and SFH of a dSph is shown in
Fig.\ref{cetus}. It is the Cetus dSph, observed with HST/ACS by the
L-CID group \citep{Gallart07} and to be published by Monelli et
al. (in preparation). Here the SF activity in the last several Gyrs is
negligible and the strongest peak occurred about 11 Gyr ago
(interestingly, not at the earliest epoch, though).

To date, a large fraction of LG galaxies have been studied to infer
the SFH of at least some of their regions with the synthetic CMD
method \citep[see][and references therein, for updated
  reviews]{Tosi09,Tolstoy09}: the two spirals, M31 and M33, the two
Magellanic Clouds, a dozen dIrrs, 5 transition type dwarfs and about
20 early-type dwarfs (dwarf spheroidals and dwarf ellipticals).  In
some of these fields the photometry has allowed to reach the oldest
MSTO, in others the HB, i.e. the unmistakable signature of SF activity
earlier than 10 Gyr ago \citep[e.g.][]{Held00,Baldacci05,Momany05}.
Attention is being payed \citep{DeJong08} also to the ultra-faint
dwarfs (uFds) recently discovered by the Sloan Digital Sky Survey
around the Milky Way \citep[e.g.][]{Belokurov08}, for their interest
as possible Galactic building blocks.  Additional efforts by various
groups are in progress to obtain deeper and more accurate photometry
in these and other galaxies and derive reliable SFHs over longer
lookback times. For instance, very interesting results are expected
from the L-CID HST program \citep{Gallart07} on the SFH of 6 LG dwarfs
of different type (two dIrrs, two dSphs and two transition type)
observed by the ACS with unprecedented depth and resolution.

\begin{figure}[htbp]
\begin{center}
\epsfxsize=6.5in
\epsfbox{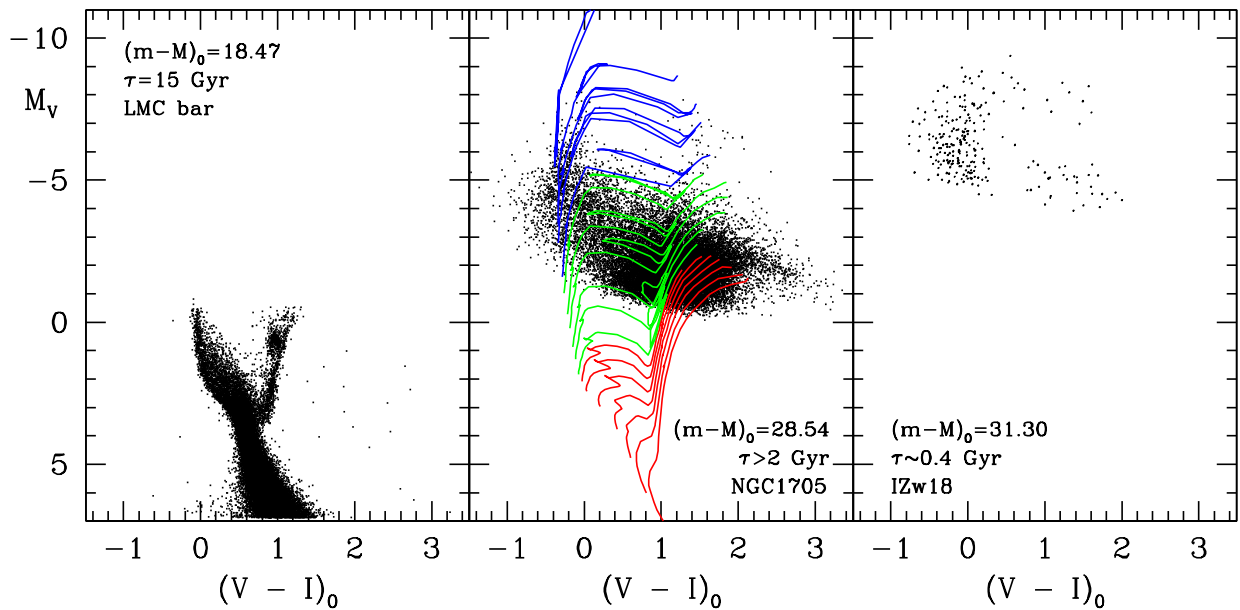}
\end{center}
\caption{Effect of distance on the resolution of individual stars and on the
corresponding lookback time $\tau$ for the SFH. CMD in absolute magnitude and
colour of systems
observed with the HST/WFPC2 and analysed with the same techniques, but
at different distances; from left to right: 50 Kpc (LMC bar), 5.1 Mpc (NGC1705) 
and 18 Mpc 
(IZw18). The central panel also shows stellar evolution tracks from
\cite{Fagotto94b}
for reference: red lines refer to low-mass stars, green lines to intermediate
mass stars, and blue lines to massive stars.
}
\label{dist}
\end{figure}

\begin{figure}[htbp]
\begin{center}
\epsfxsize=3.0in
\epsfbox{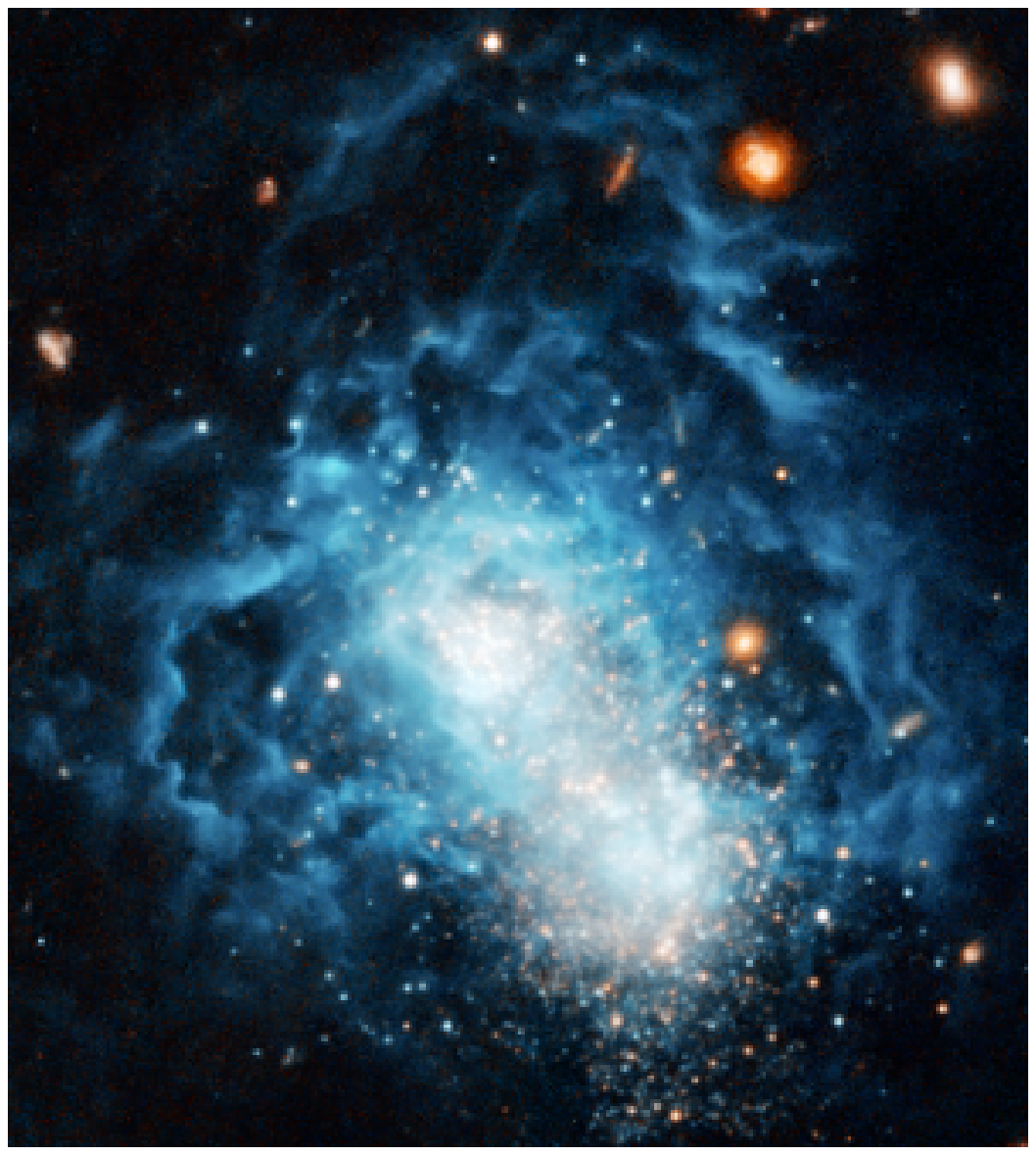}
\epsfxsize=3.0in
\epsfbox{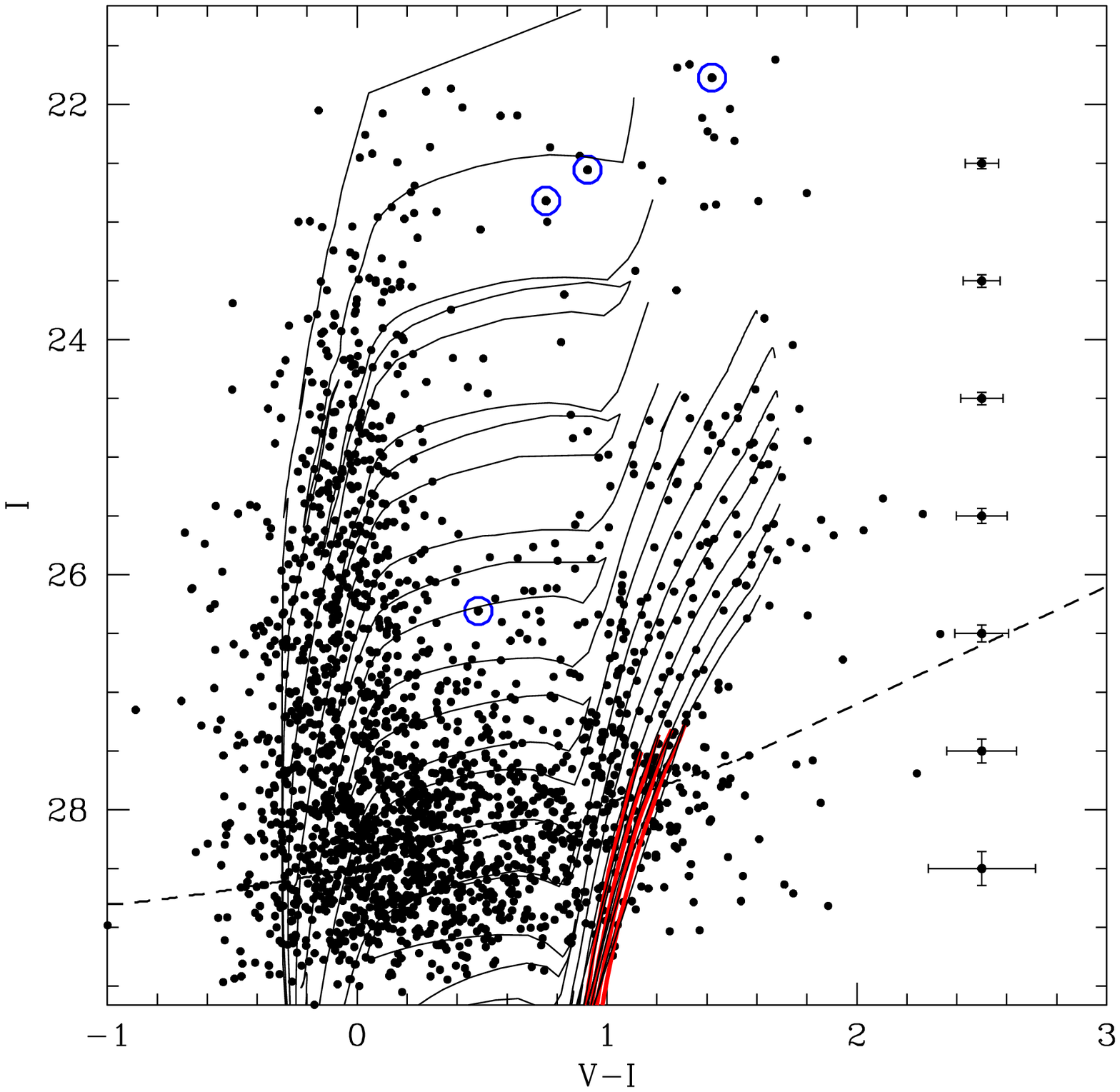}
\end{center}
\caption{Image and CMD of IZw18, obtained from HST/ACS imaging
  \citep{Aloisi07}. Overimposed on the CMD are the Z=0.0004 isochrones
  by \cite{Bertelli94} with the RGB in red. Also shown is the average
  position of the 4 classical Cepheids with reliable light-curves
  obtained from these data. Image credit: NASA, ESA and A. Aloisi
  (STScI, ESA). }
\label{IZw18_acs}
\end{figure}

In galaxies beyond the LG, distance makes crowding more severe, and
even HST cannot resolve stars as faint as the MSTO of old
populations. The higher the distance, the worse the crowding
conditions, and the shorter the lookback time $\tau$ reachable even
with the deepest, highest resolution photometry.  Depending on
distance and intrinsic crowding, the reachable lookback time in
galaxies more than 1 Mpc away ranges from several Gyrs (in the best
cases, when the RGB or even the HB are clearly identified), to several
hundreds Myr (when AGB stars are recognized), to a few tens Myr (when
only the brightest supergiants are resolved).  The effect of distance
on the possibility of resolving individual stars, and therefore on the
reachable $\tau$, is displayed in Fig.\ref{dist}, where the CMDs of
three late-type galaxies are shown, as resulting from WFPC2 photometry
in equivalent observing conditions: the LMC bar \citep{Smecker02},
with a distance modulus of 18.47 (50 kpc) and a CMD reaching a few
mags below the old MSTO; NGC1705 \citep{Tosi01}, with distance modulus
28.54 (5.1 Mpc) and a CMD reaching a few mags below the tip of the
RGB; and IZw18 \citep{Aloisi99}), with the new distance modulus 31.3
(18 Mpc) derived by \cite{Aloisi07} and AGB stars being the
faintest/older resolved stars.

Notice that the new modulus of IZw18 is inferred from the periods and
luminosities of a few classical Cepheids measured from ACS time-series
photometry, which also allowed to obtain a much deeper CMD \citep{Aloisi07}.  
The WFPC2
data shown in Fig.\ref{dist} reach only the AGB, while the CMD
obtained from the ACS (shown in Fig.\ref{IZw18_acs} together with the ACS 
image) reaches below the
tip of the RGB.  Indeed, the unique performances of the ACS have
allowed people to resolve individual stars on the RGB of some of the
most metal-poor Blue Compact Dwarfs (BCDs), such as SBS~1415+437 at 13.6 Mpc
\citep{Aloisi05} and IZw18 at 18 Mpc \citep{Aloisi07}. The discovery
of stars several Gyrs old in these extremely metal-poor systems is a
key information for galaxy formation and evolution studies. It has allowed to detect in BCDs population gradients, the central
concentration of most of the SF activity, the existence of old, metal-poor halos
\citep[e.g.][and subsequent papers]{Schulte99}.

With an amazing success rate, the ACS has allowed people to resolve
individual stars from the brightest and youngest to those as faint and
old as the red giants in an increasing number of dwarfs outside the
LG, in the distance range from 2 to 20 Mpc. This allows to study the
SFH of isolated and interacting dwarfs. People are becoming able to
compare the SFHs of LG dwarfs with those of other groups of galaxies,
such as the M81 \citep[e.g.][]{Weisz08} and the IC342 (Grocholski et
al. in preparation) groups. The resulting CMDs lead to the derivation
of their SFHs over a lookback time of at least a few Gyrs and, often,
to a more accurate estimate of their distance
\citep[e.g.][]{Aloisi07,Grocholski08}

\begin{figure}[htbp]
\begin{center}
\epsfxsize=6.5in
\epsfbox{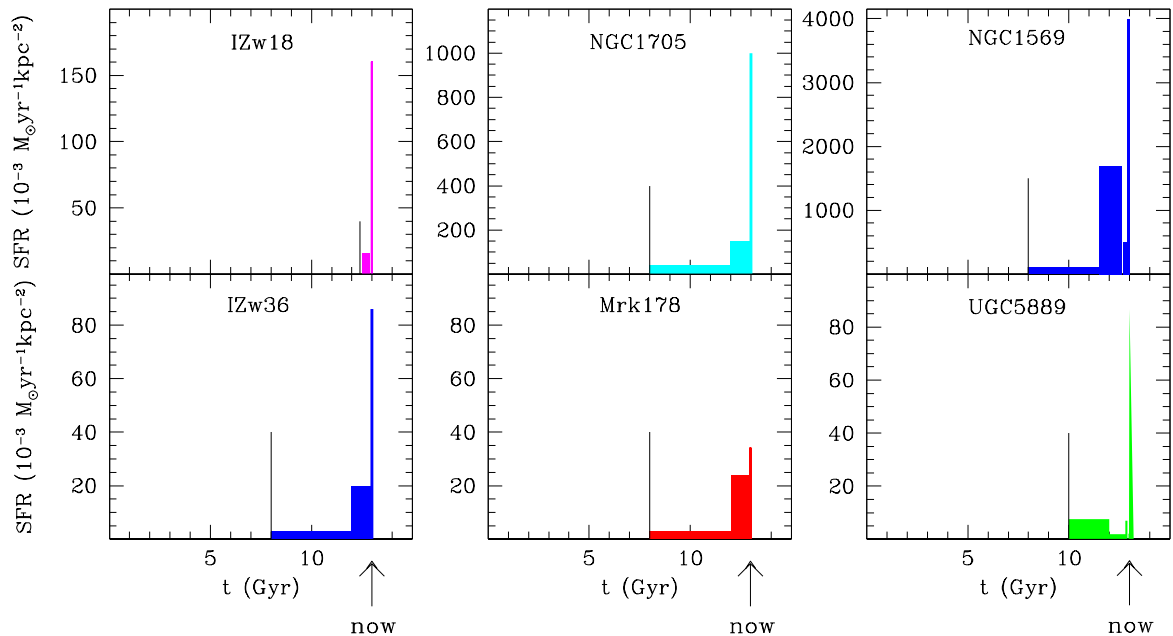}
\caption{SFHs of late-type dwarfs outside the Local Group observed with
WFPC2 or NICMOS. In all
panels the SFR per unit area as a function of time is plotted. The thin
vertical line indicates the lookback time reached by the adopted
photometry. Notice that, for those galaxies that have been subsequently 
observed also
with the ACS, the lookback time is actually quite older, and always with
indisputable evidence of SF activity already in place.
References: NGC~1569, \cite{Greggio98,Angeretti05};  
NGC~1705, \cite{Annibali03,Annibali09}; 
I~Zw~18, \cite{Aloisi99}; I~Zw~36, \cite{Schulte01};
Mrk~178, \cite{Schulte00}; UGC~5889, \cite{Vallenari05}.
      } 
\label{sfh_bcd}
\end{center}
\end{figure}

Only few groups have embarked in the more challenging application of
the synthetic CMD method beyond the LG, 
and most of them have concentrated their efforts on
starbursting late-type dwarfs
{\citep{Vallenari96,Greggio98,Lynds98,Aloisi99,Schulte99,Schulte00,Dolphin01,
Schulte01,Crone02,Annibali03,Angeretti05,Vallenari05,McQuinn09}.  However,
HST (new observations and archive) is still providing a
wealth of excellent images of external dwarfs. In particular, the 
ACS Nearby Galaxy Survey Treasury is promising an unprecedented data base
for these purposes \citep{Williams09}, and we can expect a flourishing of
applications of the synthetic CMD method to many more galaxies beyond the
LG boundaries in the near future.

All the studies performed so far have shown that all the examined galaxies 
were already active at the reached lookback time, including the BCDs that 
in the past had been suggested to be genuinely young galaxies, experiencing
now their first episode of SF. All late-type dwarfs
present a recent SF burst, which is what let people discover them in
spite of the distance, and none of them exhibits long quiescent phases within the
reached lookback time. On the other hand, the SFR differs significantly 
from one galaxy to the other. 

These results can be visualized in Fig.\ref{sfh_bcd}, where some
examples of SFH of external late-type dwarfs are given. All these SFHs
have been derived with the synthetic CMD method applied to HST/WFPC2
or NICMOS photometry.  The lookback time reached by the photometry is
indicated by the thin vertical line in each panel, and in all cases
stars of that age were detected.  As shown by \cite{Annibali09} for
NGC1705, the available data allow to rule out that these galaxies have
had short-duration, intermediate-age bursts like the current one
within the covered lookback time.  The sample of displayed galaxies
contains various types of dwarfs: UGC~5889 is a low surface brightness
galaxy (LSB), NGC~1705, IZw18, IZw36 and Mrk178 are BCDs, while
NGC~1569 is classified as dIrr. Nonetheless, they all show a
qualitatively similar behavior: a stronger current burst overimposed
on a moderate and rather continuous SF activity.  Quantitatively,
instead, the SFRs differ from each other by orders of magnitude.

The general results drawn from all the SFHs derived so far from CMDs in and
beyond the LG can be schematically summarized as follows:

\begin{itemize}
\item Evidence of long interruptions in the SF activity is found only in 
early-type galaxies;

\item Few early-type dwarfs have experienced only one episode of SF activity
concentrated at the earliest epochs: many show instead extended or
recurrent SF activity;

\item  No galaxy currently at its first SF episode has been found yet;

\item  No frequent evidence of strong SF bursts is found;

\item There is no significant difference in the SFH of dIrrs and BCDs, except
for the current SFR.
\end{itemize}

\section{Concluding remarks}

By comparing the results on the SFHs described in the previous section, one can
infer interesting conclusions and attempt some speculations.

An interesting result of the SFH studies both in the LG and beyond
is that all dwarfs have, and have had, fairly moderate SF activity. 
None of the dwarfs SFRs from the CMDs studied so far ever 
reaches values as high as
 1 M$_{\odot}$/yr, and only one (NGC~1569) gets close to it
\citep{Greggio98,Angeretti05,Grocholski08}. Since 1 M$_{\odot}$/yr is the 
minimum rate required to let a galaxy contribute to the
overabundance of faint blue objects in deep galaxy counts
\citep[see the models by][]{Babul96}, this makes it quite unlikely that
dwarfs are responsible for the blue excess.

If we look at the dwarfs shown in Fig.\ref{sfh_bcd}, we notice that the least 
active system is one of the BCDs and the most  active one is the dIrr. 
This is not inconsistent with the findings from an extensive $H_{\alpha}$ 
study of 94 late-type galaxies \cite{Hunter04}, showing that the typical 
SFR of irregular galaxies is $10^{-3}M_{\odot}yr^{-1}kpc^{-2}$ and that of 
BCDs is generally higher. From that survey, Hunter \& Elmegreen \cite{Hunter04} 
indeed conclude that NGC~1569 and NGC~1705 are among the few systems
with unusually high star formation, and that the
star formation regions are not intrinsically different in the various
galaxy types, except for a significantly  higher spatial concentration in BCDs.

In our view, this
suggests that either the morphological classification does not strictly 
correspond to the intensity of the SF activity, or that, most likely, the
traditional classification, based in most cases on photographic plates, is 
rather uncertain for systems too distant to be properly
resolved before the advent of HST. Probably an active dwarf such as NGC1569,
hosting three Super Star Clusters and a huge number of HII regions, would have
been classified as BCD, had it been just a few Mpc farther away.

If we compare the SFHs of late-type dwarfs inside and outside the LG
(e.g.  Fig.\ref{leoa} and Fig.\ref{sfh_bcd}), we see that the overall
scenario is quite similar, but the latter galaxies are always more
active than the former at very recent epochs. Part of this is
presumably due to the selection effect resulting from the difficulty
of finding distant dwarf, faint galaxies, unless currently
active. This effect is also the reason why early-type and quiescent
dwarfs are so rare in the surveys performed so far outside the LG
except for deep surveys devoted to individual galaxy clusters and
groups, where they preferentially reside in central regions.  Indeed,
most often the HII region emission is what led to the discovery of
distant dwarfs (recall that BCDs were originally called "extragalactic
HII regions") and it is thus inevitable that these sytems have recent
SF activity.  From this point of view, it is interesting to note that
the SFH of the external dwarfs of Fig.\ref{sfh_bcd} is very similar in
shape to that of the NGC~602 region in the SMC \citep{Cignoni09},
consistent with the circumstance that NGC~602 is also associated with
an HII region (N90).

How do the results on SFHs affect our understanding of galaxy formation ?

The circumstance that all dwarfs contain stars as old as the reached lookback
time, independently of their metallicity, gas content or morphological
classification suggests that their SF activity started at the earliest epochs.
This is absolutely coherent with both the hierarchical formation scenario and 
the monolithical scenario. It is 
also consistent with downsizing if their early SFR was lower than that of
more massive systems. For the (few) early-type dwarfs with studied SFH we know
that this is indeed the case; how about late-type dwarfs ? 
We don't have direct evidences, 
due to the large distance of most of these systems which prevents us to reach 
epochs older than a few Gyrs. However, all indirect arguments go in this 
direction: only with quite moderate early SF activity can dIrrs and BCDs have 
managed to remain as
metal-poor and gas-rich as they actually are. SFRs as high as the recent ones
would have inevitably consumed all their gas in much less than a Hubble time and
would have led to a significant chemical enrichment.

As mentioned in the Introduction, to select the most viable scenario
it is the combination of the SFH with the chemical and kinematic
properties of the candidate building blocks that needs to be compared
with the properties of massive galaxies. In the case of local dwarfs,
these properties have been recently reviewed by
\citep{Tolstoy09}. Stars in {\it classical} dwarfs don't resemble
those in the halo of the Milky Way, most notably their metallicity
distribution functions \citep{Venn04,Helmi06,Schoerck08} and the
abundance ratios of $\alpha$ elements over iron \citep[][and
  references therein]{Tolstoy09}. Moreover, if all early-type dwarfs
have had the relatively moderate SF activity shown in Fig. \ref{cetus}
for Cetus, with a rather long duration and the peak some Gyr after the
beginning, there is no way to let them provide the iron-poor stars
with high $[\alpha/Fe]$ typical of our halo, since the SF peak forms
most of the stars when SNeIa have already had the time to pollute the
medium with their iron.

On the other hand, the current knowledge of the outer Galaxy is far
from complete. We know that the stellar halo hosts two distinct
populations \citep[see e.g.][and references therein]{Carollo07,
  Gratton03}. \citep[][]{Roederer09} find that the so-called ``inner
halo'', selected among halo stars with prograde rotation and low
apogalactic maximum distance from the galactic center, is different
for several aspects from the ``outer halo'', selected among stars with
high retrograde rotation and high apogalactic maximum distance. In
particular: 1) the inner halo is characterized by a tight correlation
between $[\alpha/Fe]$ versus $[Fe/H]$, suggesting that either the
abundance ratios in distant regions of the inner halo are very similar
or the inner halo developed from a well homogenized interstellar
medium. In contrast, the outer halo shows a much larger scatter in
$[\alpha/Fe]$ for a given $[Fe/H]$, signature that the star formation
was spatially inhomogeneous or these stars have been accreted from
outside (from dwarf galaxies?). 2) The inner halo shows an average
$[\alpha/Fe]$ slightly higher than observed in the outer halo,
providing a clue for a more intense and short-lasting star formation
activity. 3)The inner halo is only found with metallicities in the
range $-2.5<[Fe/H]<-0.5$, while most of the outer halo is in the range
$-3.5<[Fe/H]<-1.5$.

Unfortunately the outer halo is still mostly inaccessible: current
high resolution abundances rely mainly on halo stars that pass near
the Sun. If these stars are formed in the outer halo, their selection
is biased towards higher eccentricity orbits. Avoiding this bias
implies a new class of surveys able to trace $\alpha$ variations in
situ. In this context, the Gaia mission will provide a quantum leap in
the ability to obtain highly precise astrometry, photometry and
metallicity for a volume of several Kpc.

If the outer halo is a natural place to search for possible accretion
events, the recent discovery of ultra-faint dwarfs promises to
complete the picture: containing extremely metal-poor stars, probably
with high [$\alpha$/Fe] like in our halo
\citep{Kirby08,Koch08herc,Frebel09}, these galaxies are ideal
candidates for Galactic building blocks. The problem in this case is
the extreme uncertainty still affecting their measures, due to both
faintness and high Galactic contamination. Overcoming these limits
will be a challenge only suitable for wide-field spectrographs mounted
on giant ground-based telescopes.

\acknowledgments Over the years several interesting conversations with
A. Aparicio, C. Chiosi, S. Degl'Innocenti, J. Gallagher, C. Gallart,
P.G. Prada Moroni, R. Schulte-Ladbeck, S. N. Shore, E. Skillman,
E. Tolstoy, and in particular L. Greggio, have been fruitful to dig
into the secrets of synthetic CMD building and exploitation.  We thank
A. Cole for providing the Leo A figure in appropriate format and
M. Monelli for the Cetus figure. We also thank Laura Greggio and the
anonymous referees for detailed and constructive comments helpful to
make this paper clearer. We acknowledge financial support from ASI
through contract ASI-INAF I/016/07/0 and from the Italian MIUR through
contract PRIN-2007JJC53X-001.


\begin{thebibliography}{100}

\bibitem{Lilly95}
S.~J. {Lilly}, L.~{Tresse}, F.~{Hammer}, D.~{Crampton}, and O.~{Le Fevre}.
\newblock {The Canada-France Redshift Survey. VI. Evolution of the Galaxy
  Luminosity Function to Z approximately 1}.
\newblock {\em \apj}, 455:108--+, December 1995.

\bibitem{Babul96}
A.~{Babul} and H.~C. {Ferguson}.
\newblock {Faint Blue Galaxies and the Epoch of Dwarf Galaxy Formation}.
\newblock {\em \apj}, 458:100--+, February 1996.

\bibitem{Peimbert74}
M.~{Peimbert} and S.~{Torres-Peimbert}.
\newblock {Chemical composition of H II regions in the Large Magellanic Cloud
  and its cosmological implications}.
\newblock {\em \apj}, 193:327--333, October 1974.

\bibitem{Olive97}
K.~A. {Olive}, G.~{Steigman}, and E.~D. {Skillman}.
\newblock {The Primordial Abundance of 4He: an Update}.
\newblock {\em \apj}, 483:788--+, July 1997.

\bibitem{Izotov98}
Y.~I. {Izotov} and T.~X. {Thuan}.
\newblock {The Primordial Abundance of 4He Revisited}.
\newblock {\em \apj}, 500:188--+, June 1998.

\bibitem{White78}
S.~D.~M. {White} and M.~J. {Rees}.
\newblock {Core condensation in heavy halos - A two-stage theory for galaxy
  formation and clustering}.
\newblock {\em \mnras}, 183:341--358, May 1978.

\bibitem{Frenk88}
C.~S. {Frenk}, S.~D.~M. {White}, M.~{Davis}, and G.~{Efstathiou}.
\newblock {The formation of dark halos in a universe dominated by cold dark
  matter}.
\newblock {\em \apj}, 327:507--525, April 1988.

\bibitem{Bellazzini03}
M.~{Bellazzini}, R.~{Ibata}, F.~R. {Ferraro}, and V.~{Testa}.
\newblock {Tracing the Sgr Stream with 2MASS. Detection of Stream stars around
  Outer Halo globular clusters}.
\newblock {\em \aap}, 405:577--583, July 2003.

\bibitem{Belokurov07}
V.~{Belokurov}, D.~B. {Zucker}, N.~W. {Evans}, J.~T. {Kleyna}, S.~{Koposov},
  S.~T. {Hodgkin}, M.~J. {Irwin}, G.~{Gilmore}, M.~I. {Wilkinson},
  M.~{Fellhauer}, D.~M. {Bramich}, P.~C. {Hewett}, S.~{Vidrih}, J.~T.~A. {De
  Jong}, J.~A. {Smith}, H.-W. {Rix}, E.~F. {Bell}, R.~F.~G. {Wyse}, H.~J.
  {Newberg}, P.~A. {Mayeur}, B.~{Yanny}, C.~M. {Rockosi}, O.~Y. {Gnedin}, D.~P.
  {Schneider}, T.~C. {Beers}, J.~C. {Barentine}, H.~{Brewington},
  J.~{Brinkmann}, M.~{Harvanek}, S.~J. {Kleinman}, J.~{Krzesinski}, D.~{Long},
  A.~{Nitta}, and S.~A. {Snedden}.
\newblock {Cats and Dogs, Hair and a Hero: A Quintet of New Milky Way
  Companions}.
\newblock {\em \apj}, 654:897--906, January 2007.

\bibitem{Ferguson02}
A.~M.~N. {Ferguson}, M.~J. {Irwin}, R.~A. {Ibata}, G.~F. {Lewis}, and N.~R.
  {Tanvir}.
\newblock {Evidence for Stellar Substructure in the Halo and Outer Disk of
  M31}.
\newblock {\em \aj}, 124:1452--1463, September 2002.

\bibitem{Ibata04}
R.~{Ibata}, S.~{Chapman}, A.~M.~N. {Ferguson}, M.~{Irwin}, G.~{Lewis}, and
  A.~{McConnachie}.
\newblock {Taking measure of the Andromeda halo: a kinematic analysis of the
  giant stream surrounding M31}.
\newblock {\em \mnras}, 351:117--124, June 2004.

\bibitem{Ferguson05}
A.~M.~N. {Ferguson}, R.~A. {Johnson}, D.~C. {Faria}, M.~J. {Irwin}, R.~A.
  {Ibata}, K.~V. {Johnston}, G.~F. {Lewis}, and N.~R. {Tanvir}.
\newblock {The Stellar Populations of the M31 Halo Substructure}.
\newblock {\em \apjl}, 622:L109--L112, April 2005.

\bibitem{Cowie96}
L.~L. {Cowie}, A.~{Songaila}, E.~M. {Hu}, and J.~G. {Cohen}.
\newblock {New Insight on Galaxy Formation and Evolution From Keck Spectroscopy
  of the Hawaii Deep Fields}.
\newblock {\em \aj}, 112:839--+, September 1996.

\bibitem{Mouri06}
H.~{Mouri} and Y.~{Taniguchi}.
\newblock {Downsizing of star-forming galaxies by gravitational processes}.
\newblock {\em \aap}, 459:371--374, November 2006.

\bibitem{Neistein06}
E.~{Neistein}, F.~C. {van den Bosch}, and A.~{Dekel}.
\newblock {Natural downsizing in hierarchical galaxy formation}.
\newblock {\em \mnras}, 372:933--948, October 2006.

\bibitem{Cattaneo08}
A.~{Cattaneo}, A.~{Dekel}, S.~M. {Faber}, and B.~{Guiderdoni}.
\newblock {Downsizing by shutdown in red galaxies}.
\newblock {\em \mnras}, 389:567--584, September 2008.

\bibitem{Eggen62}
O.~J. {Eggen}, D.~{Lynden-Bell}, and A.~R. {Sandage}.
\newblock {Evidence from the motions of old stars that the Galaxy collapsed.}
\newblock {\em \apj}, 136:748--+, November 1962.

\bibitem{Tolstoy09}
E.~{Tolstoy}, V.~{Hill}, and M.~{Tosi}.
\newblock {Star Formation Histories, Abundances and Kinematics of Dwarf
  Galaxies in the Local Group}.
\newblock {\em ArXiv e-prints}, April 2009.

\bibitem{Searle73}
L.~{Searle}, W.~L.~W. {Sargent}, and W.~G. {Bagnuolo}.
\newblock {The History of Star Formation and the Colors of Late-Type Galaxies}.
\newblock {\em \apj}, 179:427--438, January 1973.

\bibitem{Gallagher84a}
J.~S. {Gallagher}, III, D.~A. {Hunter}, and A.~V. {Tutukov}.
\newblock {Star formation histories of irregular galaxies}.
\newblock {\em \apj}, 284:544--556, September 1984.

\bibitem{Gallagher84b}
J.~S. {Gallagher}, III and D.~A. {Hunter}.
\newblock {Structure and Evolution of Irregular Galaxies}.
\newblock {\em \araa}, 22:37--74, 1984.

\bibitem{Fagotto94a}
F.~{Fagotto}, A.~{Bressan}, G.~{Bertelli}, and C.~{Chiosi}.
\newblock {Evolutionary sequences of stellar models with new radiative
  opacities. III. Z=0.0004 and Z=0.05}.
\newblock {\em \aaps}, 104:365--376, April 1994.

\bibitem{Fagotto94b}
F.~{Fagotto}, A.~{Bressan}, G.~{Bertelli}, and C.~{Chiosi}.
\newblock {Evolutionary sequences of stellar models with new radiative
  opacities. IV. Z=0.004 and Z=0.008}.
\newblock {\em \aap}, 105:29--38, May 1994.

\bibitem{Greggio02}
L.~{Greggio}.
\newblock {The Color-Magnitude Diagram of composite stellar populations}.
\newblock In T.~{Lejeune} and J.~{Fernandes}, editors, {\em Observed HR
  Diagrams and Stellar Evolution}, volume 274 of {\em Astronomical Society of
  the Pacific Conference Series}, pages 444--+, 2002.

\bibitem{Carlson07}
L.~R. {Carlson}, E.~{Sabbi}, M.~{Sirianni}, J.~L. {Hora}, A.~{Nota},
  M.~{Meixner}, J.~S. {Gallagher}, III, M.~S. {Oey}, A.~{Pasquali}, L.~J.
  {Smith}, M.~{Tosi}, and R.~{Walterbos}.
\newblock {Progressive Star Formation in the Young SMC Cluster NGC 602}.
\newblock {\em \apjl}, 665:L109--L114, August 2007.

\bibitem{Sabbi07}
E.~{Sabbi}, M.~{Sirianni}, A.~{Nota}, M.~{Tosi}, J.~{Gallagher}, M.~{Meixner},
  M.~S. {Oey}, R.~{Walterbos}, A.~{Pasquali}, L.~J. {Smith}, and
  L.~{Angeretti}.
\newblock {Past and Present Star Formation in the SMC: NGC 346 and its
  Neighborhood}.
\newblock {\em \aj}, 133:44--57, January 2007.

\bibitem{Cignoni09}
M.~{Cignoni}, E.~{Sabbi}, A.~{Nota}, M.~{Tosi}, S.~{Degl'Innocenti}, P.~G.~P.
  {Moroni}, L.~{Angeretti}, L.~R. {Carlson}, J.~{Gallagher}, M.~{Meixner},
  M.~{Sirianni}, and L.~J. {Smith}.
\newblock {Star Formation History in the Small Magellanic Cloud: The Case of
  NGC 602}.
\newblock {\em \aj}, 137:3668--3684, March 2009.

\bibitem{Kroupa01}
P.~{Kroupa}.
\newblock {On the variation of the initial mass function}.
\newblock {\em \mnras}, 322:231--246, April 2001.

\bibitem{Chabrier03}
G.~{Chabrier}.
\newblock {Galactic Stellar and Substellar Initial Mass Function}.
\newblock {\em \pasp}, 115:763--795, July 2003.

\bibitem{Ferraro89}
F.~R. {Ferraro}, F.~{Fusi Pecci}, M.~{Tosi}, and R.~{Buonanno}.
\newblock {A method for studying the star formation history of dwarf irregular
  galaxies. I - CCD photometry of WLM}.
\newblock {\em \mnras}, 241:433--452, December 1989.

\bibitem{Tosi91}
M.~{Tosi}, L.~{Greggio}, G.~{Marconi}, and P.~{Focardi}.
\newblock {Star formation in dwarf irregular galaxies - Sextans B}.
\newblock {\em \aj}, 102:951--974, September 1991.

\bibitem{Bertelli92}
G.~{Bertelli}, M.~{Mateo}, C.~{Chiosi}, and A.~{Bressan}.
\newblock {The star formation history of the Large Magellanic Cloud}.
\newblock {\em \apj}, 388:400--414, April 1992.

\bibitem{Greggio93}
L.~{Greggio}, G.~{Marconi}, M.~{Tosi}, and P.~{Focardi}.
\newblock {Star formation in dwarf irregular galaxies - DDO 210 and NGC 3109}.
\newblock {\em \aj}, 105:894--932, March 1993.

\bibitem{Marconi95}
G.~{Marconi}, M.~{Tosi}, L.~{Greggio}, and P.~{Focardi}.
\newblock {Star formation in dwarf irregular galaxies NGC 6822}.
\newblock {\em \aj}, 109:173--199, January 1995.

\bibitem{Gallart96a}
C.~{Gallart}, A.~{Aparicio}, G.~{Bertelli}, and C.~{Chiosi}.
\newblock {The Local Group Dwarf Irregular Galaxy NGC 6822.II.The Old and
  Intermediate -Age Star Formation History}.
\newblock {\em \aj}, 112:1950--+, November 1996.

\bibitem{Gallart96b}
C.~{Gallart}, A.~{Aparicio}, G.~{Bertelli}, and C.~{Chiosi}.
\newblock {The Local Group Dwarf Irregular Galaxy NGC 6822.III. The Recent Star
  Formation History}.
\newblock {\em \aj}, 112:2596--+, December 1996.

\bibitem{Aparicio97a}
A.~{Aparicio}, C.~{Gallart}, and G.~{Bertelli}.
\newblock {The Star Formation History of the Pegasus Dwarf Irregular Galaxy}.
\newblock {\em \aj}, 114:669--679, August 1997.

\bibitem{Tolstoy96b}
E.~{Tolstoy} and A.~{Saha}.
\newblock {The Interpretation of Color-Magnitude Diagrams through Numerical
  Simulation and Bayesian Inference}.
\newblock {\em \apj}, 462:672--+, May 1996.

\bibitem{Dolphin97}
A.~{Dolphin}.
\newblock {A new method to determine star formation histories of nearby
  galaxies}.
\newblock {\em New Astronomy}, 2:397--409, November 1997.

\bibitem{Gallart99}
C.~{Gallart}, W.~L. {Freedman}, A.~{Aparicio}, G.~{Bertelli}, and C.~{Chiosi}.
\newblock {The Star Formation History of the Local Group Dwarf Galaxy Leo I}.
\newblock {\em \aj}, 118:2245--2261, November 1999.

\bibitem{Cole99}
A.~A. {Cole}, E.~{Tolstoy}, J.~S. {Gallagher}, III, J.~G. {Hoessel}, J.~R.
  {Mould}, J.~A. {Holtzman}, A.~{Saha}, G.~E. {Ballester}, C.~J. {Burrows},
  J.~T. {Clarke}, D.~{Crisp}, R.~E. {Griffiths}, C.~J. {Grillmair}, J.~J.
  {Hester}, J.~E. {Krist}, V.~{Meadows}, P.~A. {Scowen}, K.~R. {Stapelfeldt},
  J.~T. {Trauger}, A.~M. {Watson}, and J.~R. {Westphal}.
\newblock {Stellar Populations at the Center of IC 1613}.
\newblock {\em \aj}, 118:1657--1670, October 1999.

\bibitem{Hernandez99}
X.~{Hernandez}, D.~{Valls-Gabaud}, and G.~{Gilmore}.
\newblock {Deriving star formation histories: inverting Hertzsprung-Russell
  diagrams through a variational calculus maximum likelihood method}.
\newblock {\em \mnras}, 304:705--719, April 1999.

\bibitem{Holtzman99}
J.~A. {Holtzman}, J.~S. {Gallagher}, III, A.~A. {Cole}, J.~R. {Mould}, C.~J.
  {Grillmair}, G.~E. {Ballester}, C.~J. {Burrows}, J.~T. {Clarke}, D.~{Crisp},
  R.~W. {Evans}, R.~E. {Griffiths}, J.~J. {Hester}, J.~G. {Hoessel}, P.~A.
  {Scowen}, K.~R. {Stapelfeldt}, J.~T. {Trauger}, and A.~M. {Watson}.
\newblock {Observations and Implications of the Star Formation History of the
  Large Magellanic Cloud}.
\newblock {\em \aj}, 118:2262--2279, November 1999.

\bibitem{Harris01}
J.~{Harris} and D.~{Zaritsky}.
\newblock {A Method for Determining the Star Formation History of a Mixed
  Stellar Population}.
\newblock {\em \apjs}, 136:25--40, September 2001.

\bibitem{Vergely02}
J.-L. {Vergely}, J.~{K{\"o}ppen}, D.~{Egret}, and O.~{Bienaym{\'e}}.
\newblock {An inverse method to interpret colour-magnitude diagrams}.
\newblock {\em \aap}, 390:917--929, August 2002.

\bibitem{Cignoni06}
M.~{Cignoni}, S.~{Degl'Innocenti}, P.~G. {Prada Moroni}, and S.~N. {Shore}.
\newblock {Recovering the star formation rate in the solar neighborhood}.
\newblock {\em \aap}, 459:783--796, December 2006.

\bibitem{Schroder03}
K.-P. {Schr{\"o}der} and B.~E.~J. {Pagel}.
\newblock {Galactic archaeology: initial mass function and depletion in the
  `thin disc'}.
\newblock {\em \mnras}, 343:1231--1240, August 2003.

\bibitem{Hernandez00b}
X.~{Hernandez}, D.~{Valls-Gabaud}, and G.~{Gilmore}.
\newblock {The recent star formation history of the Hipparcos solar
  neighbourhood}.
\newblock {\em \mnras}, 316:605--612, August 2000.

\bibitem{Naylor06}
T.~{Naylor} and R.~D. {Jeffries}.
\newblock {A maximum-likelihood method for fitting colour-magnitude diagrams}.
\newblock {\em \mnras}, 373:1251--1263, December 2006.

\bibitem{Skillman02}
E.~D. {Skillman} and C.~{Gallart}.
\newblock {First Results of the Coimbra Experiment}.
\newblock In T.~{Lejeune} and J.~{Fernandes}, editors, {\em Observed HR
  Diagrams and Stellar Evolution}, volume 274 of {\em Astronomical Society of
  the Pacific Conference Series}, pages 535--+, 2002.

\bibitem{cigno06}
M.~{Cignoni}, S.~{Degl'Innocenti}, P.~G. {Prada Moroni}, and S.~N. {Shore}.
\newblock {Recovering the star formation rate in the solar neighborhood}.
\newblock {\em \aap}, 459:783--796, December 2006.

\bibitem{Dolphin02}
A.~E. {Dolphin}.
\newblock {Numerical methods of star formation history measurement and
  applications to seven dwarf spheroidals}.
\newblock {\em \mnras}, 332:91--108, May 2002.

\bibitem{Aparicio09}
A.~{Aparicio} and S.~L. {Hidalgo}.
\newblock {IAC-pop: Finding the Star Formation History of Resolved Galaxies}.
\newblock {\em \aj}, 138:558--567, August 2009.

\bibitem{Gallart05}
C.~{Gallart}, M.~{Zoccali}, and A.~{Aparicio}.
\newblock {The Adequacy of Stellar Evolution Models for the Interpretation of
  the Color-Magnitude Diagrams of Resolved Stellar Populations}.
\newblock {\em \araa}, 43:387--434, September 2005.

\bibitem{Kerber09}
L.~{Kerber}, L.~{Girardi}, S.~{Rubele}, and M.~. {Cioni}.
\newblock {On the Recovery of the Star Formation History of the LMC from the
  VISTA Survey of the Magellanic System}.
\newblock {\em ArXiv e-prints}, January 2009.

\bibitem{Aparicio97b}
A.~{Aparicio}, C.~{Gallart}, and G.~{Bertelli}.
\newblock {The Stellar Content and the Star Formation History of the Local
  Group Dwarf Galaxy LGS 3.}
\newblock {\em \aj}, 114:680--693, August 1997.

\bibitem{Harris04}
J.~{Harris} and D.~{Zaritsky}.
\newblock {The Star Formation History of the Small Magellanic Cloud}.
\newblock {\em \aj}, 127:1531--1544, March 2004.

\bibitem{Salpeter55}
E.~E. {Salpeter}.
\newblock {The Luminosity Function and Stellar Evolution.}
\newblock {\em \apj}, 121:161--+, January 1955.

\bibitem{Tinsley80}
B.~M. {Tinsley}.
\newblock {Evolution of the Stars and Gas in Galaxies}.
\newblock {\em Fundamentals of Cosmic Physics}, 5:287--388, 1980.

\bibitem{Scalo98}
J.~{Scalo}.
\newblock {The IMF Revisited: A Case for Variations}.
\newblock In G.~{Gilmore} and D.~{Howell}, editors, {\em The Stellar Initial
  Mass Function (38th Herstmonceux Conference)}, volume 142 of {\em
  Astronomical Society of the Pacific Conference Series}, pages 201--+, 1998.

\bibitem{Hurley98b}
J.~{Hurley} and C.~A. {Tout}.
\newblock {The binary second sequence in cluster colour-magnitude diagrams}.
\newblock {\em \mnras}, 300:977--980, November 1998.

\bibitem{Olsen99}
K.~A.~G. {Olsen}.
\newblock {Star Formation Histories from Hubble Space Telescope Color-Magnitude
  Diagrams of Six Fields of the Large Magellanic Cloud}.
\newblock {\em \aj}, 117:2244--2267, May 1999.

\bibitem{Stanimi04}
S.~{Stanimirovi{\'c}}, L.~{Staveley-Smith}, and P.~A. {Jones}.
\newblock {A New Look at the Kinematics of Neutral Hydrogen in the Small
  Magellanic Cloud}.
\newblock {\em \apj}, 604:176--186, March 2004.

\bibitem{Skillman03}
E.~D. {Skillman}, E.~{Tolstoy}, A.~A. {Cole}, A.~E. {Dolphin}, A.~{Saha}, J.~S.
  {Gallagher}, R.~C. {Dohm-Palmer}, and M.~{Mateo}.
\newblock {Deep Hubble Space Telescope Imaging of IC 1613. II. The Star
  Formation History}.
\newblock {\em \apj}, 596:253--272, October 2003.

\bibitem{Cole07}
A.~A. {Cole}, E.~D. {Skillman}, E.~{Tolstoy}, J.~S. {Gallagher}, III,
  A.~{Aparicio}, A.~E. {Dolphin}, C.~{Gallart}, S.~L. {Hidalgo}, A.~{Saha},
  P.~B. {Stetson}, and D.~R. {Weisz}.
\newblock {Leo A: A Late-blooming Survivor of the Epoch of Reionization in the
  Local Group}.
\newblock {\em \apjl}, 659:L17--L20, April 2007.

\bibitem{Tolstoy96a}
E.~{Tolstoy}.
\newblock {The Resolved Stellar Population of Leo A}.
\newblock {\em \apj}, 462:684--+, May 1996.

\bibitem{Dolphin00b}
A.~E. {Dolphin}.
\newblock {The star formation histories of two northern LMC fields}.
\newblock {\em \mnras}, 313:281--290, April 2000.

\bibitem{Smecker02}
T.~A. {Smecker-Hane}, A.~A. {Cole}, J.~S. {Gallagher}, III, and P.~B.
  {Stetson}.
\newblock {The Star Formation History of the Large Magellanic Cloud}.
\newblock {\em \apj}, 566:239--244, February 2002.

\bibitem{Javiel05}
S.~C. {Javiel}, B.~X. {Santiago}, and L.~O. {Kerber}.
\newblock {Constraints on the star formation history of the Large Magellanic
  Cloud}.
\newblock {\em \aap}, 431:73--85, February 2005.

\bibitem{Gallart08}
C.~{Gallart}, P.~B. {Stetson}, I.~P. {Meschin}, F.~{Pont}, and E.~{Hardy}.
\newblock {Outside-In Disk Evolution in the Large Magellanic Cloud}.
\newblock {\em \apjl}, 682:L89--L92, August 2008.

\bibitem{Dolphin01b}
A.~E. {Dolphin}, A.~R. {Walker}, P.~W. {Hodge}, M.~{Mateo}, E.~W. {Olszewski},
  R.~A. {Schommer}, and N.~B. {Suntzeff}.
\newblock {Old Stellar Populations of the Small Magellanic Cloud}.
\newblock {\em \apj}, 562:303--313, November 2001.

\bibitem{Noel07}
N.~E.~D. {No{\"e}l}, C.~{Gallart}, E.~{Costa}, and R.~A. {M{\'e}ndez}.
\newblock {Old Main-Sequence Turnoff Photometry in the Small Magellanic Cloud.
  I. Constraints on the Star Formation History in Different Fields}.
\newblock {\em \aj}, 133:2037--2052, May 2007.

\bibitem{Hidalgo09}
S.~L. {Hidalgo}, A.~{Aparicio}, and C.~{Gallart}.
\newblock {Recovering the ages and metallicities of stars of a complex stellar
  population system}.
\newblock In E.~E. {Mamajek}, D.~R. {Soderblom}, and R.~F.~G. {Wyse}, editors,
  {\em IAU Symposium}, volume 258 of {\em IAU Symposium}, pages 245--252, June
  2009.

\bibitem{Dohm98}
R.~C. {Dohm-Palmer}, E.~D. {Skillman}, J.~{Gallagher}, E.~{Tolstoy},
  M.~{Mateo}, R.~J. {Dufour}, A.~{Saha}, J.~{Hoessel}, and C.~{Chiosi}.
\newblock {The Recent Star Formation History of GR 8 from Hubble Space
  Telescope Photometry of the Resolved Stars}.
\newblock {\em \aj}, 116:1227--1243, September 1998.

\bibitem{Dohm02}
R.~C. {Dohm-Palmer}, E.~D. {Skillman}, M.~{Mateo}, A.~{Saha}, A.~{Dolphin},
  E.~{Tolstoy}, J.~S. {Gallagher}, and A.~A. {Cole}.
\newblock {Deep Hubble Space Telescope Imaging of Sextans A. I. The Spatially
  Resolved Recent Star Formation History}.
\newblock {\em \aj}, 123:813--831, February 2002.

\bibitem{Seiden79}
P.~E. {Seiden}, L.~S. {Schulman}, and H.~{Gerola}.
\newblock {Stochastic star formation and the evolution of galaxies}.
\newblock {\em \apj}, 232:702--706, September 1979.

\bibitem{Young07}
L.~M. {Young}, E.~D. {Skillman}, D.~R. {Weisz}, and A.~E. {Dolphin}.
\newblock {The Aptly Named Phoenix Dwarf Galaxy}.
\newblock {\em \apj}, 659:331--338, April 2007.

\bibitem{Smecker96}
T.~A. {Smecker-Hane}, P.~B. {Stetson}, J.~E. {Hesser}, and D.~A. {Vandenberg}.
\newblock {Episodic Star Formation in the Carina dSph Galaxy}.
\newblock In C.~{Leitherer}, U.~{Fritze-von-Alvensleben}, and J.~{Huchra},
  editors, {\em From Stars to Galaxies: the Impact of Stellar Physics on Galaxy
  Evolution}, volume~98 of {\em Astronomical Society of the Pacific Conference
  Series}, pages 328--+, 1996.

\bibitem{Hurley98}
D.~{Hurley-Keller}, M.~{Mateo}, and J.~{Nemec}.
\newblock {The Star Formation History of the Carina Dwarf Galaxy}.
\newblock {\em \aj}, 115:1840--1855, May 1998.

\bibitem{Hernandez00}
X.~{Hernandez}, G.~{Gilmore}, and D.~{Valls-Gabaud}.
\newblock {Non-parametric star formation histories for four dwarf spheroidal
  galaxies of the Local Group}.
\newblock {\em \mnras}, 317:831--842, October 2000.

\bibitem{Dolphin05}
A.~E. {Dolphin}, D.~R. {Weisz}, E.~D. {Skillman}, and J.~A. {Holtzman}.
\newblock {Star Formation Histories of Local Group Dwarf Galaxies}.
\newblock {\em ArXiv Astrophysics e-prints}, June 2005.

\bibitem{Gallart07}
C.~{Gallart} and {The Lcid Team}.
\newblock {The ACS LCID project: overview and first results}.
\newblock In A.~{Vazdekis} and R.~F. {Peletier}, editors, {\em IAU Symposium},
  volume 241 of {\em IAU Symposium}, pages 290--294, August 2007.

\bibitem{Tosi09}
M.~{Tosi}.
\newblock {Star formation histories of resolved galaxies}.
\newblock In E.~E. {Mamajek}, D.~R. {Soderblom}, and R.~F.~G. {Wyse}, editors,
  {\em IAU Symposium}, volume 258 of {\em IAU Symposium}, pages 61--72, June
  2009.

\bibitem{Held00}
E.~V. {Held}, I.~{Saviane}, Y.~{Momany}, and G.~{Carraro}.
\newblock {The Elusive Old Population of the Dwarf Spheroidal Galaxy Leo I}.
\newblock {\em \apjl}, 530:L85--L88, February 2000.

\bibitem{Baldacci05}
L.~{Baldacci}, L.~{Rizzi}, G.~{Clementini}, and E.~V. {Held}.
\newblock {Variable stars in the dwarf irregular galaxy NGC 6822:
  Thephotometric catalogue}.
\newblock {\em \aap}, 431:1189--1201, March 2005.

\bibitem{Momany05}
Y.~{Momany}, E.~V. {Held}, I.~{Saviane}, L.~R. {Bedin}, M.~{Gullieuszik},
  M.~{Clemens}, L.~{Rizzi}, M.~R. {Rich}, and K.~{Kuijken}.
\newblock {HST/ACS observations of the old and metal-poor Sagittarius dwarf
  irregular galaxy}.
\newblock {\em \aap}, 439:111--127, August 2005.

\bibitem{DeJong08}
J.~T.~A. {de Jong}, J.~{Harris}, M.~G. {Coleman}, N.~F. {Martin}, E.~F. {Bell},
  H.-W. {Rix}, J.~M. {Hill}, E.~D. {Skillman}, D.~J. {Sand}, E.~W. {Olszewski},
  D.~{Zaritsky}, D.~{Thompson}, E.~{Giallongo}, R.~{Ragazzoni}, A.~{DiPaola},
  J.~{Farinato}, V.~{Testa}, and J.~{Bechtold}.
\newblock {The Structural Properties and Star Formation History of Leo T from
  Deep LBT Photometry}.
\newblock {\em \apj}, 680:1112--1119, June 2008.

\bibitem{Belokurov08}
V.~{Belokurov}, M.~G. {Walker}, N.~W. {Evans}, D.~C. {Faria}, G.~{Gilmore},
  M.~J. {Irwin}, S.~{Koposov}, M.~{Mateo}, E.~{Olszewski}, and D.~B. {Zucker}.
\newblock {Leo V: A Companion of a Companion of the Milky Way Galaxy?}
\newblock {\em \apjl}, 686:L83--L86, October 2008.

\bibitem{Aloisi07}
A.~{Aloisi}, G.~{Clementini}, M.~{Tosi}, F.~{Annibali}, R.~{Contreras},
  G.~{Fiorentino}, J.~{Mack}, M.~{Marconi}, I.~{Musella}, A.~{Saha},
  M.~{Sirianni}, and R.~P. {van der Marel}.
\newblock {I Zw 18 Revisited with HST ACS and Cepheids: New Distance and Age}.
\newblock {\em \apjl}, 667:L151--L154, October 2007.

\bibitem{Bertelli94}
G.~{Bertelli}, A.~{Bressan}, C.~{Chiosi}, F.~{Fagotto}, and E.~{Nasi}.
\newblock {Theoretical isochrones from models with new radiative opacities}.
\newblock {\em \aaps}, 106:275--302, August 1994.

\bibitem{Tosi01}
M.~{Tosi}, E.~{Sabbi}, M.~{Bellazzini}, A.~{Aloisi}, L.~{Greggio},
  C.~{Leitherer}, and P.~{Montegriffo}.
\newblock {The Resolved Stellar Populations in NGC 1705}.
\newblock {\em \aj}, 122:1271--1288, September 2001.

\bibitem{Aloisi99}
A.~{Aloisi}, M.~{Tosi}, and L.~{Greggio}.
\newblock {The Star Formation History of I ZW 18}.
\newblock {\em \aj}, 118:302--322, July 1999.

\bibitem{Aloisi05}
A.~{Aloisi}, R.~P. {van der Marel}, J.~{Mack}, C.~{Leitherer}, M.~{Sirianni},
  and M.~{Tosi}.
\newblock {Do Young Galaxies Exist in the Local Universe? Red Giant Branch
  Detection in the Metal-poor Dwarf Galaxy SBS 1415+437}.
\newblock {\em \apjl}, 631:L45--L48, September 2005.

\bibitem{Schulte99}
R.~E. {Schulte-Ladbeck}, U.~{Hopp}, L.~{Greggio}, and M.~M. {Crone}.
\newblock {A Near-Infrared Stellar Census of the Blue Compact Dwarf Galaxy VII
  ZW 403}.
\newblock {\em \aj}, 118:2705--2722, December 1999.

\bibitem{Weisz08}
D.~R. {Weisz}, E.~D. {Skillman}, J.~M. {Cannon}, A.~E. {Dolphin}, R.~C.
  {Kennicutt}, Jr., J.~{Lee}, and F.~{Walter}.
\newblock {The Recent Star Formation Histories of M81 Group Dwarf Irregular
  Galaxies}.
\newblock {\em \apj}, 689:160--183, December 2008.

\bibitem{Grocholski08}
A.~J. {Grocholski}, A.~{Aloisi}, R.~P. {van der Marel}, J.~{Mack},
  F.~{Annibali}, L.~{Angeretti}, L.~{Greggio}, E.~V. {Held}, D.~{Romano},
  M.~{Sirianni}, and M.~{Tosi}.
\newblock {A New Hubble Space Telescope Distance to NGC 1569: Starburst
  Properties and IC 342 Group Membership}.
\newblock {\em \apjl}, 686:L79--L82, October 2008.

\bibitem{Greggio98}
L.~{Greggio}, M.~{Tosi}, M.~{Clampin}, G.~{de Marchi}, C.~{Leitherer},
  A.~{Nota}, and M.~{Sirianni}.
\newblock {The Resolved Stellar Population of the Poststarburst Galaxy NGC
  1569}.
\newblock {\em \apj}, 504:725--+, September 1998.

\bibitem{Angeretti05}
L.~{Angeretti}, M.~{Tosi}, L.~{Greggio}, E.~{Sabbi}, A.~{Aloisi}, and
  C.~{Leitherer}.
\newblock {The Complex Star Formation History of NGC 1569}.
\newblock {\em \aj}, 129:2203--2216, May 2005.

\bibitem{Annibali03}
F.~{Annibali}, L.~{Greggio}, M.~{Tosi}, A.~{Aloisi}, and C.~{Leitherer}.
\newblock {The Star Formation History of NGC 1705: A Poststarburst Galaxy on
  the Verge of Activity}.
\newblock {\em \aj}, 126:2752--2773, December 2003.

\bibitem{Annibali09}
F.~{Annibali}, M.~{Tosi}, M.~{Monelli}, M.~{Sirianni}, P.~{Montegriffo},
  A.~{Aloisi}, and L.~{Greggio}.
\newblock {Young Stellar Populations and Star Clusters in NGC 1705}.
\newblock {\em \aj}, 138:169--183, July 2009.

\bibitem{Schulte01}
R.~E. {Schulte-Ladbeck}, U.~{Hopp}, L.~{Greggio}, M.~M. {Crone}, and I.~O.
  {Drozdovsky}.
\newblock {A Near-Infrared Stellar Census of Blue Compact Dwarf Galaxies: The
  Wolf-Rayet Galaxy I Zw 36}.
\newblock {\em \aj}, 121:3007--3025, June 2001.

\bibitem{Schulte00}
R.~E. {Schulte-Ladbeck}, U.~{Hopp}, L.~{Greggio}, and M.~M. {Crone}.
\newblock {A Near-Infrared Stellar Census of Blue Compact Dwarf Galaxies:
  NICMOS Detection of Red Giant Stars in the Wolf-Rayet Galaxy Markarian 178}.
\newblock {\em \aj}, 120:1713--1730, October 2000.

\bibitem{Vallenari05}
A.~{Vallenari}, L.~{Schmidtobreick}, and D.~J. {Bomans}.
\newblock {The star formation history of the LSB galaxy UGC 5889}.
\newblock {\em \aap}, 435:821--829, June 2005.

\bibitem{Vallenari96}
A.~{Vallenari} and D.~J. {Bomans}.
\newblock {Star formation history of the POST starburst galaxy NGC 1569.}
\newblock {\em \aap}, 313:713--722, September 1996.

\bibitem{Lynds98}
R.~{Lynds}, E.~{Tolstoy}, E.~J. {O'Neil}, Jr., and D.~A. {Hunter}.
\newblock {Star Formation in and Evolution of the Blue Compact Dwarf Galaxy UGC
  6456 Determined from Hubble Space Telescope Images}.
\newblock {\em \aj}, 116:146--162, July 1998.

\bibitem{Dolphin01}
A.~E. {Dolphin}, L.~{Makarova}, I.~D. {Karachentsev}, V.~E. {Karachentseva},
  D.~{Geisler}, E.~K. {Grebel}, P.~{Guhathakurta}, P.~W. {Hodge},
  A.~{Sarajedini}, and P.~{Seitzer}.
\newblock {The stellar content and distance of UGC 4483}.
\newblock {\em \mnras}, 324:249--256, June 2001.

\bibitem{Crone02}
M.~M. {Crone}, R.~E. {Schulte-Ladbeck}, L.~{Greggio}, and U.~{Hopp}.
\newblock {The Star Formation History of the Blue Compact Dwarf Galaxy UGCA
  290}.
\newblock {\em \apj}, 567:258--276, March 2002.

\bibitem{McQuinn09}
K.~B.~W. {McQuinn}, E.~D. {Skillman}, J.~M. {Cannon}, J.~J. {Dalcanton},
  A.~{Dolphin}, D.~{Stark}, and D.~{Weisz}.
\newblock {The True Durations of Starbursts: HST Observations of Three Nearby
  Dwarf Starburst Galaxies}.
\newblock {\em ArXiv e-prints}, January 2009.

\bibitem{Williams09}
B.~F. {Williams}, J.~J. {Dalcanton}, A.~C. {Seth}, D.~{Weisz}, A.~{Dolphin},
  E.~{Skillman}, J.~{Harris}, J.~{Holtzman}, L.~{Girardi}, R.~S. {de Jong},
  K.~{Olsen}, A.~{Cole}, C.~{Gallart}, S.~M. {Gogarten}, S.~L. {Hidalgo},
  M.~{Mateo}, K.~{Rosema}, P.~B. {Stetson}, and T.~{Quinn}.
\newblock {The ACS Nearby Galaxy Survey Treasury. I. The Star Formation History
  of the M81 Outer Disk}.
\newblock {\em \aj}, 137:419--430, January 2009.

\bibitem{Hunter04}
D.~A. {Hunter} and B.~G. {Elmegreen}.
\newblock {Star Formation Properties of a Large Sample of Irregular Galaxies}.
\newblock {\em \aj}, 128:2170--2205, November 2004.

\bibitem{Venn04}
K.~A. {Venn}, M.~{Irwin}, M.~D. {Shetrone}, C.~A. {Tout}, V.~{Hill}, and
  E.~{Tolstoy}.
\newblock {Stellar Chemical Signatures and Hierarchical Galaxy Formation}.
\newblock {\em \aj}, 128:1177--1195, September 2004.

\bibitem{Helmi06}
A.~{Helmi}, M.~J. {Irwin}, E.~{Tolstoy}, G.~{Battaglia}, V.~{Hill},
  P.~{Jablonka}, K.~{Venn}, M.~{Shetrone}, B.~{Letarte}, N.~{Arimoto},
  T.~{Abel}, P.~{Francois}, A.~{Kaufer}, F.~{Primas}, K.~{Sadakane}, and
  T.~{Szeifert}.
\newblock {A New View of the Dwarf Spheroidal Satellites of the Milky Way from
  VLT FLAMES: Where Are the Very Metal-poor Stars?}
\newblock {\em \apjl}, 651:L121--L124, November 2006.

\bibitem{Schoerck08}
T.~{Schoerck}, N.~{Christlieb}, J.~G. {Cohen}, T.~C. {Beers}, S.~{Shectman},
  I.~{Thompson}, A.~{McWilliam}, M.~S. {Bessell}, J.~E. {Norris},
  J.~{Melendez}, S.~{Solange Ramirez}, D.~{Haynes}, P.~{Cass}, M.~{Hartley},
  K.~{Russell}, F.~{Watson}, F.~. {Zickgraf}, B.~{Behnke}, C.~{Fechner},
  B.~{Fuhrmeister}, P.~S. {Barklem}, B.~{Edvardsson}, A.~{Frebel},
  L.~{Wisotzki}, and D.~{Reimers}.
\newblock {The stellar content of the Hamburg/ESO survey. V. The metallicity
  distribution function of the Galactic halo}.
\newblock {\em ArXiv e-prints}, September 2008.

\bibitem{Carollo07}
D.~{Carollo}, T.~C. {Beers}, Y.~S. {Lee}, M.~{Chiba}, J.~E. {Norris},
  R.~{Wilhelm}, T.~{Sivarani}, B.~{Marsteller}, J.~A. {Munn}, C.~A.~L.
  {Bailer-Jones}, P.~R. {Fiorentin}, and D.~G. {York}.
\newblock {Two stellar components in the halo of the Milky Way}.
\newblock {\em \nat}, 450:1020--1025, December 2007.

\bibitem{Gratton03}
R.~G. {Gratton}, E.~{Carretta}, S.~{Desidera}, S.~{Lucatello}, P.~{Mazzei}, and
  M.~{Barbieri}.
\newblock {Abundances for metal-poor stars with accurate parallaxes. II. alpha
  -elements in the halo}.
\newblock {\em \aap}, 406:131--140, July 2003.

\bibitem{Roederer09}
I.~U. {Roederer}.
\newblock {Chemical Inhomogeneities in the Milky Way Stellar Halo}.
\newblock {\em \aj}, 137:272--295, January 2009.

\bibitem{Kirby08}
E.~N. {Kirby}, J.~D. {Simon}, M.~{Geha}, P.~{Guhathakurta}, and A.~{Frebel}.
\newblock {Uncovering Extremely Metal-Poor Stars in the Milky Way's Ultrafaint
  Dwarf Spheroidal Satellite Galaxies}.
\newblock {\em \apjl}, 685:L43--L46, September 2008.

\bibitem{Koch08herc}
A.~{Koch}, A.~{McWilliam}, E.~K. {Grebel}, D.~B. {Zucker}, and V.~{Belokurov}.
\newblock {The Highly Unusual Chemical Composition of the Hercules Dwarf
  Spheroidal Galaxy}.
\newblock {\em \apjl}, 688:L13--L16, November 2008.

\bibitem{Frebel09}
A.~{Frebel}, J.~D. {Simon}, M.~{Geha}, and B.~{Willman}.
\newblock {High-Resolution Spectroscopy of Extremely Metal-Poor Stars in the
  Least Evolved Galaxies: Ursa Major II and Coma Berenices}.
\newblock {\em ArXiv e-prints}, February 2009.

\end{thebibliography}
\end{document}